\documentstyle[12pt]{article}

\advance \topmargin by -\headheight
\advance \topmargin by -\headsep
     
\evensidemargin \oddsidemargin
\begin{document}
\newcommand{\sect}[1]{\setcounter{equation}{0}\section{#1}}
\renewcommand{\theequation}{\thesection.\arabic{equation}}
\date{} 
\topmargin -.6in
\def\lab{\label}
\def\nonu{\nonumber}
\def\rf#1{(\ref{eq:#1})}
\def\lab#1{\label{eq:#1}} 
\def\br{\begin{eqnarray}}
\def\er{\end{eqnarray}}
\def\be{\begin{equation}}
\def\ee{\end{equation}}
\def\0{\nonumber}
\def\lb{\lbrack}
\def\rb{\rbrack}
\def\({\left(}
\def\){\right)}
\def\v{\vert}
\def\bv{\bigm\vert}
\def\lskip{\vskip\baselineskip\vskip-\parskip\noindent}
\relax
\newcommand{\nit}{\noindent}
\newcommand{\ct}[1]{\cite{#1}}
\newcommand{\bi}[1]{\bibitem{#1}}
\def\a{\alpha}
\def\b{\beta}
\def\ca{{\cal A}}
\def\cm{{\cal M}}
\def\cn{{\cal N}}
\def\cf{{\cal F}}
\def\d{\delta} 
\def\D{\Delta}
\def\eps{\epsilon}
\def\g{\gamma}
\def\G{\Gamma}
\def\grad{\nabla}
\def\h{ {1\over 2}  }
\def\hc{\hat{c}}
\def\hd{\hat{d}}
\def\hg{\hat{g}}
\def\hp{ {+{1\over 2}}  }
\def\hm{ {-{1\over 2}}  }
\def\k{\kappa}
\def\l{\lambda}
\def\L{\Lambda}
\def\lg{\langle}
\def\m{\mu}
\def\n{\nu}
\def\o{\over}
\def\om{\omega}
\def\O{\Omega}
\def\p{\phi}
\def\pa{\partial}
\def\pr{\prime}
\def\ra{\rightarrow}
\def\rh{\rho}
\def\rg{\rangle}
\def\s{\sigma}
\def\t{\tau}
\def\th{\theta}
\def\ti{\tilde}
\def\wti{\widetilde}
\def\inte{\int dx }
\def\xb{\bar{x}}
\def\yb{\bar{y}}
\def\tpsi{{\tilde \psi}}
\def\tchi{ {\tilde \chi}}
\def\btpsi{\bar {\tilde \psi}}
\def\btchi{\bar {\tilde \chi}}
\def\tr{\mathop{\rm tr}}
\def\Tr{\mathop{\rm Tr}}
\def\partder#1#2{{\partial #1\over\partial #2}}
\def\ds{{\cal D}_s}
\def\wtwo{{\wti W}_2}
\def\lie{{\cal G}}
\def\alie{{\widehat \lie}}
\def\dlie{{\cal G}^{\ast}}
\def\elie{{\widetilde \lie}}
\def\edlie{{\elie}^{\ast}}
\def\hlie{{\cal H}}
\def\wlie{{\widetilde \lie}}

\def\rlx{\relax\leavevmode}
\def\inbar{\vrule height1.5ex width.4pt depth0pt}
\def\IZ{\rlx\hbox{\sf Z\kern-.4em Z}}
\def\IR{\rlx\hbox{\rm I\kern-.18em R}}
\def\IC{\rlx\hbox{\,$\inbar\kern-.3em{\rm C}$}}
\def\one{\hbox{{1}\kern-.25em\hbox{l}}}

\def\PRL#1#2#3{{\sl Phys. Rev. Lett.} {\bf#1} (#2) #3}
\def\NPB#1#2#3{{\sl Nucl. Phys.} {\bf B#1} (#2) #3}
\def\NPBFS#1#2#3#4{{\sl Nucl. Phys.} {\bf B#2} [FS#1] (#3) #4}
\def\CMP#1#2#3{{\sl Commun. Math. Phys.} {\bf #1} (#2) #3}
\def\PRD#1#2#3{{\sl Phys. Rev.} {\bf D#1} (#2) #3}
\def\PLA#1#2#3{{\sl Phys. Lett.} {\bf #1A} (#2) #3}
\def\PLB#1#2#3{{\sl Phys. Lett.} {\bf #1B} (#2) #3}
\def\JMP#1#2#3{{\sl J. Math. Phys.} {\bf #1} (#2) #3}
\def\PTP#1#2#3{{\sl Prog. Theor. Phys.} {\bf #1} (#2) #3}
\def\SPTP#1#2#3{{\sl Suppl. Prog. Theor. Phys.} {\bf #1} (#2) #3}
\def\AoP#1#2#3{{\sl Ann. of Phys.} {\bf #1} (#2) #3}
\def\PNAS#1#2#3{{\sl Proc. Natl. Acad. Sci. USA} {\bf #1} (#2) #3}
\def\RMP#1#2#3{{\sl Rev. Mod. Phys.} {\bf #1} (#2) #3}
\def\PR#1#2#3{{\sl Phys. Reports} {\bf #1} (#2) #3}
\def\AoM#1#2#3{{\sl Ann. of Math.} {\bf #1} (#2) #3}
\def\UMN#1#2#3{{\sl Usp. Mat. Nauk} {\bf #1} (#2) #3}
\def\FAP#1#2#3{{\sl Funkt. Anal. Prilozheniya} {\bf #1} (#2) #3}
\def\FAaIA#1#2#3{{\sl Functional Analysis and Its Application} {\bf #1} (#2)
#3}
\def\BAMS#1#2#3{{\sl Bull. Am. Math. Soc.} {\bf #1} (#2) #3}
\def\TAMS#1#2#3{{\sl Trans. Am. Math. Soc.} {\bf #1} (#2) #3}
\def\InvM#1#2#3{{\sl Invent. Math.} {\bf #1} (#2) #3}
\def\LMP#1#2#3{{\sl Letters in Math. Phys.} {\bf #1} (#2) #3}
\def\IJMPA#1#2#3{{\sl Int. J. Mod. Phys.} {\bf A#1} (#2) #3}
\def\AdM#1#2#3{{\sl Advances in Math.} {\bf #1} (#2) #3}
\def\RMaP#1#2#3{{\sl Reports on Math. Phys.} {\bf #1} (#2) #3}
\def\IJM#1#2#3{{\sl Ill. J. Math.} {\bf #1} (#2) #3}
\def\APP#1#2#3{{\sl Acta Phys. Polon.} {\bf #1} (#2) #3}
\def\TMP#1#2#3{{\sl Theor. Mat. Phys.} {\bf #1} (#2) #3}
\def\JPA#1#2#3{{\sl J. Physics} {\bf A#1} (#2) #3}
\def\JSM#1#2#3{{\sl J. Soviet Math.} {\bf #1} (#2) #3}
\def\MPLA#1#2#3{{\sl Mod. Phys. Lett.} {\bf A#1} (#2) #3}
\def\JETP#1#2#3{{\sl Sov. Phys. JETP} {\bf #1} (#2) #3}
\def\JETPL#1#2#3{{\sl  Sov. Phys. JETP Lett.} {\bf #1} (#2) #3}
\def\PHSA#1#2#3{{\sl Physica} {\bf A#1} (#2) #3}
\newcommand\twomat[4]{\left(\begin{array}{cc}  
{#1} & {#2} \\ {#3} & {#4} \end{array} \right)}
\newcommand\twocol[2]{\left(\begin{array}{cc}  
{#1} \\ {#2} \end{array} \right)}
\newcommand\twovec[2]{\left(\begin{array}{cc}  
{#1} & {#2} \end{array} \right)}

\newcommand\threemat[9]{\left(\begin{array}{ccc}  
{#1} & {#2} & {#3}\\ {#4} & {#5} & {#6}\\ {#7} & {#8} & {#9} \end{array} \right)}
\newcommand\threecol[3]{\left(\begin{array}{ccc}  
{#1} \\ {#2} \\ {#3}\end{array} \right)}
\newcommand\threevec[3]{\left(\begin{array}{ccc}  
{#1} & {#2} & {#3}\end{array} \right)}

\newcommand\fourcol[4]{\left(\begin{array}{cccc}  
{#1} \\ {#2} \\ {#3} \\ {#4} \end{array} \right)}
\newcommand\fourvec[4]{\left(\begin{array}{cccc}  
{#1} & {#2} & {#3} & {#4} \end{array} \right)}

\begin{titlepage}
\vspace*{-2 cm}
\noindent


\vskip 1 cm
\begin{center}
{\Large\bf  Multicharged Dyonic Integrable Models}   \vglue 1  true cm
I. Cabrera-Carnero, { J.F. Gomes}, 
 { G.M. Sotkov} and { A.H. Zimerman}\\

\vspace{1 cm}

{\footnotesize Instituto de F\'\i sica Te\'orica - IFT/UNESP\footnote{cabrera@ift.unesp.br, 
jfg@ift.unesp.br, sotkov@ift.unesp.br, zimerman@ift.unesp.br}\\
Rua Pamplona 145\\
01405-900, S\~ao Paulo - SP, Brazil}\\

\vspace{1 cm}

\end{center}

\normalsize
\vskip 0.2cm

\begin{center}
{\large {\bf ABSTRACT}}\\
\end{center}
\noindent

We introduce and study new integrable models (IMs) of $A_n^{(1)}$-Non-Abelian Toda type which admit $U(1)\otimes U(1)$ charged
topological solitons.  They correspond to the symmetry breaking $SU(n+1) \rightarrow SU(2)\otimes SU(2)\otimes U(1)^{n-2}$
 and are conjectured to describe charged dyonic domain walls of $N=1$ $SU(n+1)$ SUSY gauge theory in large $n$ limit. 
 It is shown that this family of relativistic IMs corresponds to  the first 
 negative grade $q={-1}$ member of a dyonic hierarchy
 of generalized cKP type.  The explicit relation between the  1-soliton solutions (and the conserved charges as well )
  of the IMs of grades $q=-1$ and  $q=2$  is found.  The properties of the IMs corresponding to more general
 symmetry breaking $SU(n+1) \rightarrow SU(2)^{\otimes p}\otimes U(1)^{n-p}$  as well as IM with global $SU(2)$  symmetries
 are discussed.
\noindent

\vglue 1 true cm

\end{titlepage}

\sect{Introduction}

Topological solitons of two dimensional $N=2$ SUSY abelian affine 
$A_n$-Toda models are known to describe certain {\it
domain walls} (DW) solutions of four  dimensional  $N=1$ SUSY $SU(n+1)$ gauge theory \cite{kov}, \cite{wit1}. 
 According to  Witten's arguments
\cite{wit2}, for large $n$  this  SUSY gauge theory should coincide with the ordinary (non SUSY) QCD.  It is therefore
reasonable to relate the DW of the (non SUSY) $SU(n+1)$ Yang-Mills-Higgs type (YMH) model (for large $n$ and maximal
symmetry breaking  $SU(n+1) \rightarrow U(1)^n$ ) to the solitons of the corresponding {\it non supersymmetric } abelian
affine $A_n$-Toda theory for $n\rightarrow \infty$.  
Together with these simple DW (with no internal structure), the superstring description of
4-D gauge theories \cite{wit1}, \cite{review} provides examples of, say, $U(1)$, (or $U(1)^l, l\leq n$) charged DW of {\it dyonic }
type or DW that requires non maximal breaking of $SU(n+1)$ to $SU(l)\otimes U(1)^{n-l+1}$ \cite{gau}, etc.   Hence the
properties of the abelian affine Toda solitons (and their $N=2$ SUSY versions) are not sufficient to describe such dyonic
DW carrying internal charges - $U(1)^p$ or even non-abelian, say $SU(2)$, etc.   It is natural to expect appropriately
chosen non-abelian (NA) affine $A_n$-Toda models \cite{lez-sav} (and their $N=2$ supersymmetric 
extension) to be the main tool in the description
of DW with internal charges.  Recently    the $U(1)$
 charged
topological solitons  of certain singular affine NA-Toda models \cite{eletric}, \cite{dyonic} have been interpreted 
 as $U(1)$ charged dyonic DW of 4-D $SU(n+1)$ gauge theory.  It is expected that 
  such DW-2D soliton relation originates from  the fact that
 specific symmetry reduction (reflecting DW properties) of $SU(n+1)$ self-dual YM (SDYM) equations reproduces the equations
 of motion of certain two dimensional integrable models (2-D IMs).  
 The simplest
representative of such  dyonic  IM (with $p=1$, i.e. one  $U(1)$ symmetry) is given by the Lagrangian \cite{eletric}
\br
{\cal L}_{n}^{p=1} &=& {1\o 2} \sum_{i=1}^{n-1} 
k_{ij}\pa^{\mu} \varphi_i \pa_{\mu}\varphi_j + {{(\pa^{\mu} \psi \pa_{\mu}
\chi + \eps^{\mu \nu} \pa_{\mu} \psi \pa_{\nu}
 \chi )}\o {1+ \b^2 {{(n+1)}\o {2n}}\psi \chi e^{-\b \varphi_1}}}e^{-\b \varphi_1} - V
 \label{1.1}
 \er
with potential 
\br
V= {{\mu^2} \o {\b^2}}\( \sum_{i=1}^{n-1} e^{-\b k_{ij}\varphi_j} + e^{\b (\varphi_1 + 
\varphi_{n-1})}(1+ \b^2 \psi \chi e^{-\b
\varphi_1}) - n \)
\nonu
\er
where $k_{ij} = 2\d_{i,j} - \d_{i, j-1}- \d_{i, j+1}, i,j = 1, \cdots n-1$. It represents a pair of $U(1)$ charged fields
$\psi , \chi $ of Lund-Regge type \cite{lund} interacting with a set of neutral fields $\varphi_i$ of the $A_{n-1}$ abelian
affine Toda model.  
For imaginary coupling $\b \rightarrow i \b_0 $ its
potential has n - distinct vacua and manifests discrete $Z_2 \otimes Z_{n}$  and a global $U(1)$ symmetry. 
 As a consequence the 
 IM (\ref{1.1})
admits  both, $U(1)$ charged   and neutral, topological solitons.  The semiclassical
spectrum of such dyonic solitons turns out to be quite similar to the one of charged DW of $N=1$ $SU(n+1)$ Supersymmetric
Yang-Mills (SYM)
 theory for non-maximal breaking $SU(n+1) \rightarrow SU(2)\otimes U(1)^{n-1}$.

 Apart from the problem of exact identification of certain 2-D solitons as 4-D DWs \cite{dw} an important question to be
 addressed is whether one can construct in this manner also $U(1)^p, p \geq 2,$ multicharged DWs 
  and {\it nonabelian} (say $SU(2)$) DW as well.  Which are the 2-D IMs that admit topological solitons
 carrying $U(1)^p$ or $SU(2)$ charges? What are the main features of such solitons and how to derive their masses and
 charges?  Should we restrict ourselves in considering  relativistic 2-D IMs of NA-Toda type only,  or we also have to 
 include    certain nonrelativistic IMs, as for example the nonlinear Schroedinger type models? The present paper provides a
 partial answer to these questions.  It is devoted to the construction of both, {\it relativistic and nonrelativistic} 
 dyonic IMs of global $U(1)\otimes U(1)$, (i.e. $p=2$) symmetry and to the detailed study of their soliton spectra. 
  We also
 introduce and discuss its generalization to multicharged IMs with $U(1)^p (p=1, \cdots , n-1$) and $U(2)$ global
 symmetries.  The main purpose of the present research is  the construction of soliton  solutions of the following 
 $A_n^{(1)} (p=2)$ dyonic IM (and its nonrelativistic counterpart (\ref{5.7})):
 \br
 {\cal L}_{n}^{p=2} &=& {1\o 2} \sum_{i=1}^{n-2} 
k_{ij}\pa \varphi_i \bar \pa \varphi_j + {{1\o {\Delta}}} \( (1+ \b^2 {{n}\o {2(n-1)}}\psi_n \chi_n e^{-\b
\varphi_{n-2}})\bar \pa \psi_1 \pa \chi_1 e^{-\b \varphi_1} \right. \nonu \\
&+& \left. (1+ \b^2 {{n}\o {2(n-1)}}\psi_1 \chi_1 e^{-\b
\varphi_{1}})\bar \pa \psi_n \pa \chi_n e^{-\b \varphi_{n-2}}\right. \nonu \\
&+& \left. {{\b^2}\o {2(n-1)}} (\chi_1 \psi_n \bar \pa \psi_1 \pa \chi_n + 
\chi_n \psi_1 \bar \pa \psi_n \pa \chi_1)e^{-\b (\varphi_1 + \varphi_{n-2})} \)
 - V_n^{p=2}
 \label{1.2}
 \er
with potential
\br
V_n^{p=2} = {{\mu^2}\o {\b^2}}\( \sum_{i=1}^{n-2} e^{-\b k_{ij}\varphi_j} + e^{-\b (\varphi_1 + \varphi_{n-2})}(1+ \b^2 \psi_n
\chi_n e^{-\b\varphi_{n-2}}) (1+\b^2 \psi_1\chi_1 e^{-\b\varphi_1}) -n+1 \)
\nonu
\er
where $\varphi_0 = \varphi_{n-1} =0$ and 
\br
 \Delta = 1+ {{\b^2n}\o {2(n-1)}}(\psi_1\chi_1 e^{-\b\varphi_1}+ 
\psi_n \chi_n e^{-\b\varphi_{n-2}}) + {{\b^4 (n+1)}\o {4(n-1)}}\psi_1 \chi_1 \psi_n \chi_n e^{-\b (\varphi_1 +
\varphi_{n-2})}.
\nonu
\er
The Lagrangian (\ref{1.2}) is  invariant under global $U(1)\otimes U(1)$ transformations: $ 
\psi_a^{\pr} = e^{i \b^2 \eps_a}\psi_a $, 
$\chi_a^{\pr} = e^{-i \b^2 \eps_a}\chi_a, (a=1,n)$, $ \varphi_i^{\pr} = \varphi_i, $ 
(with charges $Q_1$ and $Q_n$) as well as under the    $Z_2\otimes Z_{n-1}$ discrete transformations.
For imaginary coupling $\b = i \b_0$, $V_n^{p=2}$ represents $(n-1)$ distinct zeros  and
 now the IM (\ref{1.2}) admits three different kinds of 1-soliton solutions, namely:
\begin{itemize}
\item Neutral (of $A_{n-2}$ abelian Toda type) solitons \cite{liao} 
\item $U(1)$ charged 1-solitons ($Q_1 \neq 0, Q_n =0$ and $Q_n \neq 0, Q_1 =0$) of $A_{n-1}(p=1)$ dyonic type
\cite{eletric}, \cite{dyonic}
\item New $U(1)\otimes U(1)$ charged 1-solitons ($Q_1 \neq 0, Q_n \neq 0$, $Q_{mag}\neq 0$) which appears as bound states 
of the above 
$(Q_1,0)$ and $(0,Q_n)$ 1-solitons.
\end{itemize}
From the Hamiltonian reduction point of view \cite{lez-sav} the $U(1)\otimes U(1)$ 
charged  IM (\ref{1.2}) appears as a natural
generalization of the $U(1)$ dyonic model (\ref{1.1}) including an extra pair of charged fields $\psi_n, \chi_n$ and a
rather complicated kinectic term ( involving quartic terms in the denominator $\Delta$) for $\psi_a, \chi_a, a=1,n$.   
Hence, the methods developed in the construction of  solitons of the IM (\ref{1.1}), \cite{eletric},\cite{dyonic}
can be extended to the case of multicharged IM (\ref{1.2}) with certain modifications concerning the genuine ($Q_1, Q_n$)
charged solitons.  An important {\it new feature} of the semiclassical spectra of these ($Q_1, Q_n$) 1-solitons (see Sect.
3.4):
\br
M_{j_{el}, j_{\varphi}}&=& {{4(n-1)\mu}\o{\b_0^2}}| \sin {{(4\pi j_{\varphi} - \b_0^2 j_{el})}\o {4(n-1)}}|, \; j_{el} = 0, \pm
1, \cdots \quad j_{\varphi} = 0, \pm 1, \cdots , \pm (n-2), \nonu \\
Q_{mag} &=& {{4\pi }\o {\b_0^2}}j_{\varphi}, \;\;\;\;  Q_1^{el} = \b_0^2 ({{j_{el}}\o 2} + \d_0 ), \; \; \;\;
Q_n^{el} = \b_0^2 (-{{j_{el}}\o 2} + \d_0 ), \quad \d_0 \in R
\label{1.5}
\er
is that their masses $M_{j_{el},j_{\varphi}}$ depend on the difference of the $U(1)$ charges  $Q_1^{el} - Q_n^{el} = \b^2_0
j_{el}$ only.  The latter is quantized semiclassically ($j_{el} \in Z$), while the sum of the charges  
$Q_1^{el} + Q_n^{el} = \b^2_0 \d_0$ remains continuous ($\d_0 \in R$).  The new kind of {\it dyonic effect} that 
 takes place for the model 
(\ref{1.2})  should be also mentioned.  Together with the standard shift  of $Q_a^{el}$ by the magnetic charge $Q_{mag}$
(see Sect. 2.5):
\br 
Q_a^{el} \rightarrow Q_a^{el} - {{\b_0^2 }\o {2\pi }}\nu^a j_{\varphi}, \quad a=1,n
\nonu
\er
which is 
induced by the topological term (\ref{2.40}) (present in the IM (\ref{1.1}) as well), one can add to the Lagrangian
(\ref{1.2})
a {\it new}
topological term 
\br
\d {\cal L}_{\theta}^{top} = {{\nu_{\theta}}\o {4 \pi^2}} 
\eps^{\mu \nu } \pa_{\mu} ln \( {{\chi_n}\o {\psi_n}}\) \pa_{\nu} ln \( {{\chi_1}\o {\psi_1}}\)
\nonu
\er
Its effect is another shift of the electric charges:
\br
Q_1^{el}  \rightarrow Q_1^{el} - {{\b_0^2 }\o {2\pi^2 }}\nu_{\theta} Q_{\theta}^{(1)}, \quad 
Q_n^{el}  \rightarrow Q_n^{el} + {{\b_0^2 }\o {2\pi^2 }}\nu_{\theta} Q_{\theta}^{(n)}
\nonu
\er
by the topological charges $Q_{\theta}^{(a)} = {{\pi}\o 2}j_{\theta}^a$.
 
Our dyonic IM (\ref{1.2}) is not the only candidate to describe the $U(1)\otimes U(1)$ charged DWs.  Answering the question
about the 
{\it nonrelativistic} 2-D IMs admiting  $U(1)\otimes U(1)$ charged topological solitons, we derive in Sect. 5, a new
family of IMs of constrained  Kadomtsev-Petviashvili (cKP) type .   They are represented by a system of integrable second order differential equations
(\ref{5.7}) for the set of 
fields $u_l(x,t), r_a(x,t), q_a(x,t) (l=1, \cdots , n-2; a=1,n)$, which is invariant under global 
$U(1)\otimes U(1)$ transformation ($\eps_a = $ const):
\br
q_a^{\pr} = e^{i\eps_a} q_a, \quad r_a^{\pr} = e^{-i\eps_a} r_a, \quad u_l^{\pr} = u_l
\nonu
\er
Although the nonrelativistic IM (\ref{5.7})  does not belong to the class of affine NA-Toda IMs, it shares (by
construction) the same algebraic structure underlining the relativistic IM (\ref{1.2}).  This fact allows us to establish
simple relations between the 1-soliton solutions of both models, as well as between the conserved charges characterizing
their soliton 
spectra, {similar to those found in  \cite{chodos} involving  charges of sine-Gordon and mKdV  and its
generalization to certain affine NA-Toda and cKP models  in ref. \cite{miramontes} }.
An interesting feature of IM (\ref{5.7}) is that the fields $u_l(x,t)$ itselves serve as densities of the  conserved
charges $\tilde Q_l = \int u_ldx$ called ``fluxes''.  They appear to be the nonrelativistic counterparts of 
the electric and
magnetic charges of the relativistic model (\ref{1.2}) as one can see from the following relation
\br
\tilde Q_l = {{1}\o {2(n-1)}}\( l (4\pi j_{\varphi} - Q_1^{el} + Q_n^{el}) + (n-1) Q_1^{el}\)
\label{carga1}
\er
The fluxes $\tilde Q_l$, together with the ``particle number'' charge $\tilde Q_0$ given by
\br
\tilde Q_0 = {{2(n-1)\kappa }\o {\g}} |\sin {{\tilde Q_1 -  \tilde Q_{n-2}}\o {2(n-3)}}|, 
\quad \kappa = {{\mu} \o {\b_0^2}} 
\label{carga2}
\er
determine the essential part of the 1-soliton spectrum of the nonrelativistic IM (\ref{5.8}). 
 An important characteristic  
 of the relation between the IMs (\ref{1.2}) and (\ref{5.7}) is that the masses of 
 the relativistic  1-solitons and the nonrelativistic ``particle'' number $\tilde Q_0$ are proportional,  
\br
\tilde Q_0 =  {1\o {2\g}} M^{rel}_{j_{el}, j_{\varphi}}
\nonu
\er
as one  can  verify by comparing eqns. (\ref{1.5}), (\ref{carga1}) and (\ref{carga2}).  It is worthwile to mention that
the existence of the nonrelativistic counterpart of the IM (\ref{1.2}) is not a specific  feature of 
 the particular IMs with
$U(1)\otimes U(1)$ symmetry we are considering.  Applying the methods presented in Sect. 5 one can construct an entire
hierarchy of nonrelativistic IMs (of higher order $q=2,3, \cdots $ differential equations) for each abelian or nonabelian
affine Toda model, including the simplest case of $U(1)$ dyonic IM (\ref{1.1}).

 This paper is organized as follows.
Sect. 2 contains together with the path integral derivation of the effective Lagrangian (\ref{1.2}) 
(and its ungauged versions
  (\ref{1.6}) and (\ref{1.8})), the corresponding zero curvature representations as well as the proof of their 
classical
integrability.  The construction of the new ($Q_1, Q_n$)-charged topological 1-soliton solutions
 of IM (\ref{1.2}) is 
approached in
two different manners.  In Sect. 3   we present the explicit 1-soliton solutions obtained by applying the 
vacua B\"acklund
transformation, i.e. by constructing and solving the corresponding soliton first order  differential equations.  
The derivation of the
semiclassical 1-soliton spectrum is given in Sect. 3.4.  Sect. 4 is devoted to the vertex operator 
($\tau$ function) method.
 Following the standard ``Heisenberg subalgebra'' constructions, 
  the explicit  form   of the four different 
 type (``neutral'' and ``charged'') soliton vertex operators is obtained. 
  For example, the ($Q_1, Q_n$)-charged 1-soliton is 
 represented by specific
 composite 4- vertex operator.  The corresponding $G_0$ -
  $\tau$-functions $\tau_0, \tau_j, j=1, \cdots n-2, \;\; \tau_a^{\pm}, \;
 a=1,n $
are realized as vacuum expectation values of appropriate soliton vertex operators. 
 It turns out that the cumbersome
algebraic manipulation to be performed in the derivation of the explicit form of the 
$\tau$-functions is 
simpler in the case of the nonrelativistic generalized constrained KP hierarchy models 
constructed in Sect.5.
They have  the same algebraic structure $A_n^{(1)}, Q, \lie_0, \eps_+$ (see Sect. 2.1) as our relativistic 
dyonic model (\ref{1.2}) but with  the negative grade $q=-1$ constant element 
$\eps_-$  replaced by a constant element 
 of grade  $q=2$. 
Another difference lies in the zero grade  parameters (i.e. the physical fields) of the 
nonrelativistic model.  They  are directly
related to the current $(\pa_x g_0)g_0^{-1}$, while for the relativistic theory they 
parametrize the group element $g_0 \in G_0$.  The precise statement presented in Sect. 5 is that the relativistic 
 dyonic IM (\ref{1.2}) appears as the first negative flow 
  of the new dyonic  hierarchy $A_n^{(1)}(p=2, q)$ given by eqns. (\ref{5.7}).  
It is a generalization of the well known
relation between the sine-Gordon and MKdV \cite{marian}, Lund-Regge and Non linear Schroedinger models \cite{aratyn2}, 
the
generalized $A_n^{(1)}(p=1)$ dyonic models (\ref{1.1}) and the constrained cKP hierarchy \cite{cabrera} and their
 supersymmetric versions \cite{emil}. 
 
Section 6 contains preliminary discussion of the algebraic structure and the effective actions of more general multicharged
dyonic models, as for example the $SL(3)$ generalization of the Lund-Regge model and 
 $SL(2)^{\otimes p} \otimes U(1)^{n-p}$ models ($p=1,2, \cdots n-1$).

\sect{Dyonic $A_n^{(1)}(p=2)$ IM as Gauged Two-loop WZW Model}

\subsection{Algebraic Structure of the Hamiltonian Reduction}

The gauged $G_0/ {G_0^0}$- WZW model 
 based on finite dimensional Lie algebra $\lie $ is given by
\br
S_{G/H}(g,A,\bar{A})&=&S_{WZW}(g)
\nonumber
\\
&-&\frac{k}{2\pi}\int d^2x Tr\( A(\bar{\partial}gg^{-1}-\epsilon_{+})
+\bar{A}(g^{-1}\partial g-\epsilon_{-})+Ag\bar{A}g^{-1} + A_0 \bar A_0 \) \nonu \\
\label{2.1}
\er
where $A = A_- + A_0, \; \bar A = \bar A_+ + \bar A_0$, 
$A_- \in  \lie_{<}, \; \bar A_+ \in { \lie}_{> },
\; A_0, \bar A_0 \in \lie_0^0, \; g \in G$  and $G_0^0 = \exp (\lie_0^0 )$. As  is well known   
  all the conformal field
theories: abelian and non-abelian Toda models, parafermions, \cite{ora}, \cite{annals}, etc.
 can be represented as
appropriate gauged WZW models.
The subgroups $H_{<}, { H}_{> } \subset G$  are generated by   positive/negative grade subalgebras $\lie_{<} $ and
$\lie_{>}$ respectively,
 according to the appropriately chosen grading operator $Q$ \footnote{see \cite{lez-sav} and  app. A of \cite{annals}
for our algebraic notation}, decomposing the Lie algebra $\lie$ in graded subspaces,
\br
Q&=& \sum_{i=1}^{n} s_i {{2\l_i \cdot H } \o {\a_i^2}},\quad  \quad [ Q, \lie_l] = l \lie_l, \; s_i \in Z, \nonu \\
\lie &=&\oplus \lie_l, \quad \quad [\lie_l, \lie_k ] \subset \lie_{l+k}, \; l,k= 0, \pm 1, \cdots 
\label{2.2}
\er
where $\l_i$ denote the fundamental weights of $\lie$.
The constant elements $\eps_{\pm} \in \lie_{\pm 1}$ of grade $\pm 1$, (say $\eps_{\pm } = \sum_{i=1}^{n} \mu_i^{\pm} E_{\pm
\a_i}^{(0)}$) indicate those  WZW currents \footnote{The WZW currents in question have the from $J_{\b}  =Tr (E_{\b}
g_0^{-1}\pa g_0 )$ and $ \bar J_{\b}= Tr (E_{\b} \bar \pa g_0 g_0^{-1})$ } that are taken to be 
 constants,  $i.e. J_{\a_i} = \mu_i^{+}, \bar J_{-\a_i} = \mu_i^{-}$.  In fact, $Q$ is providing $G$ with
specific Gauss-like decomposition $G= H_{<}G_0 H_{>}$ with the important difference that the zero grade subgroup $G_0$
(parametrized by physical fields, say $\varphi_i, \psi_a, \chi_a$ in eqn. (\ref{1.2})) is in general non-abelian.  Each
conformal model is specified by the choice  $\lie, Q, \eps_{\pm}$. 

 Similar construction takes place for $\hat
{\lie }$, an infinite dimensional Affine (Kac-Moody ) algebra \cite{lez-sav}. 
 They give rise to a larger class of (non-conformal )
integrable models which are in fact  integrable deformations of the conformal 
models described above.  The main ingredient
is now the so called {\it two-loop} gauged WZW model   
(i.e., the  model associated to an affine algebra $\hat {\lie}$)  \cite{aratyn} described 
formally by the same action (\ref{2.1})
with an important difference that all $\hat {\lie},\hat {H}_{<}, H_{>} $ are infinite 
dimensional (and so is $\hat g$ and $A,
\bar A$) \cite{dyonic}.  The zero grade subgroup  $G_0 = G/H$ is however chosen to be {\it finite dimensional}. 
 The grading operator $ Q$ is now 
given by 
\br
 Q = \tilde h d + \sum_{i=1}^{n} s_i {{2\l_i \cdot H } \o {\a_i^2}},\quad  [d, E_a^{(n)}] = n E_a^{(n)}, \quad
[d, H_i^{(n)}] = n H_i^{(n)}, \; n\in Z
\label{2.3}
\er
where $\tilde h$ is a positive integer to be specified below.  The major difference from the conformal models is in the
constant grade  element $\eps_{\pm} $ that may include extra affine 
generators of the form $E_{\mp \b}^{(\pm 1)}= \l^{\pm 1}E_{\mp \b}^{(0)}$. For instance
in the abelian $A_n$ affine Toda models, where $Q$ is given by the {\it principal} gradation we have,
\br
Q&=& (n+1)d+ \sum_{i=1}^{n}  {{2\l_i \cdot H } \o {\a_i^2}}, \quad \hat {\eps}_{\pm} = \sum_{i=1}^{n}E_{\pm \a_i}^{(0)} + 
E_{\mp (\a_1 + \cdots + \a_n) }^{(\mp 1)}
\label{2.4}
\er
Since in this case the invariant subalgebra
 $\lie_0^0 \in \lie_0$ defined as $[ \lie_0^0, \eps_{\pm}] =0$ is trivial (empty), these models have   no 
 continuous (local or global) symmetries. Their Lagrangians are 
 invariant under the discrete
permutation group $S_n$ (Weyl group of $A_n$) \cite{liao}.  

On the other hand, the so called {\it homogeneous} gradations   
\br
Q=d, \quad   {\eps}_{\pm} = \sum_{i=1}^{n} \mu_i h_i^{(\pm 1)}, \quad \lie_0^0 = \{ h_i^{(0)}, i=1, \cdots n \} = U(1)^n
\label{2.5}
\er
define a class of IMs of electric type with maximal number $n$ of local $U(1)^n$ symmetries and potential 
$V = Tr \(  {\eps}_{+}g_0  {\eps}_{-}g_0^{-1}\) $ which has no nontrivial zeros. 
 One can further gauge fix axial or
vector  local symmetries by considering the $G_0 / G_0^0$-gauged WZW model,
\br
S(g_0,{A}_{0},\bar{A}_{0} ) &=&  S_{WZW}(g_0)- 
 {{k\o {2\pi}}} \int Tr \( \eps_+ g_0  \eps_- g_0^{-1}\) d^2x\nonu \\   
  &-&{{k\o {2\pi}}}\int Tr\( \pm  A_{0}\bar{\partial}g_0
g_0^{-1} + \bar{A}_{0}g_0^{-1}\partial g_0
\pm  A_{0}g_0\bar{A}_{0}g_0^{-1} + A_{0}\bar{A}_{0} \)d^2x \nonu \\
\label{2.6}
\er
where the $\pm $ signs correspond to axial or vector gaugings respectively, $g_0^{\pr} = \a_0 g_0 \a_0^{\pr}, \;
A_0^{\pr} = A_0 - \a_0^{-1} \pa \a_0, \; \bar A_0^{\pr} = \bar A_0 - \bar  \pa \a_0^{\pr}(\a_0^{\pr})^{-1}$ and $\a_0^{\pr} =
\a_0(z, \bar z) \in \lie_0^0$ for axial and $\a_0^{\pr} =
\a_0^{-1}(z, \bar z) \in \lie_0^0$ for vector cases.  Since $G_0/U(1)^l$, $0\leq l \leq n$ is 
non abelian and define
homogeneous space,  the IMs of this type are known as Homogeneous  sine-Gordon models \cite{pousa}. 
They are  generalization
of the usual complex SG model (i.e., $n=1$, known as  Lund-Regge \cite{lund}).  Note that the corresponding 
effective action  manifest global
$U(1)^l$ symmetries.

The two examples illustrate the extreme cases, the  {\it principal} gradation, with no continuous symmetries ($\lie_0^0 = \emptyset$) but
maximal number of topological charges and 
the  {\it homogeneous} gradation with maximal number of local (or global ) $U(1)^l$ symmetries, but no topological charges.  An
interesting class of IMs arises when the following intermediate (dyonic type) gradation is considered
\br
Q =  \tilde h d 
+ \sum_{i\neq a_1, a_2, \cdots, a_p}  {{2\l_i \cdot H } \o {\a_i^2}}, 
\quad \hat \eps_{\pm} = \sum_{i\neq a_1, a_2, \cdots, a_p}^{n} E_{\pm \a_i}^{(0)} + 
\sum_{s} \mu_s^{(1)}E_{\mp \rho_s}^{(\pm 1)}
\label{2.7}
\er
where $a = (a_1, \cdots , a_p)$ denote the corresponding  set of ommited  fundamental weights $\l_a $  and simple roots
$\a_a$;   
$\tilde h =1+\sum_{i\neq a_1, a_2, \cdots, a_p}  {{2\l_i \cdot \rho_s } \o {\a_i^2}}$ and  
$\rho_s$ are appropriate composite roots of $A_n^{(1)}$ to specified below.   
The main feature of such dyonic IM is that
(i) only part of the Cartan subalgebra, say $U(1)^l, l \leq p \leq n-1$ generate local 
(or global ) Noether symmetries and (ii) their
potential for imaginary coupling constant and appropriate choice of $ \eps_{\pm}$ 
has $(n-l+1)$ distinct zeros and $Z_{n-l+1}$
as an asymptotic (discrete) symmetry group.
  Hence they admit soliton solutions carrying both $U(1)^l$ electric and topological
(magnetic)  $Z_{n-l+1}$ charges.  The simplest  is the $A_n^{(1)}(l=p=1)$ dyonic IM defined by 
\br
Q = nd +  \sum_{i =2}^{n}  {{2\l_i \cdot H } \o {\a_i^2}}, 
\quad \hat \eps_{\pm} = \sum_{i=2}^{n} \mu_i E_{\pm \a_i}^{(0)} + 
E_{\mp (\a_2 +\cdots + \a_n)}^{(\pm 1)}
\label{2.8}
\er
such that $\lie_0 = SL(2)\otimes U(1)^{n-1}$ and $\lie_0^0 = \{\l_1 \cdot H \}$.  It has been introduced and 
systematically studied in refs. 
 \cite{eletric}, \cite{dyonic}. 
 The problem we address here concerns the construction and the study of the properties of
solitons for multicharged ($p>1$) IMs.  Among the vast variety of families of p-multicharged dyonic IM we choose one simple
representative characterized by the form of the zero grade  subgroup \footnote{ Note that for given p, one has several
inequivalent choices for $s_i$ and $\mu_i, \mu_s^{(1)}$ that leads to different $\lie_0$ for example, for $p=2$, we have
$\lie_0^{(1)} = SL(2)\otimes SL(2)\otimes U(1)^{n-2}$ and $\lie_0^{(2)} = SL(3)\otimes U(1)^{n-2}$
corresponding to $Q_1 = (n-1) d + \sum_{i=2}^{n-1} \l_i \cdot H$ and 
 $Q_2 = (n-1) d + \sum_{i=3}^{n} \l_i \cdot H$, respectively.}
$\lie_0 = SL(2)^p \otimes U(1)^{n-p}$.  Although the methods we are employing in the construction of the p-charged dyonic
models  are universal (and valid for all $1\leq p \leq n-1$)  
we restrict ourselves for simplicity to consider the case $p=l=2$ only, 
\br
Q = (n-1)d +  \sum_{i =2}^{n-1}  \l_i \cdot H  , 
\quad  \eps_{\pm} = \mu \(\sum_{i=2}^{n-1} E_{\pm \a_i}^{(0)} + 
E_{\mp (\a_2 +\cdots + \a_{n-1})}^{(\pm 1)}\), 
\label{2.9}
\er
and $ \lie_0^0 =\{ \l_1\cdot H, \l_n\cdot H \}$.
The specific choice  of $s_i$ and $\mu_i$, namely, $s= (0, 1,1, \cdots 1,0)$ in eqn. (\ref{2.9}) is not a strong 
 restriction.
 Following the discussion of Sect. 2.7 of ref. \cite{eletric} (concerning the Weyl families of $A_n(p=1)$ IMs) one
 can easily verify that any other choices\footnote{except the particular case $\tilde s =  (0, 0,1, \cdots 1,1)$
 and $Q=Q_2$.  Due to the fact that now $\a_1 \cdot \a_2 \neq 0 $ ($\a_1 \cdot \a_j = 0 $ for all other cases),  this model
 is not Weyl equivalent to the original one, since it does not preserve the structure of $\lie_0$} of $s$, preserving the structure of $\lie_0$, 
  say $\tilde s =  (0, 1,1, \cdots 1,0,1)$ is related to (\ref{2.9})
 by certain Weyl transformation.  The actions for both models ($s$ and $\tilde s$) are related by change of variables
 similar to the one of $A_n(p=1)$ case presented in ref. \cite{eletric}.  Our symmetric choice (\ref{2.9}),
 represents however certain advantages in the construction of $U(1)\otimes U(1)$ topological solitons in Sect. 3 and
 4 below.

\subsection{Effective Dyonic Actions}

Given the graded structure (\ref{2.9}) of the affine group $A_n^{(1)}$, the elements of the zero grade subgroup are
parametrized as 
\br
g_0 = e^{(  \btchi_1 E_{-\a_1}^{(0)} + \btchi_n E_{-\a_n}^{(0)})}
e^{(R_1 \l_1\cdot H^{(0)} + R_n \l_n\cdot H^{(0)} + \sum_{i=1}^{n-2} \varphi_{i} h_{i+1}^{(0)})}
e^{( \btpsi_1 E_{\a_1}^{(0)} + \btpsi_n E_{\a_n}^{(0)})}
\label{2.10}
\er      
and the corresponding positive/negative grade subgroups $H_{<}$ and $H_{>}$contain all nonphysical fields and are generated by the
positive/negative grade $\lie_{>}\;  (\lie_{<})$ 
generators of the  $A_n^{(1)}$ affine
algebra with respect to $Q$ given in (\ref{2.7}).   As we have explained before, the starting point in the derivation of the $A_n^{(1)}(p=2)$ dyonic IM action is the
gauged ($H_{<},H_{>}$ invariant ) $G_0 = G/ H$ two-loop WZW model given by eqns. (\ref{2.1}) 
without the $A_0 \bar A_0$ term.
The result of the formal functional integration yields the following  $U(1)\otimes U(1)$-ungauged effective action
\br
S_{G_0}(g_0) = S_{WZW}(g_0) - {{k}\o {2\pi}} \int dz d\bar z Tr \(\eps_+ g_0 \eps_- g_0^{-1}\)
\label{2.11} 
\er
$ g_0 \in G_0 = SL(2)\otimes SL(2)\otimes U(1)^{n-2}$. Taking into account the parametrization (\ref{2.10}), 
it leads to the following action ($\b^2 = - {{2\pi}\o {k}}$):
\br
{\cal L}_{G_0} &=& {1\o 2} \sum_{i=1}^{n-2} k_{ij}\pa \varphi_i \bar \pa \varphi_j + 
\pa  \btchi_1 \bar \pa  \btpsi_1 e^{ \b( R_1-\varphi_1)} + 
\pa  \btchi_n \bar \pa \btpsi_n e^{ \b( R_n-\varphi_{n-2})} \nonu \\ 
&+& 
{{1}\o {2(n+1)}} \( n\pa R_1 \bar \pa R_1 + n\pa R_n \bar \pa R_n +
 \pa R_1 \bar \pa R_n + \pa R_n \bar \pa R_1 \) -V_0
\label{1.6}
\er
where 
\br
V_0 = {{\mu^2}\o {\b^2}}\( \sum_{i=1}^{n-2} e^{-\b k_{ij} \varphi_j}+ e^{\b (\varphi_1 + \varphi_{n-2})} (1 + 
\b^2  \btpsi_1  \btchi_1 e^{\b (R_1 -
\varphi_{1})})(1 + \b^2 \btpsi_n  \btchi_n e^{\b (R_n -
\varphi_{n-2})})-n+1\). 
\nonu
\er
As we have mentioned, the fields $\varphi_j, j=1, \cdots , n-2, R_a,  \btpsi_a,  \btchi_a, \; (a=1,n)$ parametrize 
the nonabelian zero
grade subgroup $G_0 $. The ungauged IM (\ref{1.6}) represents an integrable
deformation of the $G_0$-WZW model with the potential $V_0$.
 An important feature of (\ref{1.6}) is that it is invariant
under chiral $U(1)\otimes U(1)$ local gauge transformations $w_a(z), \bar w_a(\bar z) \; $:
\br
\varphi_i^{\pr} = \varphi_i, \quad R_a^{\pr} = R_a + w_a + \bar w_a, \quad  \btpsi_a^{\pr} = e^{-\b w_a} \btpsi_a, 
\quad \btchi_a^{\pr} = e^{-\b \bar w_a} \btchi_a
\label{1.7}
\er 
reflecting the fact that $G_0$ has two dimensional $\eps_{\pm}$-invariant subgroup  $\lie_0^0 = 
\{ \l_1 \cdot H, \l_n \cdot H \}$, such that 
$[\lie_0^0, \eps_{\pm}] =0$.

We further consider an  intermediate $A_n^{(1)}(p=2)$ dyonic IM  with one local $U(1)$ and one global (axial) $U(1)$
 (generated by
$(\l_1+\l_n)\cdot H$ ) 
symmetry.  It  is obtained from eqns. (\ref{2.6}) with $A_0= a_0(\l_1+\l_n)\cdot H, 
 \bar A_0= \bar a_0(\l_1+\l_n)\cdot H$ \footnote{$a_0(z, \bar z), \bar a_0(z, \bar z)$  are 
 arbitrary functions parametrizing the
 auxiliary fields} 
 and zero grade factor 
group element $g_0^f $ is taken in the form  
\br
g_0^f = e^{( \bar \chi_1 E_{-\a_1}^{(0)} +\bar \chi_n E_{-\a_n}^{(0)})}
e^{(\bar R (\l_1 -\l_n)\cdot H^{(0)} + \sum_{i=1}^{n-2} \varphi_{i} h_{i+1}^{(0)})}
e^{( \bar \psi_1 E_{\a_1}^{(0)} +\bar \psi_n E_{\a_n}^{(0)})}
\label{2.10f}
\er 
where $\bar \psi_a =  \btpsi_a e^{{1\o 4}( R_1+R_n)}, \; \bar \chi_a =  \btchi_a e^{{1\o 4}(R_1+R_n)}$ and 
$\bar R = {1\o 2} (R_1-R_n)$. 
 In order to construct its action,
  we first  consider the partition function for this  intermediate  $A_n(p=2)$ model,
\br
{\cal Z}_n^{p=2}= \int Dg_0 DA_0 D\bar A_0 \exp (-S^{ax}_{G_o/G_0^0}(g_0, A_0, \bar A_0))
\nonu
\er
where the action $S^{ax}_{G_0/G_0^0}$ of the axial gauge fixed model is given by (\ref{2.6}) with upper signs.
Integrating over the corresponding  auxiliary fields $A_0, \bar A_0$, we derive the effective action representing this
IM,
 \br
{\cal L}_{n, inter}^{p=2} &=& {{(n-1)}\o{(n+1)}} \pa \bar R \bar \pa \bar R  
+{1\o 2} \sum_{i=1}^{n-2} k_{ij}\pa \varphi_i \bar \pa \varphi_j  \nonu \\
&+&{1\o
{\Delta_0 }} \( (1+ {{\b^2}\o 4} \bar \psi_n \bar \chi_n e^{-\b (\varphi_{n-2}+  \bar R)})
\bar \pa \bar \psi_1 \pa \bar \chi_1 e^{\b (\bar R -  \varphi_1)}  \right. \nonu \\ 
&+ & \left.
 (1+ {{\b^2}\o 4} \bar \psi_1 \bar \chi_1 e^{-\b (\varphi_{1}-  \bar R)})
\bar \pa \bar \psi_n \pa \bar \chi_n e^{-\b (\bar R +  \varphi_{n-2})} \right. \nonu \\
&-&\left.
 {{\b^2}\o 4}(\bar \psi_n \bar \chi_1 \bar \pa \bar \psi_1 \pa \bar  \chi_n + 
\bar \psi_1 \bar \chi_n \bar \pa \bar \psi_n \pa \bar  \chi_1)e^{-\b (\varphi_1 + \varphi_{n-2})} \) - V_{inter}
\label{1.8}
\er
where
\br
 V_{inter} = {{\mu^2}\o {\b^2}}\( \sum_{i=1}^{n-2} e^{-\b k_{ij} \varphi_j } + 
e^{\b (\varphi_1+\varphi_{n-2})}(1+ \b^2 \bar \psi_1 \bar \chi_1 e^{\b (\bar R -  \varphi_1)})
(1+ \b^2\bar \psi_n \bar \chi_n e^{-\b( \bar R +  \varphi_{n-2})}) -n+1\), \nonu \\
\nonu
\er
and $\Delta_0 = 1 + {{\b^2}\o 4} (\bar \psi_1 \bar \chi_1 e^{\b (\bar R -  \varphi_1)} + 
\bar \psi_n \bar \chi_n e^{-\b( \bar R +  \varphi_{n-2})})$.
Note that the denominator $\Delta_0$ in (\ref{1.8}) is quadratic in $\psi_a, \chi_a$, i.e., similar to $A_n(p=1)$ IM
(\ref{1.1}),  but
with an extra pair of fields $\psi_n, \chi_n$  and an additional chiral local $U(1)$ gauge symmetry:
\br
\bar \psi_1^{\pr} &=& e^{-\b w}\bar \psi_1, \quad \bar \chi_1^{\pr} = e^{-\b \bar w}\bar \chi_1, \quad 
\bar \psi_n^{\pr} = e^{\b w}\bar \psi_n, \quad \bar \chi_n^{\pr} = e^{\b \bar w}\bar \chi_n, \nonu \\
\bar R^{\pr} &=& \bar R +  w + \bar w, \quad \varphi_l^{\pr}
= \varphi_l.
\nonu
\er

Finally, in order to derive the action for the completely gauged IM (\ref{1.2})
we substitute $A_0= a_{01}\l_1\cdot H + a_{0n}\l_n\cdot H, 
\bar A_0= \bar a_{01}\l_1\cdot H + \bar a_{0n}\l_n\cdot H$ in eqn. (\ref{2.6}) with the upper signs (axial gauging)
and we take $g_0^f$ in the form
\br
g_0^f = e^{(  \chi_1 E_{-\a_1}^{(0)} + \chi_n E_{-\a_n}^{(0)})}
e^{ \sum_{i=1}^{n-2} \varphi_{i} h_{i+1}^{(0)}}
e^{(  \psi_1 E_{\a_1}^{(0)} + \psi_n E_{\a_n}^{(0)})}
\label{2.10b}
\er
where $ \psi_a =  \btpsi_a e^{{1\o 2} R_a}, \;  \chi_a =  \btchi_a e^{{1\o 2} R_a}$.
Following the procedure described above (i.e. integrating over $A_0, \bar A_0$, etc), we obtain the effective Lagrangian
given by eqn. (\ref{1.2}) (see ref. \cite{cabrera} for details).

It is worthwhile to mention that the conceptual difference between  the completely gauged IM (\ref{1.2}) 
and the ungauged (\ref{1.6}) and the
intermediate models  (\ref{1.8}) is that the action (\ref{1.2})  has no local symmetries.  Instead it has 
  two global  symmetries 
described by transformations (\ref{2.31}).

Similarly to the axial and vector $A_n^{(1)}(p=1)$ dyonic IM \cite{eletric}, \cite{dyonic}, applying the method developed in ref.
\cite{tdual} one can derive the CPT-invariant vector model (\ref{1.2}) starting from eqns. (\ref{2.6}) with lower signs (-).
One can also consider the mixed case  by axial gauging one $U(1)$ and vector gauging the other, but this topic is out of
the purpose of the present paper.

\subsection{Leznov-Saveliev equations and the Zero Curvature Representation}

 The equations of motion for the models (\ref{1.2}), (\ref{1.6}) and (\ref{1.8}) can be written in the following matrix
 Leznov-Saveliev form \cite{lez-sav}
 \be 
\bar \pa (g_0^{-1} \pa g_0) + [ {\eps_-}, g_0^{-1}  {\eps_+} g_0] =0, 
\quad \pa (\bar \pa g_0 g_0^{-1} ) - [ {\eps_+}, g_0 {\eps_-} g_0^{-1}] =0
\label{2.16}
\ee
The local $U(1)\otimes U(1)$ symmetries (\ref{1.7})  reflect the fact that the traces of the above equations with
$\l_a\cdot H$ vanish, i.e., 
$\bar \pa  J_{\l_a \cdot H^{(0)}} =\pa  \bar J_{\l_a \cdot H^{(0)}} = 0$, where
\br
  J_{\l_a \cdot H^{(0)}} =
  Tr(g_0^{-1} \pa g_0 \l_a\cdot H^{(0)}),\quad 
  \bar J_{\l_a \cdot H^{(0)}} = 
  Tr(\bar \pa g_0 g_0^{-1}\l_a \cdot H^{(0)} ), \; a=1,n
\label{2.17}
\er
due to the $\lie_0^0$ property that $[\l_a \cdot H , \eps_{\pm}]=0, a=1,n$.  The axial (vector) gauge 
fixing of these $U(1)\otimes
U(1)$ symmetries is known to be equivalent to implementing the constraints
$J_{\l_a \cdot H^{(0)}} =\bar J_{\l_a \cdot H^{(0)}} = 0$, i.e., to the system of equations 
 \br
 n\pa R_1 (1+ \b^2 {{n+1}\o {2n}} \psi_1 \chi_1 e^{-\b\varphi_1}) + \pa R_n &=& (n+1)\psi_1 \pa \chi_1 e^{-\b \varphi_1},
  \nonu \\
 \pa R_1 + n \pa R_n (1+ \b^2 {{n+1}\o {2n}} \psi_n \chi_n e^{-\b\varphi_{n-2}})&=& (n+1)\psi_n \pa \chi_n e^{-\b
 \varphi_{n-2}}\nonu \\
 n \bar \pa R_1 (1+ \b^2 {{n+1}\o {2n}} \psi_1 \chi_1 e^{-\b\varphi_1}) + \bar \pa R_n &=& (n+1)\chi_1 \bar \pa \psi_1 e^{-\b \varphi_1},
  \nonu \\
\bar  \pa R_1 + n \bar \pa R_n (1+ \b^2 {{n+1}\o {2n}} \psi_n \chi_n e^{-\b\varphi_{n-2}})&=& (n+1)\chi_n \bar \pa \psi_n e^{-\b
 \varphi_{n-2}}\nonu
 \er
 The diagonalization of the above system leads to the definition of the auxiliary {\it nonlocal } fields $R_a, a=1,n$ which
are going to play an important role in the construction of the 1-soliton solutions in the next sections, 
\br
(n-1)\Delta \pa R_1&=& n \b \psi_1 \pa \chi_1 (1+ \b^2 {{n+1}\o {2n}} \psi_n \chi_n e^{-\b\varphi_{n-2}})e^{-\b \varphi_1}
 - \b \psi_n \pa
\chi_n e^{-\b \varphi_{n-2}}
 \nonu \\
(n-1)\Delta \pa R_n&=& n \b \psi_n \pa \chi_n (1+ \b^2 {{n+1}\o {2n}} \psi_1\chi_1 e^{-\b\varphi_{1}})e^{-\b \varphi_{n-2}}
 - \b \psi_1 \pa
\chi_1 e^{-\b \varphi_{1}}
 \nonu \\
(n-1)\Delta \bar \pa R_1&=& n \b \chi_1 \bar \pa \psi_1 (1+ 
\b^2 {{n+1}\o {2n}} \psi_n \chi_n e^{-\b\varphi_{n-2}})e^{-\b \varphi_1}
 - \b \chi_n \bar \pa
\psi_n e^{-\b \varphi_{n-2}}
 \nonu \\
(n-1)\Delta \bar \pa R_n&=& n \b \chi_n \bar \pa \psi_n (1+ 
\b^2 {{n+1}\o {2n}} \psi_1\chi_1 e^{-\b\varphi_{1}})e^{-\b \varphi_{n-2}} 
- \b \chi_1 \bar \pa
\psi_1 e^{-\b \varphi_{1}}
\label{2.18}
\er
where
\br
 \Delta   &=&  1 + {{n\b^2}\o {2(n-1)}} \( \psi_1 \chi_1 e^{-\b \varphi_1} 
 + \psi_{n} \chi_{n} e^{-\b \varphi_{n-2})}\)  
  +
 {{(n+1)\b^4}\o {4(n-1)}}  \psi_1 \chi_1 \psi_{n} \chi_{n} e^{ -\b (\varphi_1  +\varphi_{n-2} )  } \nonu 
\er
and  $\lambda_1 \cdot \lambda_j = {{n+1-j}\o {n+1}},\;\;
\lambda_{n} \cdot \lambda_j = {{j}\o {n+1}},\;\; j=1, \cdots , n-1 $.

The zero curvature representation 
\br
\pa \bar {\cal A}- \bar \pa  {\cal A} - [ {\cal A}, \bar {\cal A}]=0,
\label{zcc}
\er
of the $A_n^{(1)}(p=2)$  equations (\ref{2.16}) is known to be  the crucial ingredient in the proof of the classical
integrability of the model.  As it is not difficult to check the pure gauge potentials $ {\cal A},\bar {\cal A}$, has the
standard Leznov-Saveliev form \cite{lez-sav},
\be
{\cal A}= -g_0  {\eps_-}  g_0^{-1} ,\quad  
\bar {\cal A}=  {\eps_+}   + \bar \pa g_0 g_0^{-1} 
\label{2.19}
\ee
common for a large class of grades $|q|=1$, (i.e. $  q(\eps_{\pm})= \pm 1 $, where $[Q, \eps_{\pm }] = 
q(\eps_{\pm})\eps_{\pm}$ ) singular (or nonsingular) affine NA-Toda IM. 
As in the case of $A_n^{(1)}(p=1)$ dyonic models \cite{dubna} one can complete the proof of the integrability by deriving the
explicit form of the classical $R$-matrix and the infinite set of conservation laws in involution.  The calculation is
quite
strightforward and standard for all affine abelian and NA Toda models.

\subsection{Vacua Structure and Symmetries}

The existence of soliton solutions for the $A_n (p=2)$ IM (\ref{1.2}) is known to be intrinsically connected to the
following three related objects:
\begin{itemize}
\item distinct zeros of the potential  $V_n^{p=2}= {1\o {\b^2}} \(Tr (\eps_+ g_0 \eps_- g_0^{-1}) - \mu^2 (n-1)\)$

\item set of nontrivial constant solutions of its equation of motion

\item the  existence of few different boundary conditions (i.e. admissible asymptotic values at $x= \pm \infty$) for the
fields $\varphi_l, \psi_a, \chi_a, R_a$.

\end{itemize}
 The very origin  of the nontrivial vacua structure is hidden within the discrete symmetries of (\ref{1.2}) and in the
 corresponding topological charges.  In our case, for imaginary $\b = i \b_0$, the action (\ref{1.2}) is invariant under the
 following discrete group of transformations 
 \br
 \varphi^{\pr}_{l} = \varphi_l + \({{2\pi }\o{\b_0}}\) {{lN}\o {n-1}}, \quad l=1,2, \cdots n-2, \nonu \\
 \psi^{\pr}_1 = w^{{1\o 2}(N + (n-1)s_1 )}\psi_1, \quad 
\psi^{\pr}_n = w^{-{1\o 2}(N - (n-1)s_n )}\psi_n \nonu \\
\chi^{\pr}_1 = w^{{1\o 2}(N + (n-1)\tilde {s}_1 )}\chi_1, \quad 
\chi^{\pr}_n = w^{-{1\o 2}(N - (n-1)\tilde s_n )}\chi_n \nonu\\
\label{2.20}
\er
where  $w= e^{{{2\pi i}\o {n-1}}}$ and $N$ is an arbitrary integer 
(defining different $Z_{n-1} $ vacua sectors) and $s_a, \tilde s_a$ are two pairs of even
(or odd) integers (i.e., $s_a + \tilde s_a= 2S^{\pr}_a, \; \; s_a - \tilde s_a= 2L_a^{\pr}$ and 
$ S^{\pr}_a, L^{\pr}_a \in Z$). 
 The action (\ref{1.2})
is also invariant under CP (but not CPT) transformation, 
\br
\varphi_l^{\pr \pr} &=& \varphi_l, \quad \psi_a^{\pr \pr} =\chi_a,  \quad \chi_a^{\pr \pr} =\psi_a \nonu \\
 P x &=& -x, \quad \quad P \pa = \bar \pa 
\label{2.21}
\er
It is convenient to introduce the following new 
 parametrization for the field variables $\psi_a, \chi_a$, 
\br
\psi_1 = {1\o {\b_0}}e^{i\b_0 ({1\o 2}\varphi_1 - \theta_1)}\sinh (\b_0 {\cal R}_1), \quad 
\chi_1 = {1\o {\b_0}}e^{i\b_0 ({1\o 2}\varphi_1 +\theta_1)}\sinh (\b_0 {\cal R}_1), \nonu \\
\psi_n = {1\o {\b_0}}e^{i\b_0 ({1\o 2}\varphi_{n-2} - \theta_n)}\sinh (\b_0 {\cal R}_n), \quad 
\chi_n = {1\o {\b_0}}e^{i\b_0 ({1\o 2}\varphi_{n-2} +\theta_n)}\sinh (\b_0 {\cal R}_n)
\nonu
\er
As we shall see the angular variables $\theta_a, a=1,n$ make transparent the origin of the topological charges
$Q_{\theta}^a$.  They also play an important role in the discussion of T-duality transformations at the end  of this
subsection.  
The  transformations (\ref{2.20}) and (\ref{2.21}) in this parametrization acquires the form 
\br
{\cal R}_a^{\pr} ={\cal R}_a, \quad \theta_a^{\pr} = \theta_a+ {{\pi }\o {\b_0}}L_a^{\pr}, \nonu \\
{\cal R}_a^{\pr \pr } ={\cal R}_a, \quad \theta_a^{\pr \pr} = -\theta_a+ {{2 \pi }\o {\b_0}}L_a^{\pr \pr}
\label{2.22}
\er 
with $L_a^{\pr}, L_a^{\pr \pr }$ arbitrary integers.  Combining the $\theta_a^{\pr}$ and $\theta_a^{\pr \pr}$  
 transformations we find 
 \br
 \tilde \theta_a = -{{i}\o {2\b_0}} ln \( {{\tilde \chi_a}\o {\tilde \psi_a}}\) = \theta_a + {{\pi L_a}\o {2 \b_0}}, \quad 
 L_a = 2 L_a^{\pr \pr} - L_a^{\pr}
 \label{2.23}
 \er
 The conclusion is that together with the trivial vacua: $\varphi_l=0, {\cal R}_a =0, \theta_a=0$, (i.e., $N=L_a=0$) our
 model (\ref{1.2}) possess an infinite set of distinct vacua given by
 \br
 \varphi_l^{(N)} = \( {{2\pi}\o{\b_0}}\) {{lN}\o {n-1}}, \quad {\cal R}_a^{(0)} =0, \quad \theta_a^{(L)} = {{\pi L}\o {2\b_0}}
 \label{2.24}
 \er 
 defining its vacua lattice $(\varphi_l^{(N)}, \theta_a^{(L)})$. Due to the $Z_{n-1}\otimes Z_2$ symmetry, 
 
 a) $Z_{n-1}:$  
 \br
 e^{\b \varphi_l }  & \rightarrow  & w^{lN} e^{\b \varphi_l} \nonu \\
 \psi_1 &  \rightarrow & w^{{1\o 2}(N+(n-1)s_1)}\psi_1, \quad \psi_n  
 \rightarrow w^{-{1\o 2}(N-(n-1)s_n)}\psi_n, \nonu \\
 \chi_1 & \rightarrow & w^{{1\o 2}(N+(n-1)\tilde s_1)}\chi_1, \quad \chi_n  
 \rightarrow w^{-{1\o 2}(N-(n-1)\tilde s_n)}\chi_n,
\nonu
\er

 b) $Z_{2}$: $
\psi_a \rightarrow \chi_a, \quad \chi_a \rightarrow \psi_a $

\nit we have a finite set of allowed boundary conditions (at $x =\pm \infty$):
\br
\varphi_l^{(N)}(\pm \infty ) = \( {{2\pi }\o{\b_0}}\) {{lN_{\pm}}\o {n-1}}, \quad {\cal R}_a(\pm \infty ) =0, \quad
\theta_a^{(L)}(\pm \infty ) = {{\pi L_a^{\pm}}\o {2 \b_0}}
\label{2.26}
\er
(i.e. $\psi_a (\pm \infty ) = \chi_a(\pm \infty ) =0$, but $ {{\psi_a } / {\chi_a}}(\pm \infty ) \neq 0$).
  One therefore
expects the existence of finite energy topological solitons, interpolating between different vacua ($N_{\pm}, L_a^{(\pm
)}$).  The corresponding topological charges
\br
j_{\varphi} = N_+- N_- = 0, 1, \cdots n-2 \;  {\rm mod} \;(n-1), \quad j_{\theta}^a = L_a^+-L_a^- = 0, 1,
 \cdots \; {\rm mod } \; 2
\nonu
\er
are represented by the following topological currents
\br
J_l^{\mu }
= {{2(n-1)}\o {\b_0}} \eps^{\mu \nu} \pa_{\nu} \varphi_l, \quad Q_l^{top} = \int_{-\infty}^{+\infty} J_l^0 dx = lQ_{mag},
\quad Q_{mag}= {{4\pi}\o {\b_0^2}}j_{\varphi}
\label{2.27}
\er
and
\br
J_{\theta}^{\mu , a} = \b_0 \eps^{\mu \nu } \pa_{\nu } \theta_a, \quad Q_{\theta}^a = \int J_{\theta}^{0,a} dx = {{\pi
j_{\theta}^a}\o {2}}
\label{2.28}
\er
(for our $U(1)\otimes U(1)$-charged topological 1-solitons (\ref{3.19}), (\ref{3.23}), $j_{\varphi}= \pm 1, j_{\theta}^a =0$).

The transformations (\ref{2.20}) and (\ref{2.21}) that allowed us to determine the vacua lattice $(\varphi_l^{(N)},
\theta_a^{(L)})$ of our model,    does not exhaust all the discrete symmetries of the $A_n^{(1)}(p=2)$ dyonic IM
(\ref{1.2}).  There exist two other $Z_2$ discrete groups leaving the action (\ref{1.2}) invariant,
\begin{enumerate}
\item $Z_2$-reflections
\br
\varphi_l^{\pr \pr \pr} = \varphi_{n-l-1}, \quad l=1,2, \cdots n-2 \nonu \\
\psi_1^{\pr \pr \pr} = \psi_n, \quad \psi_n^{\pr \pr \pr} = \psi_1, \quad 
\chi_1^{\pr \pr \pr} = \chi_n, \quad \chi_n^{\pr \pr \pr} = \chi_1
\label{2.29}
\er

\item Composed Weyl transformation $\Pi P$:
\br
\tilde g_0 = \Pi (g_0^{-1}), \quad \Pi (\eps_+) = \eps_-, \quad \Pi ^2 =1, \nonu \\
\Pi (E_{\a_a})=  E_{\Pi (\a_a)} = E_{\a_a}, \;\; a=1,n, \quad P \pa = \bar \pa 
\label{2.30}
\er
\end{enumerate}
The $Z_2 \otimes Z_2$ symmetries (\ref{2.29}) and (\ref{2.30}) impose certain equivalence relations on the vacua lattice 
$(\varphi_l^{(N)}, \theta_a^{(L)}) \in Z_{n-1} \otimes Z_{2}$.  As a result the irreducible vacua lies in the coset $
Z_{n-1} \otimes Z_{2}/Z_2 \otimes Z_2$, i.e. (\ref{2.26}) with the identifications  (\ref{2.29}) and (\ref{2.30}).
As we shall see in the next Sect. 3 the $Z_2$ symmetry (\ref{2.29}) is crucial in the derivation of the first order
soliton equations (\ref{3.3}) and  (\ref{3.4}).  The $Z_2$ transformation $\Pi$ can be realized as certain compositions of
$A_{n-2}$-Weyl group transformations similar to the constructions of Sect. 2.6 of ref. \cite{eletric}.
It is the analog of the $S_n$ symmetries of the abelian $A_n$ Toda model.  

An important new feature of the dyonic IM (\ref{1.2}) is that, together with the above discrete symmetries, it manifest the
following continuous symmetries
\br
\psi^{\pr}_a = e^{i\b_0^2 \eps_a} \psi_a, \quad \chi_a^{\pr} =e^{-i\b_0^2 \eps_a} \chi_a, \quad \varphi_l^{\pr} =
\varphi_l
\label{2.31}
\er
with $\eps_a$ - the arbitrary real parameters of the global $U(1)\otimes U(1)$.  
Taking into account the definition (\ref{2.18}) of the nonlocal fields $R_a$, the corresponding (Noether) `` electric''
currents (derived from (\ref{1.2})) can be written in the following simple form
\br
J_1^{\mu , el} = {{2 \b_0}\o {n+1}}\eps^{\mu \nu }\pa_{\nu} \( nR_1 + R_n \), \quad 
J_n^{\mu , el} = {{2 \b_0}\o {n+1}}\eps^{\mu \nu }\pa_{\nu} \( R_1 + nR_n \)
\label{2.32}
\er
It therefore follows that the topological soliton with non trivial asymptotics of the fields $R_a$, i.e. $R_a (+ \infty )
\neq R_a (- \infty )$, will carry electric charges given by
\br
Q_1^{el} &=& \int_{-\infty}^{\infty} J_1^{0, el}dx = {{2 \b_0}\o {n+1}} \( n (R_1(\infty ) - R_1(-\infty )) + 
R_n(\infty ) - R_n(-\infty )\) , \nonu \\
Q_n^{el} &=& \int_{-\infty}^{\infty} J_n^{0, el}dx = {{2 \b_0}\o {n+1}} \( R_1(\infty ) - R_1(-\infty ) + 
n(R_n(\infty ) - R_n(-\infty )) \)
\label{2.33}
\er
As it is shown in Sect. 3. together with the neutral $(0,0 )$, 
we have $(0, Q_n )$ and $(Q_1, 0)$  and $(Q_1,Q_n )$
charged topological solitons. 

 It is worthwhile to mention the geometrical (target space $\s $-model) meaning of the 
$U(1)\otimes U(1)$-symmetry as well as the T-duality $O(2,2|Z)$ transformations behind it.  Writing (\ref{1.2}) in symbolic 
($\s $-model) form, 
\br
{\cal L} = \( g_{MN}(\rho^L) \eta^{\mu \nu} + b_{MN}(\rho^L) \eps^{\mu \nu}\) \pa_{\mu}\rho^M \pa_{\nu}\rho^N -V(\rho^L)
\label{2.34}
\er
where $\rho^L = \{\varphi_l, {\cal R}_a, \theta_a \}$ and noting that the $U(1)\otimes U(1)$ transformation 
(\ref{2.31}) acts as $\theta_a$ translations , $\theta_a \rightarrow \theta_a + \b_0^2 \eps_a,  $
we conclude that the target space geometry (described by the metric $g_{MN}$, antisymmetric tensor $b_{MN}$ and by the
tachyon potential $V(\rho^L)$) has two {\it isometries}, i.e., $ g_{MN}, b_{MN}$ and $V(\rho^L)$  
are $\theta_a$-independent.  This is an indication that the target space geometry encoded in the kinetic part of
(\ref{1.2}) allows two equivalent T-dual descriptions.  As in the case of $U(1)$ dyonic model (\ref{1.1}) (see refs.
\cite{eletric}, \cite{tdual}) one can introduce together with the axial model (\ref{1.2}) its T-dual vector model described by
 ${\cal L}_{vec}(\tilde \varphi_l, \tilde \theta_a, \tilde {{\cal R}}_a) $,
\br
{\cal L}_{vec}= {\cal L}_{axial} + {{d{\cal F}}\o {dt}}
\label{2.35}
\er
where
\br
{{d{\cal F}}\o {dt}}= {{1}\o {4(n+1)\b }} \sum_{a=1,n}\( \pa \tilde {\theta}_a \bar \pa  {\theta}_a - 
\bar \pa \tilde {\theta}_a  \pa  {\theta}_a \)\nonu 
\er
The T-duality transformations with the above generating function ${\cal F}( \tilde {\theta}_a,  {\theta}_a)$ are known to
act as the following {\it canonical transformation } $(\Pi_{\theta_a}, \theta_a )_{axial} \rightarrow 
(\tilde {\Pi}_{\tilde {\theta}_a}, \tilde {\theta}_a )_{vector} $ such that 
\br
\tilde {\Pi}_{\tilde {\theta}_a} =- \pa_x {\theta}_a
, \quad \Pi_{\theta_a} = - \pa_x \tilde {\theta}_a, 
\label{2.36}
\er
and certain point transformations  $\tilde \varphi_l = f_l (\varphi_l, R_a)$ of the ``non-isometric'' coordinates (see refs.
\cite{eletric}, \cite{tdual} for more details ).  The relation (\ref{2.35}) between ${\cal L}_{ax}$ and ${\cal L}_{vec}$
 is a simple consequence of the fact that both Hamiltonians are in fact equal, i.e., 
${\cal H}_{ax}= {\cal H}_{vec}$.  An important feature of the abelian T-duality (\ref{2.35}) and (\ref{2.36})
is that it maps the $U(1)\otimes U(1)$ charges $Q_a^{el, axial}$ (given by eqn. (\ref{2.33})) into the vector model analog 
$Q_{\tilde \theta}^{a, vector}$ of the topological charges $Q_{\theta}^{a, axial}$ (given by eqn. (\ref{2.28})).  Since we
have that (see Sect. 3 of ref. \cite{tdual})
\br
J_a^{\mu, el} (ax) &=& e_{aN}(\rho^L) \pa_{\mu} \rho^N = \eps_{\mu \nu } \pa^{\nu} \tilde \theta_a, \nonu \\
e_{aN}(\rho^L) &=& g_{aN}(\rho^L) + b_{aN}(\rho^L), \quad  \rho^L = \{\varphi_l, {\cal R}_a; \theta_a \}, \quad 
\rho^a \equiv \theta^a, \label{current}
\er
comparing with eqns. (\ref{2.32}) we conclude that the isometric coordinates $\tilde \theta_a$ of the vector model 
can be  written as  a linear combination of the axial model nonlocal fields  $R_a$, 
\br
\tilde \theta_1 = {{2\b }\o {n+1}}(nR_1 + R_n ), \quad \tilde \theta_n = {{2\b }\o {n+1}}(R_1 + nR_n )
\label{2.37}
\er
and $\tilde Q_{\tilde \theta}^{a, vec} = \int dx \pa_x {\tilde \theta}_a$, which makes the form (\ref{2.35}) of the
generating function ${\cal F}$ explicit.  The T-dual transformations of electric $Q_a^{el, ax}$ and $\theta_a$-topological
charges $Q_{\theta}^{a, ax}$ are given by the following interchange rule,
\br
(Q_a^{el, ax}, Q_{\theta}^{a, ax}) \Leftrightarrow 
(\tilde {Q}_{\tilde {\theta}}^{a, vec}, \tilde {Q}_a^{el, vec})
\nonu
\er
 
 \subsection{Dyonic Effect of the Topological $\theta$-terms}

One of the main properties of the soliton spectrum of the dyonic IM (\ref{1.1})
(with one global $U(1)$ symmetry)  is known to be its intrinsic {\it dyonic structure} \cite{eletric}. 
 Similarly to the electric and magnetic charges
of dyons in 4-d $SU(n+1)$-Yang-Mills-Higgs model \cite{wit3}, the electric charge $Q^{el}(p=1)$ 
of the $U(1)$ charged topological
solitons acquires contributions from its topological  (i.e. magnetic) charge $Q_{mag}(p=1)$ (see eqn. ({1.5}) 
of ref. \cite{eletric}):
\br
Q_{el}^{ax}(p=1) = \b_0^2 ( j_{el}^{p=1} + {{ \nu }\o {2\pi}} j_{\varphi}^{p=1} ), \; Q_{mag}(p=1) = {{4\pi }\o
{\b_0^2}} j_{\varphi}^{p=1}, \;  j_{\varphi}^{p=1}= 0, \pm 1, \cdots \pm (n-1) 
\label{2.38}
\er
where $\nu$ is an arbitrary real parameter.  
 The origin of such dyonic effect comes from the topological (total derivative) ``$\theta$-term'' (an analog of the
corresponding four dimensional topological $\theta$-term \cite{wit3})
\br
\d {\cal L}_{ax}^{top} (p=1)= {{\b}\o {8\pi^2}}\sum_{k=1}^{n-1}  \nu_k \eps^{\mu \nu} \pa_{\mu} \varphi_k \pa_{\nu}
ln ({{\chi}\o {\psi}})
\label{2.39}
\er
Since  by construction the dyonic model $A_n^{(1)}(p=2)$ (\ref{1.2}) with $U(1)\otimes U(1)$ global symmetry is an
appropriate 
generalization of the model $A_n(p=1)$ (\ref{1.1}) it is natural to expect that  similar dyonic effects to take place.
In this case one can 
 introduce the following two kinds of topological terms:
\begin{enumerate}
\item Straightforwad generalization of the $U(1)$ term (\ref{2.39}) to the $U(1)\otimes U(1)$ case
\br 
\d {\cal L}_{\varphi}^{top} (p=2)= {{\b}\o {8\pi^2}}\sum_{k=1}^{n-2}\sum_{a=1,n}  \nu_k^{(a)} \eps^{\mu \nu} 
\pa_{\mu} \varphi_k \pa_{\nu}
ln ({{\chi_a}\o {\psi_a}})
\label{2.40}
\er
\item New kind of $\theta_a$ mixed term
\br
\d {\cal L}_{\theta_a}^{top} (p=2)=   {{\nu_{\theta}}\o{4\pi^2}} \eps^{\mu \nu} 
\pa_{\mu} ln ({{\chi_n}\o {\psi_n}}) \pa_{\nu} ln ({{\chi_1}\o {\psi_1}})
\label{2.41}
\er
\end{enumerate}
where $\nu_k^{(a)}, $ and $\nu_{\theta}$ are arbitrary real constants.  Adding the ``$\theta$-terms'' (\ref{2.40}) and 
(\ref{2.41}) to the original Lagrangian (\ref{1.2}), i.e. considering
\br
{\cal L}_{impr}(p=2) = {\cal L}(p=2) + \d {\cal L}_{\varphi}^{top} + \d {\cal L}_{\theta_a}^{top}
\label{2.42}
\er
leads to the following improvements (shifts) of the corresponding electric currents
\br
J_{1,impr}^{\mu, el}(p=2) = {{2 \b_0}\o {n+1}}\eps^{\mu \nu }\pa_{\nu} \( nR_1 + R_n  - {{\b_0^2(n+1)}\o
{8\pi^2}}\sum_{l=1}^{n-2} \nu_l^{(1)} \varphi_l - {{\b_0 (n+1)}\o {8\pi^2}}\nu_{\theta}ln {{\chi_n}\o {\psi_n}} \), \nonu \\
J_{n,impr}^{\mu, el}(p=2) = {{2 \b_0}\o {n+1}}\eps^{\mu \nu }\pa_{\nu} \( R_1 + nR_n  - {{\b_0^2(n+1)}\o
{8\pi^2}}\sum_{l=1}^{n-2} \nu_l^{(n)} \varphi_l + {{\b_0 (n+1)}\o {8\pi^2}}\nu_{\theta}ln {{\chi_1}\o {\psi_1}} \). \nonu \\
\label{2.43}
\er
The corresponding new improved electric charges take the following form
\br
\int_{-\infty}^{\infty} J_{1,impr}^{0, el} dx = Q_1^{el} - {{\b_0^2}\o {2\pi}}\(j_{\varphi} \nu^{(1)} + \nu_{\theta}
{{j_{\theta}^{(1)}}\o 2}\), \nonu \\
\int_{-\infty}^{\infty} J_{n,impr}^{0, el} dx = Q_n^{el} - {{\b_0^2}\o {2\pi}}\( j_{\varphi} \nu^{(n)} - \nu_{\theta}
{{j_{\theta}^{(n)}}\o 2}\),
\label{2.44}
\er
where $\nu^{(a)} = {{1}\o {n-1}}\sum_{l=1}^{n-2} l\nu_l^{(a)}$.  For $\nu_{\theta} =0$, i.e. without the second topological
term, one recover the well known simple effect that the electric charges acquires contribution from the magnetic
($j_{\varphi}$-topological ) charge $Q_{mag} = {{4\pi }\o {\b_0^2}} j_{\varphi}$.  The {\it new dyonic}
 effect for $\nu_{\theta}
\neq 0$ is specific for the $U(1)\otimes U(1)$ case (\ref{1.2}) (improved as eq. (\ref{2.42})) and it consists in 
the fact that the electric charges $Q^{el}_{a, impr}$ gets shifted by the topological charges $j_{\theta}^{(a)}$ as well.  

An interesting physical interpretation of the parametres $\nu_l^{a}, \nu_{\theta}$ as external constant magnetic fields
$F_{MN}(\rho)$
\br
F_{\theta_a, l} = ({{\b}\o {4\pi}})^2 \nu_l^{(a)}, \quad F_{\theta_1, \theta_2} = ({{\b}\o {4\pi}})^2 \nu_{\theta}
\label{2.45}
\er
(all other components vanish)
comes from the following paralel with the open string (in curved background $g_{MN}, b_{MN}$) with two compactified
directions say, $X_{25} = {1\o {2\b}}ln ({{\psi_1}\o {\chi_1}}), \;\;  
X_{1} = {{1}\o {2\b}}ln ({{\psi_n}\o {\chi_n}})$ and boundary (Chan-Paton) term included (as in Sect. 8.6 of
\cite{polchinski}).  The string momenta $P_1 = Q_n^{el}$ and $P_{25} = Q_1^{el}$ get contributions from the boundary
(Wilson line) term $\d S_{strong} = i \oint \sum A_a(X) dX_a, \; X_a = \theta_a, \; X_i = \{ \varphi_l, {\cal R}_a,
\theta_a\}$). It is important to note that for our 1-soliton solutions (\ref{3.22}) the string coordinates 
\br
\theta_a = -{{i}\o {2\b_0}}ln ({{\chi_a}\o {\psi_a}}) \rightarrow (X_1, X_{25})
\nonu
\er
are indeed periodic (in $\rho_-$ worldsheet coordinate) with periods $T_a = {{2\pi}\o {w_a}}$
\br
w_1 = \mu \sin \a, \quad w_n = -\mu \sin \a. \nonu 
\er
Hence the paralel with open string with two compactified dimensions is not accidental.  Following such string analogy, the
topological $\theta$-terms (\ref{2.40}) and (\ref{2.41}) can be rewritten as Wilson line contributions for certain
background gauge fields $A_{\mu}(X_{\mu})$
\br
\d S_{top} &=& \int d^2z \( \d {\cal L}_{top}^{\varphi} + \d {\cal L}_{top}^{\theta_a}\) \nonu \\
&=&({{\b}\o {4\pi}})^2  \oint dt \( \sum_{a=1,n} \nu_l^a \varphi_l \pa_t \theta_a + {1\o 2}\nu_{\theta} (\theta_1 \pa_t
\theta_n - \theta_n \pa_t \theta_1 )\) \nonu \\
&=&\sum \oint d\theta^a A_{\theta_a} (X_{M})
\label{2.46}
\er
where we have denoted by $A_{\theta^a}$ the following terms 
\br
A_{\theta^1} &=& ({{\b}\o {4\pi }})^2 \( \sum_{l=1}^{n-2}\nu_l^{(1)}\varphi_l - {1\o 2}\nu_{\theta} \theta_n \)\nonu \\
A_{\theta^n} &=& ({{\b}\o {4\pi }})^2 \( \sum_{l=1}^{n-2}\nu_l^{(n)}\varphi_l + {1\o 2}\nu_{\theta} \theta_1 \)
\label{2.47}
\er
and all  other $A_{M}$ for $M \neq a$ vanish.  Therefore the magnetic fields $F_{MN} = \pa_MA_N - \pa_NA_M$ for our
$U(1)$  gauge potential  (\ref{2.47}) are given by eq. (\ref{2.45}).

\sect{Multicharged  Topological Solitons}

\subsection{Soliton equations}

The well known relation between dressing and B\"acklund transformation \cite{babelon} provides a simple derivation of 
1-soliton
equations  for a large class of grade one affine NA-Toda models.  Arguments similar to those used for the $A_n^{(1)}(p=1)$
dyonic IM \cite{eletric}, lead us to the following compact form of $1^{st}$-order soliton differential equations (DE)
\br
g_0^{-1} \pa g_0 X - [ g_0^{-1} Y g_{0,{vac}}, \eps_-] =0, \quad 
\bar \pa g_0 g_0^{-1} Y - [g_0X g_{0,{vac}}^{-1}, \eps_+]=0, 
\label{3.1}
\er
\br
[X, \eps_-] = [Y, \eps_+] =0
\label{3.2}
\er
imposed together with the image of the system (\ref{3.1}) under $Z_2$ discrete transformations (\ref{2.29}).  The constant
elements $X(\l, a_i), Y(\l, b_i)$ of the universal enveloping algebra are realized as
\br
X&=&X_{01}I +X_{02}\l_1 \cdot H + X_{03}\l_n \cdot H  + \sum_{k=1}^{n-2} a_k (\eps_-)^k, \nonu \\
Y&=&Y_{01}I +Y_{02}\l_1 \cdot H + Y_{03}\l_n \cdot H  + \sum_{k=1}^{n-2} b_k (\eps_+)^k
\nonu 
\er
where $X_{0i}(\l ), Y_{0i}(\l ), a_k(\l )$ and $b_k(\l )$ are arbitrary functions of the spectral parameter  $\l$ to be
determined 
and $g_{0,vac} = e^{i\b_0 (c_1 \l_1 \cdot H + c_n\l_n \cdot H)}$ is the constant vacuum solution of the L-S eqns.
(\ref{2.16}).  The verification of the fact that second  order DE (\ref{2.16}) are indeed  the 
integrability conditions
for the first order DE   (\ref{3.1}) (under conditions (\ref{3.2})) is quite straightforward.  We should mention that
the consistency of eqns. (\ref{3.1}) with their $Z_2$ image requires certain algebraic relations (see eqns. (\ref{3.4})
below) specific for the soliton equations far a large class of affine NA-Toda models \cite{eletric}.  With the parametrization of
$g_0^f$ given by eqs. (\ref{2.10b}) at hand and the subsidiary constraints (\ref{2.18}) for the nonlocal fields $R_a$ we derive
the following explicit system of $1^{st}$ order DE, 
\br
\pa \Phi_p &=& {{\mu\g }\o{\b}} \( e^{-\b \Phi_p} -  e^{-\b \Phi_{p-1}}+ \b^2 (\tpsi_1 \tchi_1 - \tpsi_n \tchi_n ) \d_{p,1}\), 
\nonu \\
\bar \pa \Phi_p &=& {{\mu }\o{\b \g }} \( e^{\b \Phi_p} - 
 e^{\b \Phi_{p+1}}- \b^2 (\tpsi_1 \tchi_1 - \tpsi_n \tchi_n ) \d_{p,n-1}\), \nonu \\
p&=& 1, 2, \cdots n-1, \quad \Phi_0 = \Phi_{n-1}, \quad  \Phi_1 = \Phi_{n} \nonu \\
\pa \tpsi_1 &=& -\mu \g \tpsi_1 e^{-\b \Phi_{n-1}} \(1+ \b^2 \tpsi_n \tchi_n e^{\b \Phi_{n-1}} - {{\b^2}\o {2}}\tpsi_1 \tchi_1 e^{\b
\Phi_{n-1}} \), \nonu \\
\pa \tpsi_n &=& -\mu \g \tpsi_n e^{\b \Phi_1} \(1+ \b^2 \tpsi_1 \tchi_1 e^{-\b \Phi_{1}} - {{\b^2}\o {2}}\tpsi_n \tchi_n e^{-\b
\Phi_{1}} \), \nonu \\
\pa \tchi_1 &=& -\mu \g \tchi_1 e^{\b \Phi_1} \( 1+ {{\b^2}\o 2}\tpsi_1 \tchi_1 e^{-\b \Phi_1} \), \nonu \\
\pa \tchi_n &=& -\mu \g \tchi_n e^{-\b \Phi_{n-1}} \( 1+ {{\b^2}\o 2}\tpsi_n \tchi_n e^{\b \Phi_{n-1}} \), \nonu \\
\bar \pa \tpsi_1 &=& {{\mu} \o {\g}}\tpsi_1 e^{\b \Phi_{1}} \(1+ {{\b^2}\o {2}} \tpsi_1 \tchi_1 e^{-\b \Phi_{1}}\), \nonu \\
\bar \pa \tpsi_n &=& {{\mu} \o {\g}} \tpsi_n e^{-\b \Phi_{n-1}} \(1+ {{\b^2}\o {2}} \tpsi_n \tchi_n e^{\b \Phi_{n-1}}\), \nonu \\
\bar \pa \tchi_1 &=& {{\mu} \o {\g}} \tchi_1 e^{-\b \Phi_{n-1}} \( 1+ {{\b^2}}\tpsi_n \tchi_n e^{\b \Phi_{n-1}}
- {{\b^2}\o 2} \tpsi_n \tchi_n e^{\b \Phi_{n-1}} \), \nonu \\
\bar \pa \tchi_n &=&  {{\mu} \o {\g}}\tchi_n e^{\b \Phi_{1}} \( 1+ {{\b^2}}\tpsi_1 \tchi_1 e^{-\b \Phi_{1}}
- {{\b^2}\o 2} \tpsi_n \tchi_n e^{-\b \Phi_1} \), 
\label{3.3}
\er
where $\g $ is the B\"acklund transformation parameter,
\br
\g = {{\l b_2}\o {a_2}} = {{e^{c_1 -c_n} (Y_{01} - Y_{02})}\o {a_1}} = {{b_1  e^{c_1 -c_n}}\o {X_{01}-X_{02}}} = \cdots 
=e^{-b}
\nonu
\er
  Another consequence of eqns. (\ref{3.1}) 
is  the following chain of algebraic relations,
\br
e^{-\b \Phi_1} - e^{\b \Phi_2} &=& e^{-\b \Phi_2} - e^{\b \Phi_3} = \cdots = 
e^{-\b \Phi_{n-2}} - e^{\b \Phi_{n-3}} \nonu \\
&=& e^{-\b \Phi_{n-1}} - e^{\b \Phi_{1}} + \b^2 (\tpsi_n \tchi_n - \tpsi_1 \tchi_1)
\label{3.4}
\er
which should be satisfied together with the first order DE (\ref{3.3}).
The new  variables $\Phi_p, \tpsi_a$ and $\tchi_a $  
\br
\Phi_p = \varphi_p -\varphi_{p-1} -{{1}\o {n+1}}R, \quad R= R_1 - R_n, \;\; \varphi_0 = \varphi_{n-1} =0, \nonu \\
\Phi_1 + \Phi_2 + \cdots + \Phi_{n-1} = -{{n-1}\o {n+1}}R, \quad \varphi_p = {{p}\o {n+1}}R + \sum_{k=1}^{p} \Phi_k, \nonu
\\
\tpsi_n = \psi_n e^{{{R}\o {2(n+1)}}}, \quad  \tchi_n = \chi_n e^{{{R}\o {2(n+1)}}}, 
\quad\tpsi_1 = \psi_1 e^{-{{R}\o {2(n+1)}}}, \quad \tchi_1 = \chi_1 e^{-{{R}\o {2(n+1)}}}
\label{3.5}
\er
introduced in (\ref{3.3}) have the advantage to simplify the construction of the solutions (1-soliton) of eqn. (\ref{3.3}).  The nonlocal fields $R_a, \;
a=1,n$ satisfy the following set of $1^{st}$-order DE,
\br
\pa R_1&=& {{\mu\g \b}\o {(n-1)}} \( \tpsi_n \tchi_n - n \tpsi_1 \tchi_1 \)\nonu \\
\pa R_n&=&{{\mu\g \b}\o {(n-1)}} \( \tpsi_1 \tchi_1 - n \tpsi_n \tchi_n \)\nonu \\
\bar \pa R_1&=& {{\b \mu }\o {\g (n-1)}} \( n\tpsi_1 \tchi_1 -  \tpsi_n \tchi_n \)\nonu \\
\bar \pa R_n&=&{{\b \mu}\o {\g (n-1)}} \( n\tpsi_n \tchi_n -  \tpsi_1 \tchi_1 \)
\label{3.6}
\er
as a consequence of their defining eqns (\ref{2.18}).

We next derive few consequences of the system (\ref{3.3}), (\ref{3.4}) and (\ref{3.6}) representing the following
 {\it solitonic conservation
laws}, 
\br
&\bar \pa  \( \g e^{-\b \Phi_p} \) + \pa \( {1\o {\g}} e^{\b \Phi_{p+1}}\)=0 
\nonu \\
&\( \g \bar \pa + {1\o {\g}} \pa \) e^{\pm \b \Phi_p}=0, \quad p=1,2,\cdots n-1 \nonu \\
&\( \g \bar \pa - {1\o {\g}} \pa \)ln ({{\tpsi_a}\o {\tchi_a}})=0, \quad a=1,n \nonu \\
&\( \g \bar \pa +{1\o {\g}} \pa \){{\tpsi_a}{\tchi_a}}=0, \quad a=1,n \nonu \\
&\( \g \bar \pa +{1\o {\g}} \pa \)(R_1\pm R_n) =0 \nonu \\
&\nabla ln \( {{\tpsi_1}\o {\tchi_1}}e^{{{(n-1)}\o {2(n+1)}}R} \) = 
\nabla ln \( {{\tpsi_n}\o {\tchi_n}}e^{-{{(n-1)}\o {2(n+1)}}R} \), \quad \nabla = \pa, \bar \pa
\label{3.12}
\er
They play an important role in the construction of the solutions (1-soliton) of the system (\ref{3.3}) and 
(\ref{3.4}).

\subsection{Soliton Solutions}

The strategy we employ in solving the complicated $1^{st}$ order system (\ref{3.3}) consists in:
\begin{enumerate}
\item To derive  all first integrals of eqns. (\ref{3.3});
\item To diagonalize the system of eqns. (\ref{3.3}) by using these $1^{st}$ integrals and the solitonic conservation laws
(\ref{3.12}), i.e. to separate the eqns. for the variables $\Phi_p, \tpsi_a, \tchi_a, R_a$;
\item To solve each one of the corresponding individual $1^{st}$ order (one variable ) equations

\end{enumerate}

The first integrals for the system (\ref{3.3}) can be easily found by simple manipulations of eqns. 
(\ref{3.3}), (\ref{3.12}).  The result is:
\br
e^{-\b \Phi_p}- e^{\b \Phi_{p+1}} = 2i \sin \a = c_0, \quad p=1, \cdots n-1
\label{3.13}
\er
\br
{{\tpsi_n}\o {\tchi_1}}e^{-\b {{(n-1)}\o {2(n+1)}}R} = e^{-d_2}, \quad 
{{\tpsi_1}\o {\tchi_n}}e^{\b {{(n-1)}\o {2(n+1)}}R} = e^{d_1}, \quad d_1 + d_2 =2\d
\label{3.14}
\er
\br
e^{\d + {{\b}\o {n+1}}(nR_1+R_n)} - e^{-\d + {{\b}\o {n+1}}(R_1+nR_n)} = De^{\d}
\label{3.15}
\er
where $c_0, d_1, d_2$ and $D$ are arbitrary complex constants.  We next substitute eqns. (\ref{3.13}) and (\ref{3.4})
 in the first two equations of the system (\ref{3.3}), thus obtaining the desired individual equations for each $\Phi_p, \b
 =i\b_0$, 
 \br
 \pa_{\rho_+} e^{i\b_0 \Phi_p} &=& \mu \(1- e^{2i\b_0\Phi_p} - 2i\sin \a e^{i\b_0 \Phi_p}\) \nonu \\
 \pa_{\rho_-} e^{i\b_0 \Phi_p} &=&0
 \label{3.16}
 \er
 where we have introduced  new variables $\rho_{\pm}$,
 \br
 \rho_+ &=& x \cosh( b) - t \sinh (b), \quad \pa_{\rho_+} = \sinh (b) \pa_t + \cosh (b) \pa_x \nonu \\
 \rho_- &=& t \cosh (b) - x \sinh (b), \quad \pa_{\rho_-} = \cosh (b) \pa_t + \sinh (b )\pa_x, \quad b = -ln \g 
 \nonu
 \er
The solutions of eqns. (\ref{3.16}) have the following form
\br
e^{-i\b_0 \Phi_p} = e^{i\a} {{S_p e^{-2i \a + 2 \mu \rho_+ \cos (\a)} -1}\o {S_p e^{ 2 \mu \rho_+ \cos (\a)} +1}}
\label{3.17}
\er
Due to eqns. (\ref{3.13}) and (\ref{3.4}) the integration constants $S_p$ should satisfy the following recurence relations
\br
S_{p+1} = e^{-2i\a \pm i \pi } S_p, \quad p=1,2, \cdots n-1
\nonu
\er
and therefore we find
\br
S_p = (-1)^{p-1} e^{-2i \a (p-1)}S_1 = (-1)^{p-n}e^{-2i \a (p -n)} X_0
\nonu 
\er
where $X_0$ is an arbitrary constant  that determines the center of mass of the soliton.
Finally, taking into account the definitions (\ref{3.5}) of the original fields $\varphi_p$ and $R=R_1 - R_n$ in terms of
$\Phi_p$ we find 1-soliton solutions in the form:
\br
e^{i\b_0 {{(n-1)}\o {(n+1)}}R} = e^{-i (n-1)(\a -\pi Sign (\a))}{{e^{-f} + (-1)^{n-1} e^{2i(n-1)\a} e^{f}}\o
{e^{-f}+e^{f}}}
\label{3.18}
\er
\br
e^{i\b_0 \varphi_l} = {{e^{-f} + (-1)^{l-n+1} e^{2i(n-1-l)\a} e^{f}}\o
{(e^{-f}+e^{f})^{{{l}\o {n-1}}} ( e^{-f} + (-1)^{n-1} e^{2i(n-1)\a} e^{f} )^{{{n-1-l}\o {n-1}}}}}, \quad l=1,2, \cdots n-2
\label{3.19}
\er
where $f=\mu\rho_+ \cos (\a ) + {1\o 2} ln X_0$.  We next calculate $R_1$ and $R_n$ from eqns. (\ref{3.15}) and 
(\ref{3.18}), 
\br
e^{i\b_0 R_1} &=&  {{De^{i {{n\b_0}\o {(n+1)}}R} }\o {e^{i\b_0 {{(n-1)}\o {(n+1)}}R} - e^{-2\d }}} =
D e^{2 \d - i \pi -i n (\a - \pi Sign (\a ) } {{(e^{-f} + (-1)^{n-1} e^{2i(n-1)\a} e^{f})^{{n}\o {n+1}}}\o
 {(e^{-f}+e^{f})^{{{1}\o {n-1}}}(A_1 e^{-f} + A_2 e^{f})}}, \nonu \\
e^{i\b_0 R_n} &=& {{De^{-i\b_0 {{\b_0}\o {(n+1)}}R} }\o {e^{i\b_0 {{(n-1)}\o {(n+1)}}R} - e^{-2\d }}}=
 {{D e^{2 \d - i \a}(e^{-f}+e^{f})^{{{1}\o {n-1}}}}\o {(e^{-f} + (-1)^{n-1} e^{2i(n-1)\a} e^{f})^{{1}\o {n+1}}
(A_1 e^{-f} + A_2 e^{f})}}
\label{3.20}
\er
where $A_1, A_2 $ are given by
\br
A_1 = 1- e^{-i(n-1)(\a -\pi Sign (\a) + 2 \d }, \quad A_2 = 1- e^{i(n-1)(\a -\pi Sign (\a) + 2 \d }. \nonu 
\er
It remains to derive the solutions for $\tpsi_a, \tchi_a$.  The algebraic relation (\ref{3.4}) together with eqns.
(\ref{3.14}), (\ref{3.18}) and  (\ref{3.19}) allows us to calculate the product $\psi_a \chi_a, \; a=1,n$,
\br
\b_0^2 \psi_1 \chi_1 = {{N^2 (e^{-f}+e^{f})^{{-{1}\o {n-1}}}(e^{-f} + (-1)^{n-1} e^{2i(n-1)\a} e^{f})^{-{(n-2)}\o {(n-1)}}}\o
 {(A_1 e^{-f} + A_2 e^{f})}}
 \label{3.21}
 \er
 where $N^2 = (1+e^{-2i \a})( (-1)^{n-1}e^{2i(n-1)\a}-1)$.  For the ratio ${{\psi_a}\o {\chi_a}}$ we obtain the simple 
 equations
 \br
 \pa_{\rho_+} ln ({{\psi_1}\o {\chi_1}})&=& \pa_{\rho_+} ln ({{\psi_n}\o {\chi_n}})=0, \nonu \\
 \pa_{\rho_-} ln ({{\psi_1}\o {\chi_1}})&=& -\pa_{\rho_-} ln ({{\psi_n}\o {\chi_n}})=-2i \mu \sin (\a )
 \nonu 
 \er
and therefore we find 
\br
{{\psi_1}\o {\chi_1}}= e^{-2i (\mu \rho_- \sin (\a ) + p_1)}, \quad \quad 
{{\psi_n}\o {\chi_n}}= e^{2i (\mu \rho_- \sin (\a ) + p_n)}.
\label{3.22}
\er
where $p_1, p_n$ are arbitrary constants.  
The explicit expression for $\psi_a, \chi_a$ completes the final form of our $(Q_1, Q_n)$-charged topological 1-soliton
solution
\br
\psi_1 &=& {1\o {\b_0}} e^{-i (\mu \rho_- \sin (\a ) + p_1)} \sqrt {\psi_1 \chi_1}, \quad 
\chi_1 = {1\o {\b_0}} e^{i (\mu \rho_- \sin (\a ) + p_1)} \sqrt {\psi_1 \chi_1}
\nonu 
\er
\br
\psi_n 
&=& {{N}\o {\b }} {{e^{-d_2 - {{i}\o {2}}(n-3)(\a - \pi Sign (\a))}
e^{i(\mu \rho_- \sin (\a ) + p_1 )} ( e^{-f} + (-1)^{n-1} e^{2i(n-1)\a} e^{f} )^{{{-1}\o {2(n+1)}}}\o 
{(A_1 e^{-f} + A_2 e^{f})^{1\o 2}(e^{-f}+e^{f})^{{(n-2)}\o {2(n-1)}}}}}
, \nonu \\
\chi_n 
&=& {{N}\o {\b }} {{e^{-d_1 - {{i}\o {2}}(n-3)(\a - \pi Sign (\a))}
e^{-i(\mu \rho_- \sin (\a ) + q_1 )} ( e^{-f} + (-1)^{n-1} e^{2i(n-1)\a} e^{f} )^{{{-1}\o {2(n+1)}}}\o 
{(A_1 e^{-f} + A_2 e^{f})^{1\o 2}(e^{-f}+e^{f})^{{(n-2)}\o {2(n-1)}}}}}
\label{3.23}
\er
i.e., the $(Q_1, Q_n)$ 1-soliton of the dyonic model $A_n^{1)}(p=2)$ IM (\ref{1.2}) is presented by the set of functions  
(\ref{3.19}), (\ref{3.21}) and (\ref{3.23})  of  space time variables $\rho_+$ and  $\rho_-$.

It is worthwhile to mention that both, the Lagrangian (\ref{1.2}) and the $1^{st}$-order system (\ref{3.3})-(\ref{3.4})
have two interesting limits:
\begin{itemize}
\item $\psi_1 = \chi_1 =0, (R_n= \; {\rm{ const.}})$ or $\psi_n = \chi_n =0, (R_1= \; {\rm{ const.}})$ leading to the
$A_{n-1}^{(1)}(p=1)$ dyonic IM (\ref{1.1}) (see ref. \cite{eletric} for corresponding $1^{st}$-order system)  and their 1-soliton
solutions \cite{eletric}, \cite{dyonic} represent one charge $(0, Q_n)$  or $(Q_1, 0)$ topological solitons of our
$A_n$(p=2) model (\ref{2.31}).

\item $\psi_1 = \chi_1 = \psi_n = \chi_n =0$ gives rise to the $A_{n-2}^{(1)}$ abelian Affine Toda model and its neutral 
$(0, 0)$ 1-solitons are particular cases of our $(Q_1, Q_n)$-solitons.
\end{itemize}

\subsection{Soliton Spectrum}
 One of the main properties of the soliton solutions is that they carry
  finite energy and nontrivial topological charge (and
 electric charges $Q_1, Q_n$  in our case).  Thus, it  manifests a particle-like spectrum, $M,E, Q_1^{el}, Q_n ^{el}, Q_{mag},
 Q_{\theta}^a$.  The soliton charges  $Q_a ^{el}, Q_{mag}, Q_{\theta}^a$ as well 
 as their energy $E$ (and mass $M$) are known
 to depend on the boundary conditions, i.e. asymptotics of fields $\varphi_l, \psi_a, \chi_a \; 
 ({\rm at} \;  x\rightarrow \pm \infty )$, 
 and on the symmetries of the $1^{st}$ order (BPS-like) equations (\ref{3.3}).  In fact, the explicit form of the soliton
 spectrum can be derived without the knowledge of the exact 1-soliton solution (\ref{3.19}) - (\ref{3.23}).  The arguments
 are quite similar to the abelian affine Toda \cite{liao} and $A_n^{(1)}(p=1)$ dyonic IM \cite{eletric} cases. 
  One first verify the
 following ``solitonic conservation law''\br
 \bar \pa F^- &=& \pa F^+, \nonu \\
 F^- = -{{\mu \g }\o {\b^2}} (\sum_{k=1}^{n-1} e^{-\b \Phi_k } + \b^2 \tpsi_n \tchi_n ), & & \quad 
 F^+ = {{\mu  }\o {\g \b^2}} (\sum_{k=1}^{n-1} e^{\b \Phi_k } + \b^2 \tpsi_1 \tchi_1 )
\label{3.24}
\er
and as a consequence the $A_n(p=2)$  potential $V_n^{(p=2)} $ (\ref{1.2}) can be written in the form
\br
V_n^{(p=2)} = -{{1\o 2}} (\bar \pa F^- + \pa F^+)\nonu 
\er
Next step is to demonstrate (by using the 1-soliton eqns. (\ref{3.3}),(\ref{3.4}) and (\ref{3.6})) that the 
$A_n(p=2)$ stress-tensor components 
$T_{00} = T^+ + T^- +2V, \quad T_{01} = T^+- T^-$,
\br
2T^{+} &=& {1\o 2} k_{ij} \pa \varphi_i  \pa \varphi_j + {{1}\o { {\Delta}}} \(e^{-\b \varphi_1}\Delta_n \pa \psi_1 \pa
\chi_1 + e^{-\b \varphi_{n-2}}\Delta_1 \pa \psi_n \pa \chi_n \right. \nonu \\
&+& \left. \b^2 {{e^{-\b (\varphi_1 + \varphi_{n-2})}}\o {2(n-1)}} (\chi_1 \psi_n \pa
\chi_n \pa \psi_1 + \chi_n \psi_1 \pa \chi_1 \pa \psi_n ) \)
\label{3.25}
\er
and $T^- = T^+ (\pa \rightarrow \bar \pa )$ where
\br
\Delta_1 = 1+ \b^2 {{n}\o {2(n-1)}}\psi_1 \chi_1 e^{-\b \varphi_1}, \quad  
\Delta_n = 1+ \b^2 {{n}\o {2(n-1)}}\psi_n \chi_n e^{-\b \varphi_{n-2}} 
\nonu 
\er
are total derivatives, i.e., 
\br
T_{00} = \pa_x \(F^- - F^+ \), \quad T_{01} = \pa_x \(F^- + F^+ \)
\nonu 
\er
Therefore the energy and the momentum of our ($Q_1, Q_n$) 1-soliton  (\ref{3.19}) - (\ref{3.23}) gets contribution from
boundary terms only, 
\br
E= \int_{-\infty}^{+\infty} T_{00}dx = (F^- - F^+ )|_{-\infty}^{+\infty}, \quad 
P= \int_{-\infty}^{+\infty} T_{01}dx = (F^- + F^+ )|_{-\infty}^{+\infty}
\label{3.26}
\er
The last step is to calculate $E,P$ and $M= \sqrt{E^2 - P^2}$ in terms of the asymptotics of the fields $\varphi_l, \psi_a,
\chi_a, R_a$.  The analysis of the classical vacua structure of the $A_n(p=2)$ IM (\ref{1.2}) (i.e., nontrivial constant
solutions corresponding to multiple zeros of the potential $V_n^{(p=2)} $, with $\b =i\b_0$) presented in Sect. 2.4
 provides
the complete list of the admissible solitonic b.c.
\br
\varphi_l (\pm \infty) = \({{2\pi}\o {\b_0}}\) {{lN^{\pm}}\o {n-1}}, & \;  & \( \psi_a \chi_a\) (\pm \infty)= 0, \; 
 R_a (\pm \infty)= {{2\pi}\o {\b_0}} f_{\pm}^{(a)} \nonu \\
\quad 
R(\pm \infty)= {{2\pi}\o {\b_0}}(f_{\pm}^{(1)} - f_{\pm}^{(n)})&=& {{2\pi}\o {\b_0}}f_{\pm}, \quad 
i ln {{\psi_a }\o {\chi_a}} (\pm \infty)= \pi L^a_{\pm}, 
\label{3.27}
\er
where $N_{\pm}, L_{\pm}^a$ are arbitrary integers and $f_{\pm}^a$ are real numbers.   Substituting  the above asymptotic
values in eqn. (\ref{3.26}) we derive the following mass formula,
\br
M&=& {{4\mu (n-1)}\o {\b_0^2}} | \sin \({{\pi}\o {n-1}}(j_{\varphi} - (f_+-f_-){{n-1}\o {n+1}})\) |,
   \nonu \\
   j_{\varphi}&=&
N_+-N_- = 0, \pm 1, \cdots \pm (n-2)
\label{3.28}
\er
We can further simplify eq. (\ref{3.28}) by noting that according to eqns. (\ref{2.27}) and (\ref{2.33}) we have
\br
Q_1^{el} - Q_n^{el}= 4 \pi {{n-1}\o {n+1}} (f_+ - f_-), \quad Q_{mag} = {{4\pi }\o {\b_0^2}}j_{\varphi}
\label{3.28a}
\er
Therefore the 1-soliton masses are independent of $Q_{\theta}$ and  $Q_1^{el} + Q_n^{el}$, 
\br
M = {{4\mu (n-1)}\o {\b_0^2}}| \sin ( {{\b_0^2 Q_{mag} - Q_1^{el} + Q_n^{el} }\o {4(n-1)}}) |
\label{3.29}
\er
As in the case of the axial and vector $A_n(p=1)$ dyonic IM \cite{eletric}, one expects the 1-solitons of the corresponding vector
gauged $A_n(p=2)$ IM to exhibit mass spectrum with $Q_a^{el}$ replaced by $\tilde Q_{\theta}^a$.
An important question to be answered concerns the relation between $f_{\pm}^{(a)}, N_{\pm}, L_{\pm}$ introduced in
(\ref{3.27}) and the parameters that appear in the soliton solution 
(\ref{3.19})-(\ref{3.23}), and further the question of
whether  we can have $f_+^a \neq f_-^a,\; N_+ \neq N_-$, etc.
  The analysis  is quite similar to the $A_n(p=1)$ case
presented in ref. \cite{eletric}, and is based on ${\bf a)}$ the soliton conservation laws in the form (\ref{3.13})-(\ref{3.15})
 and ${\bf b)}$ the direct evaluations of the asymptotics at $x \rightarrow \pm \infty$ of the fields $\varphi_l, \psi_a,
\chi_a, R_a$ from eqns. (\ref{3.19})-(\ref{3.23}).  The result is as follows,
\br
\varphi_l (\pm \infty) &=& {{2\pi }\o {\b_0}} ({{l}\o {n-1}}N_{\pm}^l +K_{\pm}^l), \nonu \\
\theta_a (-\infty) &=& \theta_a (+\infty), \quad i.e. , \quad j_{\theta}^{(a)} = Q_{\theta}^a = 0 \nonu \\
R(+ \infty) &=&  {{2\pi  }\o {\b_0}} f_+ = {{1 }\o {\b_0}}({{(n+1)\a }} + 2\pi S_+ {{(n+1)}\o {n-1}}), \nonu \\
R(- \infty) &=&  {{2\pi }\o {\b_0}} f_- = {{1 }\o {\b_0}}({{2\pi S_-{{(n+1)}\o {n-1}} -(n+1)(\a - \pi Sign (\a )) }} )
\label{3.30}
\er
and from (\ref{3.13})-(\ref{3.15}) one gets further restrictions, 
\br
N_{\pm}^l = N_{\pm}\; \;  ({\rm mod} \; (n-1)), \quad S_{\pm } = N_{\pm} \; \; ({\rm mod} \; (n-1)),  \; \;
K_{\pm}^l =0, \quad N_+ \neq N_-
\nonu 
\er
as one can see from eqns. (\ref{3.15}) and (\ref{3.20}).  The $R_1+R_n$ asymptotics are certain functions of $\a$ and $\d$,
but they do not enter directly in the computation of the 1-soliton spectrum.

We should mention that the origin of the factor $\pi Sign (\a )$ in the explicit form of the soliton solutions (\ref{3.18}),
(\ref{3.20})  and (\ref{3.23}) as well as in the boundary values $R(\pm \infty )$ in eqn. (\ref{3.30}) comes from  the
consecutive application of eqn. (\ref{3.13}) for $x \rightarrow \pm \infty $ and $t = t_0$ fixed. This leads to the following
equation for $f_{\pm}$:
\br
\sin {{2\pi} \o {n-1}}\(  {{n-1}\o {n+1}}f_{\pm} -N_{\pm} \) = \sin {\a}
\nonu 
\er
Its general solution turns out to depend on the sign of the angle $\a$.  The same argument takes place in the determination
of the constants of integration $S_p$ (see eqn. (\ref{3.17}))  from eqn. (\ref{3.13}). 


One might wonder whether    massless 1-solitons exists.  Indeed for specific values of $Q_{mag}$ and $Q_a^{el}$, when  
the following relation 
\br
\b_0^2 Q_{mag} -Q_1^{el}+ Q_n^{el} = 4\pi (n-1) s, \quad s \in Z
\label{novaeq}
\er
takes place, we have  $M=0$ as one can see from eq. (\ref{3.29}).  However by substituting the explicit form of $Q_a^{el}$
in terms of $f_{\pm}$ and $\a$, we conclude that eq. (\ref{novaeq}) is satisfied for $\a= \a_s = \pi (s -{1\o 2})$ only.  Since
$\cos (\a_s ) =0 $ leads to $N(\a_s ) =0$ and $f(\a_s ) = \mu \rho_+ \cos (\a_s ) + {1\o 2} ln X_0 = const$ and therefore
the solutions for such values of $\a = \a_s$ coincide with  
the trivial constant vacua solution.  Hence, they do not describe 1-solitons.  In conclusion, the charges of the proper
1-soliton should satisfy the following selection rule,
\br
\b_0^2 Q_{mag} -Q_1^{el}+ Q_n^{el} \neq 4\pi (n-1) s. 
\label{neq}
\er
\subsection{Semiclassical Quantization}

An important feature of the $(Q_1, Q_n)$ charged topological 1-soliton we have constructed  is that at the rest frame it
represents periodic particle-like motion (due to the $\rho_-$ dependence of $\psi_a, \chi_a$) with period $\tau = {{2\pi }\o
{\mu \sin (\a )}}$.  Then, similarly to the SG breather \cite{coleman}, NLS and Lund-Regge 1-solitons \cite{lund} as well as
$U(1)$-charged 1-solitons of the $A_n(p=1)$ IM, one can apply the field theoretic analog of the Bohr-Sommerfeld quantization
rule
\br 
S^{ax} + E(v=0) \tau &=& \int_{0}^{\tau} dt \int_{-\infty}^{\infty} dx \Pi_{\rho_M}\dot {\rho}_M = 2\pi j_{el}, 
\quad j_{el} \in Z
\nonu \\
\Pi_{\rho_M}&=& {{\d {\cal L}}\o {\d \dot {\rho}_M}}, \quad {\cal H}^{ax} = \Pi_{\rho_M}\dot {\rho}_M- {\cal L}^{ax}
\label{3.31}
\er
in order to derive the semiclassical 1-soliton spectrum.  Taking into account that $
\Pi_{\theta_a} = {{1}\o {\b_0}} J_a^{0, el} $
(see eqns. (\ref{2.33}), (\ref{current}), (\ref{2.37})) and the following simple t-dependence of our charged 1-soliton solution
(\ref{3.19})-(\ref{3.23}) at rest frame ($ \g = e^{-b}, \cosh (b) =1, \sinh (b) =0$):
\br
\pa _t \varphi_l = \pa _t \Pi_{\varphi_l}= \pa _t  \( \psi_a  \chi_a \) = \pa _t \Pi_{{\cal R}_a}=0,  \quad 
 \pa _t ln {{\psi_1 }\o {\chi_1}} =  -\pa _t ln {{\psi_n }\o {\chi_n}} = 2i \mu \sin (\a )
 \nonu
 \er
we find that
\br 
\int_{0}^{\tau} dt \int_{-\infty}^{\infty} dx \Pi_{\rho_M}\dot {\rho}_M =
\int_{0}^{\tau} dt \int_{-\infty}^{\infty} dx \Pi_{\theta_a}\dot {\theta}_a ={{2\pi }\o {\b_0^2}}(Q_1 - Q_n)
\label{3.33}
\er
Therefore the Bohr-Sommerfeld  quantization rule  implies
\br
Q_1 - Q_n = \b_0^2 j_{el}, 
\nonu
\er
or equivalently
\br
Q_1 = \b_0^2 ({{j_{el}}\o 2} + \d_0), \quad Q_1 = \b_0^2 (-{{j_{el}}\o 2} + \d_0), \quad \d_0 = \d_0 (\a , \d )
\label{3.34} 
\er
where $\d_0$ is an arbitrary parameter, i.e., only the difference of the charges is quantized.  Their sum
 remains continuous.  
Taking into account the improvements (\ref{2.43}), (\ref{2.44}) of the electric charges $Q_a^{el}$ 
introduced by topological
``$\theta$-terms''  (\ref{2.40}), (\ref{2.41}) and the corresponding topological shifts in the momenta $\Pi_{\theta_a}$, 
(i.e. of
the electric currents $J_a^{0, el }$ eqns. (\ref{2.43}) we obtain the following (improved ) charge  quantization 
\br
\b_0^2 j_{el} = Q_1^{impr}- Q_n^{impr}= Q_1- Q_n- {{\b_0^2}\o {2\pi}} \( (\nu^{(1)} - \nu^{(n)}) j_{\varphi} +
{{\nu_{\theta}}\o 2} (j_{\theta}^{(1)} -  j_{\theta}^{(n)})\)
\nonu 
\er
and therefore
\br \Delta Q = Q_1- Q_n= \b_0^2 \( j_{el} +{{(\nu^{(1)} - \nu^{(n)})}\o {2\pi }}j_{\varphi} + {{\nu_{\theta}}\o {4\pi
}}(j_{\theta}^{(1)} - j_{\theta}^{(n)}) \)
\label{3.35}
\er
Since both $\d_0$ and $\nu_{\theta}$ are arbitrary reals we can choose $\nu_{\theta}$ such that $2\d_0 = {{\nu_{\theta}}\o
{4\pi }} ( j_{\theta}^{(n)}- j_{\theta}^{(1)})$.  Then the charge spectrum takes the form,
\br
Q_1 &=& {{\b_0^2}\o 2} \( j_{el} + {{1}\o {\pi}} \nu^{(1)} j_{\varphi} + 
{{\nu_{\theta}}\o {4\pi}}(j_{\theta}^{(1)} + j_{\theta}^{(n)})\), \nonu \\
Q_n &=& {{\b_0^2}\o 2} \( -j_{el} + {{1}\o {\pi}} \nu^{(n)} j_{\varphi} - 
{{\nu_{\theta}}\o {4\pi}}(j_{\theta}^{(1)} + j_{\theta}^{(n)})\)
\label{3.36}
\er
and the 1-soliton improved mass formula (\ref{3.29}) becomes
\br
M_{j_{el}, j_{\varphi}} = {{4\mu (n-1)}\o {\b_0^2}}| \sin {{(4\pi j_{\varphi} - \b_0^2 j_{el})}\o {4(n-1)}} |
\label{3.37}
\er
The semiclassical version of the selection rule (\ref{neq}) that  singles out the true charged topological solitons takes
now the form $(\b_0^2 = -{{2\pi }\o k})$:
\br
j_{el} \neq {{4\pi }\o {\b_0^2}} (j_{\varphi} - s(n-1))= 2k (s(n-1) - j_{\varphi}), 
\label{sel}
\er
where $j_{el} \in Z, \quad j_{\varphi} =
0, 1, \cdots n-2 \; mod \; (n-1)$.  
Therefore for {\it integer} $k$, we have to exclude certain $j_{el}$ according to eqn. (\ref{sel}), i.e. for fixed
topological charge $j_{\varphi}$, not all electric charges $j_{el}$ are allowed.  As a consequence the mass formula
(\ref{3.37}) makes sence for this specific set of allowed $(j_{el}, j_{\varphi})$ only.  

Similarly to the $A_n(p=1)$ model (\ref{1.1}) (see Sect 2 and 5 of ref. \cite{eletric}) due to T-duality that maps electric charges
of the axial model into the $Q_{\theta}^{a}$- topological charges of the vector model (and vice-versa), the topological
charges $Q_{\theta}^{a}$ (\ref{2.28}) ($Q_{\theta}^{a}=0 $ for our 1-soliton solution)  also acquires improvements
\br
Q_{\theta}^{a, impr} = f_a(\nu^a) j_{\varphi}
\label{3.38}
\er
$f_a(\nu^a)$ is certain function of $\nu^a$ determined  by T-duality (see for example eqns. (\ref{2.38}) of ref.
\cite{eletric}). 
The complete {\it semiclassical spectrum } of our $(Q_1, Q_n) $- charged 1-solitons (\ref{3.19})- (\ref{3.23}) is
represented by eqns. (\ref{3.36})- (\ref{3.38}).

\sect{Soliton  Vertex Operators}

\subsection{Dressing Transformations}
The first order soliton equations (\ref{3.3}), (\ref{3.4}) are by no means an effective tool for deriving the explicit form 
(\ref{3.19})-(\ref{3.23}) of the multicharged topological 1-solitons of the $A_n(p=2)$ dyonic IM.  The systematic construction of
the different species of N-solitons and breathers however, requires more powerful group theoretical methods as the
$A_n^{(1)}$-vertex operators and the corresponding $\tau$-functions \cite{liao}.
An important advantage of such method is that they make transparent the algebraic structure underlying the soliton solutions,
namely, the level one representation of the twisted $A_n$ affine  Kac-Moody algebra \cite{liao}, \cite{aratyn1}.  The dressing transformation
\cite{babelon} are known to be the origin of all those methods (including the vacuum B\"acklund transformation 
(\ref{3.1}) of Sect. 3).
 Let us consider two arbitrary solutions $B_s \in \hat G_0, \; s=1,2 $ of eqns. (\ref{2.16}) 
 written for the case of $A_n^{(1)}$
 extended by $d$ and the central term $c$, i.e.
 \br
 B_s = g_{0s} e^{\nu_s c + \eta_s d}, 
 \nonu
 \er
 The corresponding Lax (L-S) connections (\ref{2.19}) ${\cal{A}}(s) = {\cal{A}}(B_s), \; {\bar {\cal{A}}}(s) = 
 {\bar {\cal{A}}}(B_s) $ are related by gauge (dressing) transformations $\theta_{-,+} = \exp {\lie_{<,>}}$,
 \br
 {\cal{A}}_{\mu}(2)= \theta_{\pm} {\cal{A}}_{\mu}(1)\theta_{\pm}^{-1} + \( \pa_{\mu}\theta_{\pm}\) \theta_{\pm}^{-1}
 \label{4.1}
 \er
 They leave invariant the equations of motion (\ref{2.16}) as well as the auxiliar linear problem, i.e. the pure gauge 
${\cal{A}}_{\mu}$ defined in terms of the monodromy matrix $T(B_s)$,
\br
\(\pa_{\mu} - {{\cal{A}}(B_s)}_{\mu}\)T_s (B_s) = 0
\label{4.2}
\er
The consistency of equations (\ref{4.1}) and (\ref{4.2}) imply the  following relations 
\br
T_2 = \theta_{\pm} T_1,\; \; i.e. \; \; \theta_+ T_1 = \theta_- T_1 g^{(1)}
\label{4.3}
\er
where $g^{(1)} \in \hat {G}$ is an arbitrary constant element of the corresponding affine group.  Suppose $T_1 =
T_0(B_{vac})$ is the vacuum solution, 
\br
B_{vac} \eps_- B_{vac}^{-1} &=& \eps_-, \quad \bar \pa B_{vac}  B_{vac}^{-1}= \mu^2 z c, \nonu \\
{\cal {A}}_{vac} &=&-\eps_-, \quad \bar {{\cal {A}}}_{vac} =\eps_+ +\mu^2z c,
\label{4.4}
\er
and $T_0 = \exp (-z \eps_-) \exp (\bar z \eps_+)$ as one can easily check by using the fact 
that $[ \eps_+, \eps_-] = \mu^2
c$.  According to eqns. (\ref{4.1}) and (\ref{4.3}), every solution $T_2 = T(B)$ can be obtained 
from the vacuum configuration
(\ref{4.4}) by an appropriate gauge transformation $\theta_{\pm}$.  In fact, eqns. (\ref{4.1}) with
${\cal {A}}_{vac}$ and $\bar {{\cal {A}}}_{vac}$ as in eqn. (\ref{4.4}) and 
\br
{\cal {A}}(B) = -B \eps_- B^{-1}, \quad \bar {{\cal {A}}}= \eps_+ + \bar \pa B  B^{-1}
\nonu 
\er
allows to derive $\theta_{\pm}$ as functionals of $B$, i.e. $\theta_{\pm} = \theta_{\pm} (B)$.  We next apply eqns.
(\ref{4.3}), 
\br
\theta_-^{-1} \theta_+ = T_{vac} g^{(1)} T_{vac}^{-1}
\label{4.5}
\er
in order to obtain a non trivial field configuration  $B$  in terms of $g^{(1)}\in G$ and certain highest weight
(h.w.) representation  of the twisted algebra $A_n^{(1)}$ as we shall see in Sect. 4.2.  The first step consists in
substituting ${\cal {A}}_{vac},   \bar {{\cal {A}}}_{vac}$ and 
${\cal {A}}(B),   \bar {{\cal {A}}}(B)$ in eqn. (\ref{4.1}) and then solving it grade by grade remembering that
$\theta_{\pm}$ may be decomposed in  the form of infinite products
\br
\theta_- = e^{t(0)}e^{t(-1)}\cdots , \quad \quad \theta_+ =  e^{v(0)}e^{v(1)}\cdots
\nonu 
\er
where $t(-i)$ and $v(i), i=1,2, \cdots $ denote  linear combinations of grade $q=\mp i$ generators. 
 For grade zero we find 
 \br
 t(0) = H(\bar z), \quad \quad e^{v(0)} = B e^{G(z)-\mu^2 z \bar z c} \nonu 
 \er
 where the arbitrary functions $H(\bar z), G(z) \in \lie_0^0$ and are fixed to zero due to the subsidiary constraits
 (\ref{2.17}), (\ref{2.18}), i.e., $H(\bar z) = G(z)=0$.  The equations for $v(1), t(-1)$  
appears to be of the form
\br
B^{-1} \pa B - \mu^2 \bar z c = [ v(1), \eps_-], \quad \bar \pa B B^{-1} = [t(-1), \eps_+]+ \mu^2 z c
\nonu 
\er
The next step is to consider certain matrix elements (taken for the h.w. representation $|\l_l >$) of eqn. (\ref{4.5}). 
Since $v(i)|\l_l > = 0$ and $<\l_l | t(-i) =0, i>0$., we conclude that 
\br
<\l_l | B |\l_l >e^{-\m^2z \bar z} = <\l_l | T_0 g^{(1)} T_0^{-1} |\l_l >
\label{4.6}
\er
Taking into account the explicit parametrization of the zero grade subgroup element $B$ (\ref{2.10}) in terms of the
physical fields, $\nu, \varphi_i, \psi_a, \chi_a$ and choosing specific matrix elements we derive their
explicit  space-time dependence, 
\br
\tau_0 \equiv e^{\nu -  \mu^2 z \bar z} &=& < \lambda_0|T_{0}g^{(1)}T_{0}^{-1}|\lambda_0 >,
\nonu \\
\tau_{R_1} \equiv e^{{{nR_1}\o {n+1}} + {{R_{n}}\o {n+1}} +\nu -  \mu^2 z \bar z} &=& < \lambda_1|T_{0}g^{(1)}T_{0}^{-1}|\lambda_1 >,
\nonu \\
\tau_{R_{n}} \equiv e^{{{R_1}\o {n+1}} + {{nR_{n}}\o {n+1}} +\nu -  \mu^2 z \bar z} &=& <
\lambda_{n}|T_{0}g^{(1)}T_{0}^{-1}|\lambda_{n} >,
\nonu \\
\tau_j \equiv e^{\lambda_1 \cdot \lambda_j R_1 +\lambda_{n} \cdot \lambda_j R_{n}
 + \varphi_{j-1} +
\nu -  \mu^2 z \bar z} &=& < \lambda_j|T_{0}g^{(1)}T_{0}^{-1}|\lambda_j >,\;\;
j=2, \cdots n-1
\nonu 
\er
\br
\tau_{\psi_{l}} & \equiv & e^{(\l_1 \cdot \l_{l}  + {1\o 2}\d_{l,1} )R_1+
({1\o 2} \d_{l,n}+
\l_{n} \cdot \l_{l}) R_{l}+   
+\nu -  \mu^2 z \bar z}\psi_{l}, \;\; l=1,n \nonu \\
&=&
 < \lambda_{l}|T_{0}g^{(1)}T_{0}^{-1}E_{-\a_{l}}^{(0)}|\lambda_{l} >,
\nonu \\
\tau_{\chi_{l}} & \equiv & e^{(\l_1 \cdot \l_{l} + {1\o 2}R_1 \d_{l,1})R_1+
({1\o 2} \d_{l,n}
+ \l_{n} \cdot \l_{l}) R_{n}+  
+\nu -  \mu^2 z \bar z}\chi_l \nonu \\
&=&
 < \lambda_{l}|E_{\a_{l}}^{(0)}T_{0}g^{(1)}T_{0}^{-1}|\lambda_{l} >, l=1,  n
 \label{4.7}
\er
where for $A_n$ we have $\l_1 \cdot \l_j = {{n+1-j}\o {n+1}}, j=1, 2, \cdots n-1$. 
 In order to make the construction of the
solution solution (\ref{4.7})  complete it remains to specify the constant affine 
group element $g^{(1)}$.  The constant element  $g^{(1)}$ encodes
the information (including topological properties) about the N-soliton structure 
of eqns. (\ref{2.16}).  We shall see
 in the next Sect 4.2  that all 1-soliton solutions of the $A_n^{(1)}(p=2)$ dyonic IM
(\ref{1.2}) -
  the neutral $(0,0)$, $U(1)$-charged, 
$Q_1, 0)$ and $(0, Q_n)$, and the $U(1)\otimes U(1)$-charged  $(Q_1, Q_n)$ -  are 
 represented by certain  level one vertex
operators (i.e., h.w. representations) of the twisted $A_n^{(1)}$ algebra.  The 
twist being determined by the corresponding  
$A_n^{(1)}(p=2)$ graded structure $Q, \eps_{\pm}, \lie_0^0$.  The multi soliton 
solutions and the breathers are associated
to appropriate tensor products of these basic vertices.

\subsection{Vertex Operators}
Given $A_n^{(1)}$ affine algebra, 
\br
\lb H^m_i \, , \, H^n_j \rb &=& cm\,\d_{m+n,0}\,\d_{i,j} 
\qquad i,j = 1, \ldots , \mbox{\rm rank $\lie$}
\nonu \\
\lb H^{m}_{i} \, , \, E^n_{\a} \rb &=& (\a )^{i} E_{\a}^{m+n}
\nonu \\
\lb E^m_{\a} \, ,\, E^n_{\b} \rb  &=& \left\{ 
\begin{array}{ll}
\epsilon (\a , \b ) E_{\a + \b }^{m+n} 
& \, \mbox{\rm if $\a + \b$ is a root} \\
 \a \cdot  H^{m+n} +cm\,\d_{m+n,0}& \,\mbox{if $\a + \b = 0$} \ \\
  0 & \, {\rm otherwise}
  \end{array} \right. 
\label{4.9}
\er
($m,n \in Z$ and $c$ is its level to be taken to $1$ in the vertex operator representations),
provided with the graded structure $\{ Q, \eps_{\pm}, \lie_0^0 \}$ as in eqns. (\ref{2.9}). It is important to mention that the chiral two loop Kac-Moody algebra spanned by
$J^a (z;\l ) = \sum_{n,s \in Z} J^a_{n,s}z^{-n-1} \l ^{-s-1}$ (where $J^a_{0,m}$ denote  $H^{m}_i$ and $E^{m}_{\a}$) 
has two independent central extensions:
\br
[J^a_{s_1}(z_1), J^b_{s_2}(z_2)] &=& if^{abc}J^c_{s_1+s_2}(z_1)\d
             (z_1-z_2) \nonu \\ 
             &+& i{{k}\o {2\pi}}g^{ab} \d ^{\prime} (z_1-z_2)\d_{s_1+s_2,0} + {{c}\o {2\pi}}
             s_1 g^{ab}\d (z_1-z_2)\d_{s_1+s_2,0} \nonu
\er
The label $k$ that appear as the coupling $\b^2 ={{2\pi} \o {k}}$ in the WZW action is related to the mode expansion 
 in the space time variables $z, \bar z$. The central term $c$ appearing in (\ref{4.9}) is related to the mode expansion in
 the spectral parameter $\l$. In the vertex operator construction below, the only $\l$-part of the algebra spanned by 
 $J^a_{0,m}$, representing the dressing symmetry takes place.

 Since $\eps_{\pm}$  form a
Heisenberg subalgebra, 
\br
[\eps_+, \eps_- ]= \mu^2 c
\nonu
\er
and we have to calculate the matrix elements ($\tau$-functions ), say
\br
< \l_0 | e^{-z \eps_-} e^{\bar z \eps_+} g^{(1)} e^{-\bar z \eps_+}e^{z \eps_-} | \l_0 >,
\nonu 
\er
it is instructive to introduce a new basis for the $A_n^{(1)}$, such 
that $\eps_{\pm}$ appear as part of its Cartan
affine subalgebra.  Let us first observe that the following specific 
linear combination of $ H_i^{(l)}, E_{\b}^{(k)}$ \cite{aratyn1},
\br
{b}^a_{a+m(n-1)} &=& \sum _{i=1}^{n-1-a}
E^{(m)}_{\a _{i+1}+ \a _{i+2} + \cdots +\a _{i+a}}  
 + \sum _{i=1}^{a} E^{(m+1)}_{-(\a _{i+1}+ \a _{i+2} 
+ \cdots +\a _{i+n-1-a})}, \quad a=1, \cdots ,n-2 \nonu \\
{b }^0_{m(n-1)} &=& \sqrt{ {{(n-1)}\o 2}} \( \l_1 + \l_{n}\) \cdot H^{(m)},\quad 
{ b}_{m} =  \sqrt { {{(n+1)}\o {2(n-1)}}}\( \l_1 - \l_{n}\) \cdot H^{(m)}
\label{4.10}
\er
close new $n$-dimensional Heisenberg subalgebra,
 \br
 [b^a_{a+m(n-1)}, \( {{b^b}}_{b+l(n-1)}\) ^{\dagger} ] &=& (a+m(n-1))\d_{m,l}\d_{a,b},  \nonu 
 \er
 \br
 [b^a_{a+m(n-1)},b_{m}] &=& 0,\quad a=0,1, \cdots ,n-1,\nonu 
 \er
 \br
 [ b_{m} ,b_{l}] &=& m \d_{m+l,0}
 \label{4.11}
\er 
Therefore we can consider the generators $b^a_{a+m(n-1)}, a=0, 1, \cdots, n-1$ together with $b_{m}$ (and their conjugate)
as the generators of the affine Cartan subalgebra in the new basis for the twisted $A_n^{(1)}$ \footnote{ Note that  
$b^a_{a+m(n-1)}, a\neq 0$ has no zero modes, i.e., only two of the generators of the new Cartan subalgebra,
 namely $b^0_{m(n-1)}, b_{m}$ are untwisted.}.  Next step is to complete this basis with the corresponding new step
 operators $e_{\tilde \a}^{(l)}$.  Since we have by definition that 
 \br
 \eps_+ = \mu b_1^1, \quad \quad \eps_- = \mu (b_1^1)^{\dagger}
 \nonu 
 \er
 commute with $b_0^0$ and $b_0$, in the vanishing central charge case,   they all commute and hence share the same set of
 eigenvectors $F(\g)$,
 \begin{equation}
\lbrack \eps ^{\pm },F(\gamma )]=f^{\pm }(\gamma )F(\gamma ), \; \;\;
\lbrack b_0^0,F(\gamma )]=f_0^0(\gamma )F(\gamma ), \; \; \;\lbrack b_0,F(\gamma )]=f_0(\gamma )F(\gamma )
 \nonu
\end{equation}
Following ref. \cite{aratyn1} we find  four different types of eigenvectors 
\br  
F_{a,j}(\g ) &=& {{\hat c}\o {(w^a -1)}} 
+ \sum_{m\in Z} \gamma^{-m(n-1)} \sum_{i=1}^{n-2}
h_{i+1}^{(m)} \sum_{p=1}^{i} w^{a(i-p)} \nonu \\
&+& \sum_{b=1}^{n-2}
\sum_{m\in Z} w^{bj} \gamma^{-(b+m(n-1))}\(\sum_{i=1}^{n-1-b} w^{a(i-1)}
E^{(m)}_{\a_{i+1} +\a_{i+2} + \cdots +\a_{i+b}}\right. \nonu \\  
& +& \left. \sum_{i=1}^{b} w^{a(i+n-2-b)}E^{(m+1)}_{-(\a _{i+1} + \a _{i+2}
+ \cdots + \a_{i+n-1-b})}\) 
\label{4.12}   
\er
where $j=1,...,n-1, a=1,...,n-2$ and
$w=\exp \( \frac{2\pi i}{(n-1)}\)$, together with
\br
\bar F_{1,l} (\g ) &=& \sqrt{2} \sum_{m\in Z} \gamma^{-m(n-1)} \sum_{p=0}^{n-2} w^{pl} \g ^{-p} E_{\a_1 + \cdots + \a_{p+1}}^{m}, \nonu \\
\tilde F_{1,l} (\g ) &=& \sqrt{2}\sum_{m\in Z} \gamma^{-m(n-1)} \sum_{p=0}^{n-2} w^{pl} \g ^{-p} E_{\a_n + \cdots + \a_{n-p}}^{m}, \quad 
l=0, \cdots ,n-2 \nonu \\
 F^{\pm} (\g ) &=&\sum_{m\in Z} \gamma^{-(m+1)(n-1)+1} E_{\pm (\a_1 + \cdots + \a_{n})}^{m}
\label{4.13}
\er
 Their eigenvalues are obtained from
\br
\lbrack \eps ^{\pm },F_{a,j}(\gamma )] & =& \mu w^{\mp j}(w^{\pm a}-1)\gamma^{\pm}
F_{a,j}(\gamma )\nonu \\
 \lbrack \eps ^{\pm },\bar F_{1,l}(\gamma )] & =& - \mu\g ^{\pm 1}w^{\mp l} \bar F_{1,l}(\gamma )
 \nonu \\
\lbrack \eps ^{\pm },\tilde F_{1,l}(\gamma )] & =& \mu\g ^{\pm 1}w^{\mp l} \tilde F_{1,l}(\gamma ) \nonu \\
\lbrack \eps ^{\pm },F^{\pm }(\gamma )] & =& 0
\label{4.14}
\er
In fact, $F_{a,j}, \bar F_{1,l}, \tilde F_{1,l},  F^{\pm}$  are eigenvectors of all oscillators 
$b^a_{a+m(n-1)}$ and $b_{m}$, in particular,
\br
\lbrack b^{a_1}_{a_1+l(n-1)},F_{a_2,j}(\gamma )] & =&  w^{-a_1 j}(w^{a_1a_2}-1)\gamma^{a_1+l(n-1)}F_{a_2,j}(\gamma )
\nonu \\
\lbrack b_{l},F_{a,j}(\gamma )] & =&0, 
\label{4.15}
\er
We can therefore identify the step operators in the new basis as follows
\br
e_{\tilde {\b}_j}^{(l)} = \oint {{d\g }\o {2\pi i \g}} \g^{a+l(n-1)}F_{a,j}(\gamma ), 
\nonu 
\er
With the new basis at hand, we define the $A_n^{(1)}$ affine group element $g^{(1)}$ as
\br
g^{(1)}= e^{dF(\g )}, \quad F(\g ) =F_{a,j}, \bar F_{1,l}, \tilde F_{1,l},  F^{\pm}
\nonu 
\er
or more generally as a product including all the $A_n^{(1)}$ generators,
\br
g^{(1)}= e^{d_1F_1(\g_1 )}e^{d_2F_2(\g_2 )} \cdots e^{d_NF_N(\g_N )}
\label{4.16}
\er
The new basis introduced above  drastically simplifies the calculation of the $\tau$-functions (\ref{4.7}),
\br
T_0 g^{(1)} T_0^{-1} = \exp (d \rho (\g) F(\g )) = 1 + d \rho (\g) F(\g )
\label{4.17}
\er
where $\rho (\g) = \exp (-z f^-(\g) + \bar z f^+ (\g) )$.
 The last equality in (\ref{4.17}) reflects the fact that $F^2(\g ) =0$ for $A_n^{(1)}$ due to the well known properties of
the short distance OPE (see ref \cite{aratyn1}).

An important question concerns the specific choice of the form (\ref{4.16}) of $g^{(1)}$ that leads to different species of
neutral and charged solitons and breathers.  As in the $A_n^{(1)}(p=1)$ dyonic IM case \cite{eletric}, \cite{dyonic}  taking 
\br
g^{(1)}_a (\g)  = e^{dF_{a,n-1}(\g)}
\label{4.18}
\er
we reproduce the abelian affine $A_{n-2}^{(1)}$-Toda neutral 1-soliton solutions,
since $\tau_{\psi_a} = \tau_{\chi_a}=0$.  The corresponding $U(1)$ charges vanish identically, 
i.e. $Q_1 = Q_n =0$.   Another indication of the neutrality of such solution is that 
$F_{a, n-1}(\g)$ has zero eigenvalue
with respect to the generators of the $U(1)\otimes U(1)$ charges, namely, $\l_1 \cdot H$ and 
$\l_n \cdot H$.

The choice 
\br
g^{(1)}_1 (\g_1, \g_2)  = e^{\bar d_1\bar F_{1,1}(\g_2)}e^{\bar d_2\bar F_{1,1}^{\dagger}(\g_1)}, \quad 
g^{(1)}_n (\g_1, \g_2)  = e^{\tilde d_1 {\tilde F_{1,1}}^{\dagger}(\g_2)}e^{\tilde d_2\tilde  F_{1,1}(\g_1)}
\label{4.19}
\er
for $\g_1 = e^{B-i(\a +\pi)}, \; \g_2 = e^{B+i\a }$ (and further conditions on $\bar d_1, \bar d_2, 
\tilde d_1, \tilde d_2$ as in ref. \cite{dyonic}) leads to nontrivial $U(1)$ charged $\tilde \a_1-(Q_1, 0)$ or 
$\tilde \a_n-(0, Q_n)$ 1-soliton solutions. They provide  1-soliton solutions with $\psi_n = \chi_n =0$ or 
 $\psi_1 = \chi_1 =0$ coinciding with the electrically charged 1-soliton solution of the 
$A_n^{(1)}(p=1)$ dyonic IM \cite{eletric}, \cite{dyonic}.  Similarly, the 3-vertex solution generated by 
\br
g^{(1)}_{1a} (\g_1, \g_2, \g_3)  =g^{(1)}_1 (\g_1, \g_2)g_a^{(1)} (\g_3), \quad 
g^{(1)}_{na} (\g_1, \g_2, \g_3)  = g^{(1)}_n (\g_1, \g_2)g_a^{(1)} (\g_3)
\label{4.20}
\er
for an appropriate choice of $\g_i, i=1,2,3$ and $d, \bar d_1, \bar d_2, 
\tilde d_1, \tilde d_2$ (as in ref. \cite{dyonic}) give rise to $U(1)$-charged $\tilde \a_1$- and $\tilde \a_n$-breathers.

The simplest genuine $A_n^{(1)}(p=2)$ nontrivial $(Q_1, Q_n)$-charged 1-soliton solution (\ref{3.19})-(\ref{3.23}) 
correspond to 
\br
g^{(1)}_{1n}(\g_1, \g_2, \g_3, \g_4) = e^{ d_1\bar F_{1,1}(\g_2)}e^{ d_2\bar F_{1,1}^{\dagger}(\g_1)}
e^{ d_4\tilde F_{1,1}^{\dagger}(\g_4)}e^{ d_3\tilde  F_{1,1}(\g_3)}
\label{4.21}
\er
for $\g_1 = e^{B-i(\a +\pi)}, \; \g_2 = e^{B+i\a }$ and $d_1d_2 =- d_3 d_4 =d_0$.
The $(Q_1, Q_n)$-charged breather can be generated by taking 
\br
g^{(1)}_{1n}(\g_1, \g_2, \g_3, \g_4, \g_5) = g^{(1)}_{1n}(\g_1, \g_2, \g_3, \g_4)g_a^{(1)}(\g_5)
\label{4.22}
\er
and special choice of $\g_l, l=1,2,3,4,5$ and the coefficients $d_l$.  The multi-soliton solutions are indeed represented by
$g^{(1)}$ that includes products of the corresponding 1-soliton vertices $g^{(1)}_{1n}$ and $g^{(1)}_{a}$, say
\br
g^{(1)} = \prod_{s=1}^{N} g^{(1)}(\g_1^s, \g_2^s)
\nonu
\er
again with specific relations among the parameters $ \g_1^s, \g_2^s, \bar d_1^s, \bar d_2^s, \tilde d_1^s, \tilde d_2^s$.

\sect{Dyonic $A_n^{(1)}(p=2, q)$ hierarchy}

\subsection{Constrained KP hierarchies of $A_n^{(1)}(p=2,q=2)$ dyonic type} 

 Our motivation to introduce $A_n^{(1)}(p=2, q)$ dyonic hierarchy (of which
  $q=-1$ integrable model (\ref{1.2}) is a member) and to study the relation 
  between the $1-$solitons of $q=-1$ and say $q=2$, dyonic integrable models is
  twofold:
  
  (a) as we shall show in  section $5.3$, the knowledge of $(Q_1,Q_n)$-charged
 $1$-soliton (\ref{3.19}), (\ref{3.23}) of  $q=-1$ model (\ref{1.2}) allows to 
 construct the corresponding $1-$solitons of all $q \geq 2$ models by simple 
 change of variables (both - the field ones: $r_a(x,t)$, $q_a(x,t) $, $u_l(x,t)$   $\rightarrow \psi_a, 
  \chi_a, 
  \varphi_l$ - and the space-time ones: $\bar{z} 
  \rightarrow x, \frac{z}{\gamma} \rightarrow t\gamma^q )$;
  
   (b) since the equations of motion of $q \geq 2$ models are much simpler 
   than the $q=-1$ ones (that follows from (\ref{1.2})), the Hirota method
    \cite{aratnp}, \cite{liao} (and the explicit derivation of $G_0-\tau$- functions) works  
    more effectively in say, $q=2$ case than in the $q=-1$ one. 
    
    This suggests that in construction of multisolitons and breathers  to first
elaborate them for $q=2$ model and then to apply the above change of variables
 in order to find the corresponding solutions of the $q=-1$ model.
  The origin of the $A_n^{(1)}(p=2, q)$ dyonic hierarchy is in the fact that
   keeping a part of $q=-1$ graded structure (\ref{2.9}) : $Q, 
  {\epsilon_+}, \lie_0^0$ and replacing $\eps_-$
   (of grade $q=-1$) by an arbitrary constant element $D_q^q$ of $Q$-grade 
   $q \geq -1$ such that 
   \footnote{The additional condition $[D_q^q,
   {\epsilon}_-]=0$ is imposed in order to obtain local equations of
    motion   and allows to relate the $1-$solitons of q-model to the
     ones of $q=-1$ model}
   \be
   D_q^q \in Ker \( ad {\epsilon}_{\pm}\), \quad (D_{-1}^{-1}={\epsilon}_-)
   \label{5.1}
   \ee
   one can generate an infinite hierarchy of new integrable models.
    Their equations of motion have the following compact zero curvature form
    \cite{aratyn1}, \cite{aratyn2}:
    \be 
  \pa _{t_q} D_0 - \pa_{x} D_q - [ D_q, D_0 ]=0
  \label{5.2} 
 \ee
 where
 \be
 D_0=\bar{\cal A}(\bar{\partial}=\partial_x)={\epsilon}_+ 
 +(\partial_x g_0)g_0^{-1}, \quad D_q=\sum_{l=0}^q D_q^{(l)}, \, \, \, D_q^{(l)} \in \lie_l 
 \label{5.3}
 \ee 
   An important difference with $q=-1$ case is that now the "physical fields"
    $r_a(x,t), q_a(x,t)$ and $u_l(x,t)$
 parametrize the zero grade algebra $\lie_0$ (and not the zero grade group 
 element $g_0 \in G_0$ as in $q=-1$ case):
 \be
 {\cal A}_0=(\partial_x g_0)g_0^{-1}=\sum_{a=1, n} (r_a E_{-\alpha_a}^{(0)}+
 q_a E_{\alpha_a}^{(0)}) + \sum_{l=1}^{n-2}u_l h_{l+1}^{(0)}
 \label{5.4}
 \ee 
 The remaining grade $l=0, 1, \dots, q-1$ ``gauge potentials'' $D_q^{(l)}$ 
 introduce a set of ``auxiliary fields'' $V_{k}^{(0)}(x,t), V_{k}^{(1)}(x,t), etc$
 with $k=1, 2,\dots, n+4, \; n\geq 4$:
 \br
 D_q^{(0)} &=&\sum_{i=1}^{n} V^{(0)}_{i}h_i + V^{(0)}_{n+1} E_{\a_1}^{(0)} + V^{(0)}_{n+2}E_{\a_n}^{(0)}
 + V^{(0)}_{n+3}E_{-\a_1}^{(0)}+ V^{(0)}_{n+4}E_{-\a_n}^{(0)}, \nonu \\
 D_q^{(1)} &=& \sum_{i=1}^{n-2} V^{(1)}_{i}E^{(0)}_{\a_{i+1}} + V^{(1)}_{n-1} E_{\a_1+\a_2}^{(0)} 
 + V^{(1)}_{n}E_{\a_{n-1}+\a_n}^{(0)}
 + V^{(1)}_{n+1}E_{-(\a_2+\cdots + \a_{n-1})}^{(1)} \nonu \\
 &+&V^{(1)}_{n+2}E_{-(\a_1 + \cdots +\a_{n-1})}^{(0)} +
 V^{(1)}_{n+3}E_{-(\a_2 + \cdots +\a_{n})}^{(0)}+
 V^{(1)}_{n+4}E_{-(\a_1 + \cdots +\a_{n})}^{(0)}, \nonu \\
 \er
 etc.  Writting eqs. (\ref{5.1}) for each grade $0, 1, \dots, q+1$ we get the following
system of equations:
\br
[D_q^{(q)}, \eps_+] &=& 0 \nonu \\
\pa_{x} D_q^{(q)}  + [D_q^{(q)}, {\cal A}_0 ] + [D_q^{(q-1)}, \eps_+ ] &=& 0 \nonu \\
\pa_{x} D_q^{(q-1)}  + [D_q^{(q-1)},{\cal A}_0 ] + [D_q^{(q-2)}, \eps_+ ] &=& 0 \nonu \\
\vdots &=& \vdots \nonu \\
\pa_{x}D_q^{(1)} + [D_q^{(1)}, {\cal A}_0 ] + [D_q^{(0)}, \eps_+ ] &=& 0 \nonu \\
\pa_{t_q}{\cal A}_0 -\pa_{x} D_q^{(0)}  - [D_q^{(0)}, {\cal A}_0 ] &=& 0
\label{5.5}
\er
 The recipe in deriving the system of differential equations (of order $q$ in $x$
) for $r_a, q_a, u_l$ consists in first determining $D_q^{(q)} \in Ker \( ad 
{\epsilon}_{\pm}\) $ (i.e. starting from the first equation). For, say, 
$q=2$ we obtain ($n\geq 4)$
\br
D_2^{(2)}= -b_2^{(2)}=-\sum_{i=1}^{n-3} E_{\alpha_{i+1}+\alpha_{i+2}}^{(0)}
 -\sum_{i=1}^{2} E_{-(\alpha_{i+1}+\dots+\alpha_{i+n-3})}^{(1)}
 \label{5.6}
\er
where $b_2^{(2)} $ is a particular case of $b^a_{a+m(n-1)}, a=2, m=0$ given by eqn. (\ref{4.10}). 
Substituting eq. (\ref{5.6}) in the second of the equations (\ref{5.5}) we find
 a part of the auxiliary fields  $V_{k}^{(q-1)}$ parametrizing $Im (ad 
 {\epsilon}_+)$ in terms of $r_a, q_a, u_l$. The third 
 equation determines the  remainning part of $V_{k}^{(q-1)}$ lying in the
  $Ker \( ad{\epsilon}_{+} \) $ and those fields $V_{k}^{(q-2)}$ lying in  the $Im \(ad (\eps_{+} )\)$,
   etc. Applying this procedure for the case
   $q=2$ we find the following system of equations defining the  
   $A_n^{(1)}(p=2, q=2)$ nonrelativistic dyonic integrable model:
  \br
 && \pa_{t_2} q_1 - \pa^2_x q_1 + {{2q_1}\o {n-1}}\(  \sum_{k=1}^{n-2} \pa_x u_k +
   {1\o 2}\sum_{k,j=1}^{n-2} k_{kj} u_k u_j + r_1 q_1 + r_nq_n \) =0 \nonu \\
 && \pa_{t_2} r_1 + \pa^2_x r_1 -
   {{2r_1}\o {n-1}}\( (1-n)\pa_x u_1 + \sum_{k=1}^{n-2} \pa_x u_k +
   {1\o 2}\sum_{k,j=1}^{n-2} k_{kj} u_k u_j + r_1 q_1 + r_nq_n \) =0 \nonu \\
  &&\pa_{t_2} q_n + \pa^2_x q_n - {{2q_n}\o {n-1}}\(  \sum_{k=1}^{n-2} \pa_x u_k +
   {1\o 2}\sum_{k,j=1}^{n-2} k_{kj} u_k u_j + r_1 q_1 + r_nq_n \) =0 \nonu \\
 && \pa_{t_2} r_n - \pa^2_x r_n +
   {{2r_n}\o {n-1}}\( (1-n)\pa_x u_{n-2} + \sum_{k=1}^{n-2} \pa_x u_k +
   {1\o 2}\sum_{k,j=1}^{n-2} k_{kj} u_k u_j + r_1 q_1 + r_nq_n \) =0 \nonu \\
  &&\pa_{t_2} u_l + \pa_x V^{(0)}_l= 0, \quad l=1,2, \cdots n-2,
\label{5.7}
\er
where
\br
V_l^{(0)} &=& {{1}\o {n-1}} \(\pa_x L_l +M_l \)\nonu \\
L_l &=&2(n-l-1)\sum_{k=1}^{l-1} u_k + (n-2l-1) u_l -2l\sum_{k=l+1}^{n-2}u_k, \nonu \\
M_l &=&2\((n-l-1)r_1q_1 -l r_nq_n \) + (n-l-1) \sum_{i,j=1}^{l}k_{ij}u_iu_j - l \sum_{i,j=l}^{n-2} k_{ij}u_i u_j -
(n-2l-1)u_l^2
\nonu 
\er
We take the $q=2$ $A_n^{(1)}(p=2)$ integrable model defined by the above 
   system of equations as a representative of $A_n^{(1)}(p=2)$ positive grade
   $q \geq 2$ dyonic hierarchy to further study the relation between its soliton
   solutions with the $1$-solitons (\ref{3.19})-(\ref{3.23}) of $q=-1$ dyonic 
   model.
    
   The simplest case $n=3$ should be treated separately. Taking $D_2^{(s)} (n=3), \; s=0,1,2$ in the form
   \br
  D_2^{(2)} (n=3)&=& (\l_3 -\l_1)\cdot H^{(1)}\nonu \\
  D_2^{(1)} (n=3)&=& q_1E_{\a_1+\a_2}^{(0)} + r_1 E_{-\a_1-\a_2}^{(1)} + q_3E_{\a_2+\a_3}^{(0)} - 
  r_3 E_{-\a_2-\a_3}^{(1)} \nonu \\
  D_2^{(0)} (n=3)&=& (\pa_x q_1 +q_1 u) E_{\a_1}^{(0)} - (\pa_x q_3 +q_3 u) E_{\a_3}^{(0)}-
  (\pa_x r_1 -r_1 u) E_{-\a_1}^{(0)} \nonu \\
  &+& (\pa_x r_3 -r_3 u) E_{-\a_3}^{(0)}
  +{1\o 2}(r_1 q_1 - r_3q_3)h_2^{(0)}
  \label{xx}
  \er
 we get the explicit form (\ref{5.12}) of the equations of motion for the $A_2^{(1)}(p=q=2)$ nonrelativistic IM.

\subsection{Charged 1-solitons of $q=2$ model}

The construction of the soliton solutions for each member of $A_n (p=2, q)$
 hierarchy by vertex operator method is quite similar to the $q=-1$ case 
 described in  Section $4$. The main difference is in the form of the vacua 
 (constant) solutions:
 \br
D_q^{(v)}= D_q^{(q)}, \, \, \, D_0^{(v)}={\epsilon}_+, \quad  T_q^{(v)}=e^{-t_qD_q^{(q)}}e^{x{\epsilon}_+}
\label{5.8}
 \er
 (for $q=2$ $D_2^{(2)}$ is given by grade $2$ constant element (\ref{5.6})) and in the
  choice of new set of $T_vg^{(1)}T_v^{-1}$-matrix  elements (different from 
  (\ref{4.7}) for $q=-1$ case) representing "physical fields" $r_a, q_a, u_l$. Note that the
 form of $g^{(1)}$ (say (\ref{4.18}-\ref{4.21})) remains unchanged since 
 $[D_q^{(q)}, \eps_{\pm}]=0$, henceforth they share the same set of eigenvectors
 (\ref{4.12}), (\ref{4.13}). Following the dressing procedure of Sect. $4.1$ 
 we find for $q=2$ model:
 \br
 e^{v{(0)}}=B, \, \, \, (\partial_x B)B^{-1}= {\cal A}_0= [t{(-1)},{\epsilon}_+]
 \nonu
 \er
 \br
 t{(-1)}&=&-\sum_{l=1}^{n-2}(u_l+\partial_x \tilde \nu)E_{-\alpha_{l+1}}^{(0)} 
       -r_1E_{-\alpha_{1}-\alpha_{2}}^{(0)}+r_nE_{-\alpha_{n-1}-\alpha_{n}}^{(0)} \nonumber \\
       &+&q_1E_{\alpha_{1}+\dots+\alpha_{n-1}}^{(-1)}-q_nE_{\alpha_{2}+\dots+\alpha_{n}}^{(-1)} 
       + \partial_x \tilde \nu E_{\alpha_{2}+\dots+\alpha_{n-1}}^{(-1)}+
    WE_{\alpha_{1}+\dots+\alpha_{n}}^{(-1)}
     \label{5.9}
 \er
 where $W{(t, x)}$ is an arbitrary function since 
 $E_{\alpha_{1}+\dots+\alpha_{n}}^{(-1)} \in Ker \( ad {\epsilon}_+\) $;
  $\tilde {\nu}=\nu-\mu^2xt$. Taking into account the explicit form (\ref{5.9}) of
   $v{{(0)}}, t{(-1)}$ and $\theta_{\pm}$ (and the definition of the highest 
 weight states $|\lambda_j>, <\lambda_j|$) we consider appropriate matrix elements of 
 eqs. (\ref{4.5}) in order to extract the $\tau-$functions representing 
 $r_a, q_a, u_l$:
 \br
\tau_0 \equiv e^{\tilde \nu} &=& < \lambda_0|T_{v}{g^{(1)}T_{v}}^{-1}|\lambda_0 >,
\nonu \\
\tau_{1,(n-1)} \equiv -q_1e^{\tilde \nu} &=& < \lambda_0|E_{-\alpha_{1}-\dots-\alpha_{n-1}}^{(1)}T_{v}g^{(1)}T_{v}^{-1}|\lambda_0 >,
\nonu \\
\tau_{2n} \equiv q_ne^{\tilde \nu} &=& <
\lambda_{0}|E_{-\alpha_{2}-\dots-\alpha_{n}}^{(1)}T_{v}g^{(1)}T_{v}^{-1}|\lambda_{0} >,
\nonu \\
\tau_{12} \equiv r_1 e^{\Phi(\lambda_1) } &=& < \lambda_1|E_{\alpha_{1}+\alpha_{2}}^{(0)}T_{v}g^{(1)}T_{v}^{-1}|\lambda_1 >
\nonu
\er
\br
\tau_{n-1, n}  \equiv  -r_ne^{\Phi(\lambda_n)}&=&
 < \lambda_{n}|E_{\alpha_{n-1}+\alpha_{n}}^{(0)}T_{v}g^{(1)}T_{v}^{-1}|\lambda_{n} >,
\nonu \\
\tau_{l} \equiv e^{\Phi(\lambda_l)}&=&
 < \lambda_{l}|T_{v}g^{(1)}T_{v}^{-1}|\lambda_{l} >, l=2, 3, \cdots, n-1
 \label{5.10}
\er
where 
\br
\Phi(\lambda_1)&=&\lambda_1^2 R_1 +\lambda_1\cdot\lambda_n R_n +\tilde \nu, \quad  
\Phi(\lambda_n)=\lambda_n^2 R_n +\lambda_1\cdot\lambda_n R_1 +\tilde \nu  \nonumber\\
\Phi(\lambda_l)&=&\varphi_{l-1}+\lambda_l\cdot\lambda_1 R_1 +
\lambda_l\cdot\lambda_n R_n +\tilde \nu , l=2,\cdots, n-1 \nonumber\\
\lambda_l\cdot\lambda_n &=& \frac{l}{n+1}, \quad  \lambda_1\cdot\lambda_l=\frac{n+1-l}{n+1}
\er
Therefore the soliton solutions of $A_n^{(1)}(p=2, q=2)$ IM (\ref{5.7}) can be written 
in terms of the above $\tau$-functions as follows
\br
r_1 = -\frac{\tau_{12}}{\tau_1}; \quad r_n = \frac{\tau_{n-1,n}}{\tau_n}; \quad 
q_1 = \frac{\tau_{1, (n-1)}}{\tau_0}; \quad q_n = -\frac{\tau_{2n}}{\tau_0};\nonu \\
u_l = \pa_x ln \frac{\tau_{l+1}}{\tau_0}, \, \, \, l=1, 2, \dots, n-2.
\label{5.11}
\er
It remains to calculate the matrix elements (\ref{5.10}) for say, $g^{(1)}$ 
represented  by $4$-vertex describing $(Q_1, Q_n)$-charged $1-$soliton. We shall present here the 
explicit form of this ($4-$vertex) solution for the simplest case $n=3, q=2$ only,
since the general case of $U(1) \times U(1)$-charged $1$-soliton of $A_n(p=2, q=2)$
system (\ref{5.7}) $n=3, 4, \dots$ will be derived in Sect. $5.3$ by applying the 
vacua B\"acklund transformation method of Sect. $3$. For $n=3$ the system 
(\ref{5.7}) gets the following simple form ($t=t_2$):
  \br
\pa_{t} q_1-  \pa^2_x q_1 + q_1\pa_x   u   + q_1 u^2 + r_1 q_1^2 + r_3 q_1 q_3 &=&0 \nonu \\
\pa_{t} r_1 + \pa^2_x r_1 + r_1 \pa_x  u   - r_1 u^2 - r_1^2 q_1 - r_1 r_3 q_3 &=&0 \nonu \\
\pa_{t} q_3 + \pa^2_x q_3 - q_3 \pa_x  u   - q_3 u^2 - r_3 q_3^2 - r_1 q_1 q_3 &=&0 \nonu \\
\pa_{t} r_3 - \pa^2_x r_3 - r_3\pa_x  u   + r_3 u^2 + r_3^2 q_3 + r_3 r_1 q_1 &=&0 \nonu \\
\pa_{t} u + \pa_x \( r_1 q_1 \) -  \pa_x \( r_3 q_3 \) &=&0
\label{5.12}
\er
Its charged $1-$solution ($4$-vertex) solution is obtained by simple algebraic 
manipulations (involving $A_3^{(1)}-$algebra)
\br
\tau_0 &=& 1 + \(c_{12}b_1b_2+c_{34}b_3b_4 \) \rho_1  \rho_2 \nonu \\
\tau_2 &=& 1 + \(d_{12}b_1b_2 + d_{34}b_3b_4 \)\rho_1 \rho_2 \nonu \\
\tau_{1, n-1} &=& b_1 f_1\rho_1 \quad
 \tau_{1, 2} = b_2 e_2 \rho_2 \nonu \\
\tau_{2n} &=& b_4 i_4\rho_2 \quad 
\tau_{n-1, n}= b_3 g_3 \rho_1
\label{5.13}
\er
where
\br
\rho_1(\gamma_1)=e^{-x\gamma_1+t\gamma_1^2}, \quad \rho_2(\gamma_2)=e^{x\gamma_2-t\gamma_2^2}
\er
\br
c_{12}= {{2\g_1 \g_2^2}\o {{(\g_1-\g_2 )^2(\g_1 +
\g_2)}}}=d_{34}, \quad c_{34}= {{2\g_1^2 \g_2}\o {{(\g_1-\g_2 )^2(\g_1 +
\g_2)}}}=d_{12}\nonumber 
\er 
\br
f_1= \sqrt{2} \g_1=g_{3}, \quad e_2= \sqrt{2} \g_2=i_4
\label{5.14}
\er
The above solution is a particular case of the more general ($4$-parameter 
$\gamma_i$) $4$-vertex ($2$-soliton) solution, given by eq. (\ref{A.1})-
(\ref{A.3}) of the Appendix. The restrictions
\br
\gamma_1=\gamma_3, \quad \gamma_4=\gamma_2 \quad \gamma_1=-\gamma_2^*=e^{B+i\alpha} 
\label{5.15}
\er
we have imposed on it, ensure that the solution (\ref{5.13}-\ref{5.15}) 
represents finite (positive) energy $(Q_1,Q_n)$-charged topological $1-$soliton,
as we shall show in next section $5.4$. One can easily verify by susbstituting
$r_a, q_a, u$ (\ref{5.11}) (written in terms of $\tau-$functions (\ref{5.13})
 or (\ref{A.1})) in the system (\ref{5.12}) that they indeed satisfy
 equations (\ref{5.12}), i. e. that $\tau-$function method provides solutions of this 
 system.
 It is important to note that for $n=3$ and $q>2$ arbitrary integer (i. e. $A_3^{(1)}(p=2, 
 q)$ IM) the form of the corresponding $1-$soliton solutions is the same as in 
 $q=2$ case given by eqs. (\ref{5.13}), (\ref{5.14}) with the only difference
 that $\rho_1^{(q)}$ and $\rho_2^{(q)}$ have been changed to:
 \br
 \rho_1^{(q)}(\gamma)=e^{-x\gamma+t_q\gamma^q}, \quad  \rho_2^{(q)}(\gamma)=e^{x\gamma-t_q\gamma^q}
 \er
 The system of equations for $A_3^{(1)}(p=2, q=2)$ model is indeed of order
 $q$ in $\partial_x$ (i. e. the higher derivative is $\partial_x^q$) and involves
  more products of derivatives terms then $q=2$ system (\ref{5.12}).

 \subsection{$1$-Soliton relations: $q=2$ vs. $q=-1$ solitons}

The common graded algebraic structure of $q=-1$ and $q=2$  $A_n^{(1)}(p=2)$ integrable models
  address the question whether one can construct $q=2$ solitons in terms of corresponding  
  $q=-1$ ones. By comparing $q=-1$ and $q=2$ dressing transformations (\ref{4.1}),
   the form of $T_vg^{(1)}T_v^{-1}$ matrix  elements
\br
T_v(q)e^{bF(\gamma)}T_v^{-1}(q)&=&e^{b\rho_q(\gamma)F(\gamma)}, \quad q=-1, 2 \nonu \\
T_v(q=-1)&=&e^{-z\epsilon_-}e^{\bar{z}\epsilon_+}, \quad
T_v(q=2)=e^{-t_2 B_2^{(2)}}e^{x\epsilon_+} \nonu \\
\rho_{(-1)}(\gamma)&=&\exp{\( -\frac{z}{\gamma}+\bar{z}\gamma \) },  \quad
\rho_{(2)}(\gamma)=\exp{\( -t_2\gamma^2+x\gamma \) }
\label{5.16}
\er
 and remembering that $F(\gamma)=\{\bar F_{1, l},\tilde{F}_{1, l}, F_{a, \gamma}, F^{\pm}\}$ do not depend on $q$, 
 we observe that the only difference between $q=-1$ and  $q=2$ $\tau-$functions (\ref{4.7}), (\ref{5.10}) 
 is in the factors $\rho_q(\gamma)$. Therefore the simple change of variables
 \br
 \bar{z}\rightarrow x, \quad \frac{z}{\gamma}\rightarrow t_2\gamma^2
 \label{5.17}
 \er 
transforms the $q=-1$ $\tau-$functions in the corresponding $q=2$ ones. The problem however is that $q=2$ fields 
$r_a(x,t), q_a(x,t), u_l(x,t)$ and $q=-1$ ones $\psi_a(z,\bar{z}), \chi_a(z,\bar{z}), \varphi_l(z,\bar{z})$ are represented
by different sets of $G_0-$ $\tau-$functions. This reflects the fact that by definition $q=2$ field variables 
parametrize the $\lie_0$-algebra
\be
 {\cal {A}}_0=(\partial_x B)B^{-1}=\sum_{a=1, n} (r_a E_{-\alpha_a}^{(0)}+
 q_a E_{\alpha_a}^{(0)}) + \sum_{l=1}^{n-2}u_l h_{l+1}^{(0)}+\partial_x \tilde{\nu} {c}
\ee 
while the $q=-1$ fields are related to $G_0-$group element 
\begin{eqnarray} 
B=\exp ( {\sum_{a=1,n}
\tilde{\chi}_{a}} E_{-\alpha_{a}}^{(0)}) \exp ( \sum_{l=1}^{n-2} \varphi_l h_{l+1}^{(0)}+
\lambda_1\cdot H R_1+\lambda_n\cdot H R_n+\tilde{\nu} {c}) \exp (\sum_{a=1,n}\tilde{\psi}_{a} E_{\alpha_{a}}^{(0)}) 
\end{eqnarray}
In order to make this relation explicit we introduce the $q=-1$ $\lie_0$-WZW currents 
$\tilde {r}_a(z,\bar{z})$, $\tilde{q}_a(z,\bar{z})$, $\tilde{u}_l(z,\bar{z})$:
\be
 (\bar{\partial} B)B^{-1}=\sum_{a=1, n} (\tilde{r}_a E_{-\alpha_a}^{(0)}+
\tilde{q}_a E_{\alpha_a}^{(0)}) + \sum_{l=1}^{n-2}\tilde{u}_l h_{l+1}^{(0)}+\bar \partial \tilde{\nu} {c}
\ee 
and realize them in terms of $\psi_a(z,\bar{z}), \chi_a(z,\bar{z}), \varphi_l(z,\bar{z}), R_a(z,\bar{z})$:
\br
\tilde{u}_l&=&\bar{\partial} \{\varphi_l +\frac{n-l}{n+1}R_1+\frac{l+1}{n+1}R_n \}, \,\, l=1,\dots, n-2 \nonu \\
\tilde{q}_1&=&e^{\frac{1}{2}R_1-\varphi_1}(\bar{\partial} \psi_1-\frac{1}{2}\psi_1\bar{\partial} R_1) \nonu \\
\tilde{q}_n&=&e^{\frac{1}{2}R_n-\varphi_{n-2}}(\bar{\partial} \psi_{n}-\frac{1}{2}\psi_n\bar{\partial} R_n)\nonu \\
\tilde{r}_1&=&e^{-\frac{1}{2}R_1}\{\bar{\partial} \chi_1-\chi_1\bar{\partial}\varphi_1 \nonu \\
&-&e^{-\varphi_1}\chi_1^2\bar{\partial}\psi_1+\frac{\chi_1}{2}(1+\psi_1\chi_1e^{-\varphi_1})\bar{\partial} R_1\} \nonu \\
\tilde{r}_n&=&e^{-\frac{1}{2}R_n}\{\bar{\partial} \chi_n-\chi_n\bar{\partial}\varphi_{n-2} \nonu \\
&-&e^{-\varphi_{n-2}}\chi_n^2\bar{\partial}\psi_n+\frac{\chi_n}{2}(1+\psi_n\chi_ne^{-\varphi_{n-2}})\bar{\partial} R_n\}
\label{5.18}
\er
  For $(Q_1,Q_n)$-$1-$solitons (i.e. $4$-vertex solution) the complicated  $\tilde{r}_a, \tilde{q}_a$ expressions
   considerably simplifies, due to $1^{st}$ order soliton equations (\ref{3.3}) :
\br
\tilde{q}_1&=&\frac{1}{\gamma}\psi_1e^{\frac{1}{2}R_1-\frac{1}{n+1}(R_1-R_n)}, \quad
\tilde{q}_n=\frac{1}{\gamma}\psi_ne^{\frac{1}{2}R_n+\frac{1}{n+1}(R_1-R_n)}\nonu \\
\tilde{r}_1&=& \frac{1}{\gamma}\chi_1e^{-\varphi_1-\frac{1}{2}R_1+\frac{1}{n+1}(R_1-R_n)}, \quad
\tilde{r}_n=\frac{1}{\gamma}\chi_ne^{-\varphi_{n-2}-\frac{1}{2}R_n+\frac{1}{n+1}(R_1-R_n)}
\label{5.19}
\er
Since by definition we have
\br
q_a(x,t)&=&\tilde{q}_a(\bar{z}\rightarrow x, \frac{z}{\gamma} \rightarrow t\gamma^2)\nonu \\
r_a(x,t)&=&\tilde{r}_a(\bar{z}\rightarrow x, \frac{z}{\gamma} \rightarrow t\gamma^2)\nonu \\
u_l(x,t)&=&\tilde{u}_l(\bar{z}\rightarrow x, \frac{z}{\gamma} \rightarrow t\gamma^2)\nonu
\er
once the $q=-1$ one-soliton solution is known (\ref{3.19}), (\ref{3.20}),  (\ref{3.23}) , then 
eqs. (\ref{5.19}) (together with the change of variables (\ref{5.17})) determine the corresponding
$q=2$ one-soliton solution. The same is true for generic (say, multisoliton) solutions , but
 in this case one has to use more complicated relations (\ref{5.18}) between 
 $\lie_0$-currents $\tilde{r}_a, \tilde{q}_a, \tilde u_l$ and $G_0$-fields $\psi_a, \chi_a, \varphi_l, R_a$ .
 Let us consider the simplest case of $n=3$ in order to demostrate the relation (\ref{5.19}) between 
 $q=2$ and $q=-1$ one-solitons established above. We take $q=2$ ($n=3$) $1-$soliton 
in the form (\ref{5.11}) with $\tau-$functions given by eqs. (\ref{5.13}), (\ref{5.14}), (\ref{5.15}). 
We next change the variables $x, t_2$ to $z, \bar{z}$ (and $\rho_{\pm}$ that appears 
in $\rho(\gamma)$ according to eqs. (\ref{5.17})). Finally we apply eqs. (\ref{5.19}) replacing $\psi_a, 
\chi_a, \varphi_1=\varphi, R_a, \, \,(a=1, 3)$  by its explicit form of $q=-1$ model 
one-solitons (\ref{3.18})-(\ref{3.23}). As a result we find the following expresions of the  $q=2$ 
one-soliton parameters $b_i (i=1, 2, 3, 4)$ in terms of the $q=-1$ ones $D, d_1, d_2, 
2\delta=d_1+d_2, \alpha, p_1$:
\br
b_1=b_3e^{d_1} \quad b_2=b_4e^{d_2} \nonu
\er
\br
b_1&=&\frac{D^{\frac{1}{2}} N e^{-ip_1+\delta}}{\sqrt{2}(1-e^{-2\delta+2i\alpha})}, \quad
b_2=\frac{D^{\frac{1}{2}} N e^{i(p_1-\alpha)+\delta}}{\sqrt{2}(1-e^{-2\delta+2i\alpha})} 
\label{5.20}
\er
and the inverse relations:
\br
b_{\pm}&=&b_1b_2(1 \pm e^{2\delta}) ,\quad b_{\pm}=b_1b_2 \pm b_3b_4 \nonu \\
e^{-2\delta}&=&\frac{b_++b_-}{b_+-b_-}, \quad D=\frac{b_++b_-}{b_+-b_-}e^{i\alpha}-e^{-i\alpha}, \quad  
e^{-2ip_1}=\frac{b_1}{b_2}e^{i\alpha}
\er
and the appropiate choice of the soliton center of mass $X_0$.
Similar relations between the charged $1-$solitons of $q=2$ and $q=-1$ $A_n(p=1)$ 
dyonic models (with one U(1) symmetry) have been established in our recent paper \cite{cabrera}.

\subsection{Conserved charges}
As in the relativistic $q=-1$ $A_n^{(1)}(p=2)$ model (\ref{1.2}) (see Section $3.3$) the soliton spectrum of
 $q=2$ $A_n^{(1)}(p=2)$ IM (\ref{5.7}) is determined by the first few 
conserved charges (out of the infinite set of commuting charges) namely-"topological charges (fluxes)" $Q_l$, 
"particle density" $\tilde{Q}_0$, momenta ${\cal P}$, energy ${\cal E}$.
They all indeed depend on the asymptotic values of $G_0$-group element fields:$\tilde{\psi}_a, \tilde{\chi}_a, 
\tilde{\varphi}_l, \tilde{R}_a$ at $x \rightarrow \pm \infty$ only. Since
the boundary values of $ \tilde{\varphi}_l$ and $\tilde{R}_a, \, (\tilde{\psi}_a(\pm \infty )=
\tilde{\chi}_a(\pm \infty)=0)$ 
and $\tilde {r}_a(\pm \infty)=\tilde q_a(\pm \infty)=0$ define completely
the topological charges  $\tilde{Q}_l$, as in the $q=-1$ case all the other 
charges appears to be certain functions of these 
$\tilde{Q}_l$ and the soliton velocity $\gamma$. The problem we address in this section is
(a) to calculate these charges for our  $1-$soliton solutions and (b) to find the explicit relation between the
 conserved charges of  $q=-1$ and $q=2$ IM's. First step is to derive the $A_n(p=2, q=2)$
IM conserved currents. According to its equations of motion (\ref{5.7}) the fields $u_l$ represents the simplest
 conserved currents
\br
T^{(l)}&=&u_l, \quad \partial_{t}u_l=-\partial_{x} V_l^{(0)}, \quad
\tilde{Q}_l=\int dx u_l
\label{5.21}
\er
Taking into account the eqs. (\ref{5.7})  once more one can easily verify that
\br
T_0&=&r_1q_1+r_nq_n+\frac{1}{2}k_{ij}u_iu_j, \quad
\tilde{Q}_0=\int dx T_0
\label{5.22}
\er
defines the "particle density" conserved current. The momenta and energy densities have more 
complicated form and for $n=3$ model (see eqs.(\ref{5.12})), they are given by:
\br
T_x &=& r_1\partial_x q_1-q_1\partial_x r_1-r_3\partial_x q_3+q_3\partial_x r_3-2u(r_1q_1-r_3q_3), \quad 
{\cal P}= \int dx T_x 
 \nonu \\
T_t &=& 2(\partial_xr_1\partial_xq_1+\partial_xr_3\partial_xq_3)+\frac{1}{2}(\partial_x u)^2 
       + \frac{1}{2} (u^2+r_1q_1+r_3q_3)^2+2r_1q_1r_3q_3 \nonu \\
       &+&u(r_1\partial_x q_1-q_1\partial_x r_1+r_3\partial_x q_3-q_3\partial_x r_3), 
       \quad {\cal E}=\int dx T_t
        \label{5.23} 
\er 
The systematic way of deriving all the conserved currents (higher Hamiltonians) as well as their representations as 
total $\partial_x$-derivatives consists in applying the 
dressing transformations (\ref{4.1}) to calculate certain traces of $t{(-k)}$ (remember $\theta_-=\Pi_{i=0}^{\infty} 
e^{t(-i)}$). Starting from the vacua dressing:
\be
\epsilon_++\bar{\partial}B B^{-1}=\theta\epsilon_+\theta^{-1}+(\bar{\partial}\theta_-)\theta_-^{-1}
\label{5.24}
\ee
and writting it grade by grade (i.e. expanding in $\lambda^{-k}$) we obtain:
\br
\bar{\partial}B B^{-1}&=&\left[t{(-1)},\epsilon_+\right]={\cal A}_0 , \label{5.25}\\
\left[\epsilon_{+},t{(-2)}\right]&=&-\frac{1}{2}\left[{\cal A}_0,t{(-1)}\right]+\bar{\partial}t{(-1)},  \label{5.26}\\
\left[\epsilon_+,t{(-3)}\right]&=&-\left[{\cal A}_0,t{(-2)}\right]
 -\frac{1}{6}\left[\left[{\cal A}_0,t{(-1)}\right],t{(-1)}\right] \nonu \\
                    &-& {1\o 2}\left[\bar{\partial}t{(-1)},t{(-1)}\right] +\bar{\partial}t{(-2)} ,\label{5.27}
\er
etc. We next multiply eq. (\ref{5.26}) by $\epsilon_+$ and by taking trace we get:
\be
T_0=\frac{1}{2} tr {\cal A}_0^2=\partial_x Tr (\epsilon_+t{(-1)})
\ee
which reproduces eq. (\ref{5.22}). Taking into account the explicit form (\ref{5.9})  of $t{(-1)}$ 
 we further conclude that
\be
T_0=-\partial_x (\sum_{l=1}^{n-2} u_l)-(n-1) \partial_x^2 \tilde{\nu}
\label{5.28}
\ee 
One can continue this procedure and to determine $t{(-2)}$ from (\ref{5.26}), substituting it in 
eq. (\ref{5.27}) and by taking trace with $b_2^{(2)}$ from eq. (\ref{5.6}) (and eqn. (\ref{xx}) for $n=3$ case)
 to derive the explicit
form of $T_x$, etc. 

We next consider the problem of calculation of $q=2$ model (\ref{5.7}) conserved charges $\tilde{Q}_l$ and $\tilde{Q}_0$ and their relation to $q=-1$ model (\ref{1.2}) charges
 $Q_{mag}, Q_1, Q_n$ and $E, { P}, { M}$. According to eq. (\ref{5.25}) we have
\be
u_l=\beta\bar{\partial}(\tilde{\varphi}_l+\frac{n-l}{n+1}\tilde{R}_1+
\frac{l+1}{n+1}\tilde{R}_n), \, \, \, \bar{\partial}=\partial_{x}
\label{5.29}
\ee
and therefore the corresponding charges $\tilde{Q}_l$ are determined by the asymptotic values 
of the fields $\tilde{\varphi}_l, \tilde{R}_a $
\br
\tilde{\varphi}_l(\pm \infty)=\frac{2\pi l}{\beta(n-1)}\tilde{N}_{\pm} , 
\quad \tilde{R}_a(\pm \infty)=\frac{2 \pi}{\beta} \tilde{f}_{\pm}(a)
\er
 We have assumed that for $t$ (and $t_2$) fixed (say to zero) the $x \rightarrow \pm \infty$ limits 
 of the fields $\tilde{\varphi}_l(x_2, t_2),\tilde{R}_a(x_2, t_2)$ and 
$\varphi_l(z, \bar{z}), R_a(z, \bar{z})$ do coincide. As a result we realize $\tilde{Q}_l$ in terms of 
the $q=-1$ topological and electric charges $Q_{mag}, Q_1, Q_n$ (\ref{2.27}),
(\ref{2.33}) as follows
\be
\tilde{Q}_l=\frac{1}{2(n-1)}\left[l(4 \pi j_{\varphi}-Q_1+Q_n)+(n-1)Q_1\right]
\label{5.30}
\ee  
In order to calculate $\tilde{Q}_0$ and its relation with $E, { P}, { M}$ of the
 $q=-1$ model, we represent $T_0$ (\ref{5.22}) as total ($\partial_x \equiv \bar{\partial}$) 
 derivative by applying $1^{st}$ order soliton equations(\ref{3.3}) and 
$r_a, q_a, u_l \leftrightarrow \tilde{\psi}_a, \tilde{\chi}_a, \Phi_l$ relations
(\ref{5.18}), (\ref{5.19}):
\br
T_0=\frac{1}{\gamma^2}V_{rel}&=&\frac{1}{\gamma} \frac{\mu}{\beta^2} 
\bar{\partial}(\sum_{k=1}^{n-1}e^{-\beta \Phi_k}+\beta^2\tilde{\psi}_n \tilde{\chi}_n) 
= \frac{1}{\gamma} \frac{\mu}{\beta^2}\partial_{x_2} \tilde{F},
\label{5.31} 
\er
where $V_{rel}$ is the potential $V_{n}^{(p=2)}$.
Note that the $q=2$ coupling constant $\kappa =\frac{\mu}{\beta^2}$ appears 
in (\ref{5.31}) as a result of simple rescaling
\be
(r_a, q_a, u_l) \rightarrow (\sqrt{\kappa}r_a,\sqrt{\kappa}q_a,\sqrt{\kappa}u_l )
\ee
We next remind the form of the $q=-1$ model stress tensor $(T_{00}, T_{01})$ derived in Sect. $3$:
\br
T_{00}&=& -(\gamma+\frac{1}{\gamma})\frac{\mu}{\beta^2}\partial_x(\sum_{k=1}^{n-1}e^{-\beta \Phi_k}+\b^2 \tilde{\psi}_n
 \tilde{\chi}_n) \nonu \\
T_{01}&=& (-\gamma+\frac{1}{\gamma})\frac{\mu}{\beta^2}\partial_x(\sum_{k=1}^{n-1}e^{-\beta \Phi_k}+\b^2 \tilde{\psi}_n 
\tilde{\chi}_n) 
\label{5.32}
\er
where we have used the relation
\be
\frac{1}{\gamma}F^{-}+\gamma F^{+}=(n-1)c_0
\ee
(that follows from eqs.(\ref{3.4}) and $F^{\pm}$ definitions).
By comparing (\ref{5.31}) to eq.(\ref{5.32}) we conclude that:
\br
E &=&\int T_{00}dx=-\gamma(\gamma+\frac{1}{\gamma})\tilde{Q}_0, \quad \quad
{ P}=\int T_{01}dx=-\gamma(\gamma-\frac{1}{\gamma})\tilde{Q}_0 \nonu \\
\tilde{Q}_0 &=&\int dx_2 T_0=\frac{1}{\gamma}\frac{\mu}{\beta^2} \int dx_2 \partial_{x_2} \tilde{F}, \quad  \quad 
\tilde F = 
\sum_{k=1}^{n-1}e^{-\beta \Phi_k}+\b^2 \tilde{\psi}_n 
\tilde{\chi}_n
\label{5.33}
\er
where $\gamma=e^{-b}$ is related to the $1-$soliton velocity of  $q=-1$ model by
\be
\gamma=( \frac{1-v_{rel}}{1+v_{rel}})^{1/2}, \quad v_{rel}= th b
\ee
Therefore the $q=2$ charge $\tilde{Q}_0$ is proportional to the $q=-1$ soliton mass $M$:
\br
\tilde{Q}_0 &=&\frac{1}{2}(\frac{1+v_{rel}}{1-v_{rel}})^{1/2} M 
=\frac{2\mu (n-1)}{\beta_0^2\gamma}|\sin\frac{\left[\beta_0^2Q_{mag}-Q_1+Q_n\right]}{4(n-1)}|
\label{5.34}
\er
The form (\ref{5.33}) of $\tilde{Q}_0$-charge
\be
\tilde{Q}_0=\frac{\kappa}{\gamma}\tilde{F}|_{-\infty}^{\infty}
\ee
suggests that it can be realized in terms of the $u_l-$charges $\tilde{Q}_l$ only. We next observe that
\be
\tilde{Q}_1-\tilde{Q}_{n-2}=\frac{n-3}{2(n-1)}(\beta_0^2Q_{mag}-Q_1+Q_n)
\ee
as one can see from eq. (\ref{5.30}). Substituting it in eq.(\ref{5.34}) we find 
($n \neq 3$)
\be
\tilde{Q}_0=\frac{2(n-1)\kappa}{\gamma}|\sin\frac{\tilde{Q}_1-\tilde{Q}_{n-2}}{2(n-3)}|
\label{5.35}
\ee
For particular case $n=3$ the analog of the above relation include also the additional non-local charge 
$Q_1=\frac{4\pi}{\beta}(f_{+}^{(1)}-f_{-}^{(1)})$ :
\be
\tilde{Q}_0(n=3)=\frac{4 \kappa}{\gamma}|\sin(\frac{\tilde{Q}_1}{2}+\frac{Q_1}{4})|
\ee
The complete $1-$soliton spectrum $(\tilde{Q}_l, \tilde{Q}_0,{\cal P}, {\cal E})$ of $q=2$ model
 (\ref{5.7}) requires the evaluation of the soliton momenta ${\cal P}$ and energy ${\cal E}$ 
 (\ref{5.23}) 
 as well. The same procedure ($1^{st}$ order soliton equations  etc.) we have used in $\tilde{Q}_0$ 
 calculation can be applied to the $T_x$ and $T_t$. 
 We shall  present the complete soliton spectrum 
 of our  $q=2$  $A_n^{(1)}(p=2)$ IM's in future publication.

\sect{ Dyonic  IMs with non-abelian symmetries}

We have established in Sect.1 and 2, that the specific choice of the $A_n^{(1)}$ graded 
structure ($Q, \eps_{\pm}, \lie_0^0$
) (\ref{2.9}) giving rise to the dyonic $A_n^{(1)}(p=2)$ IM (\ref{1.2}) led to topological
 solitons carrying internal
(Noether) charges $U(1)\otimes U(1)$.  They appear as natural 
generalization of the simplest 
$A_n^{(1)}(p=1)$  dyonic IM (\ref{1.1}) (see refs. \cite{eletric}, \cite{dyonic})  and  correspond to 
a particular member of a
$U(1)^p$-dyonic  family of IMs  $A_n^{(1)}(p), \; p=1, 2, \cdots n-1$ .
Consider for example the NA-Toda models defined by the graded structure similar to (\ref{2.7})  i.e.
\br
Q= (n-2p)d + \sum_{i=1}^{p} \l_{2i}\cdot H + \sum_{j=2p+1}^{n-1}\l_{j}\cdot H
\lab{6.1}
\er
and 
\br
\eps_{\pm} = \sum_{i=2p+1}^{n-1} E_{\pm \a_i}^{(0)} + E_{\mp (\a_{2p} + \a_{2p+1} + \cdots + \a_{n-1})}^{(\pm 1)}
\lab{6.2}
\er
Since the zero grade subalgebra $\lie_0$ is given by
\br
\lie_0 = SL(2)^{p+1} \otimes U(1)^{n-p-1}, \quad \quad \lie_0^0 = U(1)^{p+1}
\lab{6.3}
\er
the field content of the corresponding $\lie_0^0$-gauged IM consists in $(p+1)$-pairs of charged fields $\psi_a, \chi_a,
a=1, 2, \cdots p+1$ and $n-p-1$ neutral fields $\varphi_l, l=1,2, \cdots n-p-1$.  The effective action for these models
 can be
derived using the methods described in Sect.2.  The potential have $n-2p-1$ distinct zeros (for imaginary coupling)
and the IM admits  $U(1)^{p+1}$-charged topological solitons. 

Having constructed the $U(1)^{p}$ dyonic IM $A_n^{(1)}(p), 1\leq p \leq n-1$ with {\it abelian} internal symmetry, 
 the  question of constructing IM with  {\it non-abelian} internal structure arises naturaly.  It is clear that one has to 
  choose $Q$
  and $\eps_{\pm}$ such that $\lie_0^0 \in \lie_0$ to be  non-abelian.  For an appropriate choice of the 
   potential (i.e. of $\eps_{\pm}$ and  $g_0$, since $V = Tr (\eps_+ g_0 \eps_- g_0^{-1})$),  the model
  admits topological solitons carrying non-abelian charges (isospin).  The simplest case is to consider $\lie_0^0 =
  U(2)\in \lie_0 = SL(3)\otimes U(1)^{n-2}$
obtained by decomposing the $A_n^{(1)}$ algebra by the grading operator,
\br
Q = (n-1)d + \sum_{i=3}^{n} \l_i \cdot H, \quad \quad \eps_{\pm} = \sum_{i=3}^{n}E_{\a_i}^{(0)} + E_{\mp (\a_3+\a_4 + \cdots
\a_n)}^{(\pm 1)}
\label{6.3}
\er
where $\lie_0^0 = \{ E_{\pm \a_1}^{(0)}, \l_1 \cdot H,\l_3 \cdot H \} $.  The model represents an
 $SL(3)/SL(2)\otimes U(1)$ analog of the
complex sine-Gordon  ($SL(2)/U(1)$) interacting with $A_{n-2}^{(1)}$ abelian affine Toda model. 
 Its  main ingredient is the new $SL(3)/U(2)$ integrable model.  It is defined by 
 the homogeneous gradation of
 $A_2^{(1)}$, 
 \br
 Q=d, \quad \quad \eps_{\pm } = \l_2 \cdot H^{(\pm 1)}, \quad \quad
 \lie_0^0 = \{ E_{\pm \a_1}^{(0)}, h_1, h_2\}
 \lab{6.4}
 \er
The corresponding 
 effective action can be obtained by considering the $A_2^{(1)}$ two-loop gauged WZW model and applying the methods
described in Sect. 2.1.  We take eqns. (\ref{2.1})with $A = A_1 + A_0, \;\; 
\bar A = \bar A_1 + \bar A_0$ where
\br
A_1 = a_1 E_{\a_1}^{(1)}, & \quad  \bar A_1 = \bar a_1 E_{-\a_1}^{(1)}, \nonu \\
A_0 = a_{01}\l_1 \cdot H + a_{02} (\l_2 - \l_1 )\cdot H,  &\quad  \bar A_0 = \bar a_{01}\l_1 \cdot H +
\bar a_{02} (\l_2 - \l_1 )\cdot H
\lab{6.5}
\er
and $g_0^f = e^{\chi_1 E_{-\a_2}+ \chi_2 E_{-\a_1-\a_2}}e^{\psi_1 E_{\a_2}+ \psi_2 E_{\a_1-\a_2}}$.
We next perform the Gaussian functional integrals in the definition of the partition function of the model,
\br
{\cal Z}(SL(3)/U(2) ) = \int DB DA D\bar A e^{-S_{G/H}} = \int DB e^{-S_{eff}}
\nonu
\er
The effective action for the $SL(3)/U(2)$- model obtained in this way has the form 
\br
S_{eff} = -{{k}\o {2\pi }} \int dz d\bar z & \( {1\o {\Delta}}( {{\bar \pa \psi_1 \pa \chi_1 }} (1 + 
\psi_1\chi_1 + \psi_2 \chi_2 )  + 
{{\bar \pa \psi_2 \pa \chi_2} }(1 + \psi_1 \chi_1 )  \right. \nonu \\
& \left. - {1\o 2}\( \psi_1 \chi_2 {{\bar \pa \psi_2 \pa \chi_1 }}+ \chi_1 \psi_2 {{\bar \pa \psi_1 \pa \chi_2 }}\)) -V \)
\label{6.6}
\er
where $V = {{2\pi \mu }\o {k}}\( {2\o 3} + \psi_1 \chi_1 + \psi_2 \chi_2 \)$ and $ \Delta = 
(1+\psi_1 \chi_1 )^2 + \psi_2 \chi_2 (1+ {3\o 4}\psi_1
\chi_1 )$.

As in the complex sine-Gordon case, the form of the potential $V$ is an indication that the IM (\ref{6.6}) can have 
only {\it non
topological} solitons carrying $U(2)$ Noether charges.  Hence this model is {\it not} of dyonic type. 
 The $A_n^{(1)}(U(2)), \; n>3$  IMs defined by grading (\ref{6.3}) representing the $A_{n-2}^{(1)}$ abelian affine Toda model
 interacting with
the model (\ref{6.6}) are  however of dyonic type. Their potential is given by
\br
V_n = {{\mu}\o {\b^2}}\( \sum_{i=1}^{n-2} e^{-\b k_{ij}\varphi_j} + e^{\b (\varphi_1 + \varphi_{n-2})} + e^{\b \varphi_1} (\psi_1
\chi_1 + \psi_2 \chi_2) -n+1\)
\nonu
\er
and for imaginary coupling
 $\b \rightarrow  i\b_0$ shows $n-1$ distinct  zeros.   Therefore as in $A_n^{(1)}(p)$ dyonic models
one expects to have topological solitons but now with $U(2)$ internal degrees of freedom.  The explicit construction of such
topological solitons with $U(2)$ isospin can be obtained by applying the methods of Sect. 3 and 4.

The {\it nonabelian} dyonic models $A_n^{(1)}(U(2))$ described above represents the simplest member of the vast family of dyonic
models $A_n^{(1)}(A_s), \; 1\leq s \leq n-1$, i.e. with $G_0^0 = U(s+1)$, defined by the graded structure
\br
Q_s = (n-s)d + \sum_{i=s+2}^{n} \l_i \cdot H, \quad \eps_{\pm} = \sum_{i=s+2}^{n}E_{\pm \a_i}^{(0)} + 
E_{\mp (\a_{s+2} + \cdots \a_n)}^{(\pm 1)}
\lab{6.8}
\er
and $
G_0 = SL(s+2)\otimes U(1)^{n-s-1}, \;\; G_0^0 = U(s+1)$.  The effective actions for these 
non-abelian dyonic models can be
derived by the methods of Sect. 2 and appear to be a straightforward generalization of the 
simplest $A_2^{(1)}(U(2))$ action
(\ref{6.6}) we have constructed above.  

We should note that the discussion of different dyonic IMs introduced in the present section is far 
from an effective recipe for their classification.  Our purpose is to discuss the properties of a vast family of dyonic IMs
according to their symmetries: abelian (i.e. $U(1)^{p}$), non-abelian , global or local, etc.
The selection of the models was done by first chosing the invariant subalgebra (i.e. symmetries)  $\lie_0^0 \in \lie_0$ and
next seeking for appropriate  grading operator $Q$ and constant grade $\pm 1$ elements $\eps_{\pm}$ (such that $[\lie_0^0,
\eps_{\pm}] = [Q, \lie_0]=0$) leading to consistent  integrable models of dyonic type.  
It is clear that the systematic classification program consists in a) listing the admissible gradings for all the affine
algebras, b) to define the constant elements of grade $\pm 1$, $\eps_{\pm}$, c) to impose the dyonic conditions, i.e.
 to separate $\eps_{\pm}$ and $Q$ such that $\lie_0^0$ is nontrivial and the potential $tr (\eps_+g_0 \eps_- g_0^{-1})$ has
 nontrivial distinct zeroes. Its realization is out of the scope of this paper.

\section{Concluding remarks}

Motivated by the problem of description of Domain Walls in 4-D $SU(n+1)$ SDYM we have introduced and studied in the present
paper a pair of $A_n^{(1)}$ dyonic IMs (one relativistic and one nonrelativistic ) that admit $U(1)\otimes U(1)$
charged topological  solitons.  The 1-soliton solutions of both models have been explicitly constructed.  Since the DWs
tensions are known to be proportional to soliton masses, the derivation of the semiclassical spectrum (\ref{1.5}) of these
solitons  should be considered as the main result of this paper.  Although the soliton mass formulae for $U(1)$- and 
$U(1)\otimes U(1)$- integrable models are quite similar, the semiclassical quantization of the electric charges of the
second  model 
represents new feature. Namely,  only their difference $Q_1 -Q_n = \b_0^2 j_{el}$ is quantized.  Therefore one can
expect that the masses of the  multicharged topological solitons of generic $U(1)^p, p \geq 3$ dyonic IMs are simple
generalization of the known particular cases $p=1,2$.  The real problem to be solved, however is the semiclassical
quantization of the corresponding electric charges $Q_a^{el}, a=1,2, \cdots p$.   The methods to be used in the
construction of these multicharged solitons are indeed an apropriate  extension of those presented in Sects. 3 and 4.
Whether and how these methods work in the case of non-abelian $SU(2)$-dyonic IM (\ref{6.6}) and what is the spectrum
of the solitons carrying $SU(2)$ isospin are interesting open questions, which require further investigation.

Together with the dyonic IMs with global $U(1)^p$ symmetry we have derived in Sect. 2  integrable models 
with {\it local} symmetries: the so called ungauged IM (\ref{1.6})  
with local $U(1)\otimes U(1)$  and the intermediate model (\ref{1.8}) with one local and one global $U(1)$
symmetries.  It is therefore interesting to know how their solitons or more general 
string-like finite energy solutions looks like. 
By construction  both models have the 1-soliton solutions of the completely gauged IM (\ref{1.2}) as particular solutions. 
Our preliminary analysis shows that their soliton spectra is more complicated and also includes massless solitons and
solitonic strings.  The complete discussion will be presented elsewhere.

The $\s$-model form of the Lagrangians (\ref{1.1}), (\ref{1.2}), (\ref{1.6}) and (\ref{1.8})  constructed in the present
paper suggests that they might have intersting string application.  
Although they are 
{\it not} conformal
invariant, it is natural to interpret their conformal affine versions \cite{bon}, \cite{aratyn} (called CAT)
 as representing strings on curved
background $g_{MN}(X),  b_{MN}(X)$ and $V$:
\br
{\cal L} = \( g_{MN}(X) \eta^{\mu \nu} + b_{MN}(X) \eps^{\mu \nu}\) \pa_{\mu}X^M \pa_{\nu}X^N -V(X^L)
\label{aaa}
\er
where $X^L = (\psi_a, \chi_a, R_a, \varphi_l, \nu, \eta )$, $L=1,2, \cdots D=n+6$  are the string coordinates and $g_{MN}(X)$ is the target space
metric,   $b_{MN}(X)$ - the antisymmetric tensor and $V(X)$ - the tachyonic potential.  For each {\it nonconformal} IM 
one introduces its  CAT counterpart by adding a pair of fields $(\nu, \eta)$, that restores conformal invariance, i.e. by
extending the $(n+4)$-dimensional off-critical string to $(n+6)$-critical one, for say $n=20$.
  In  the particular case of ungauged
model (\ref{1.6}) we have
\br
{\cal L}_{G_0}^{CAT}&=& {1\o 2} \sum_{i=1}^{n-2} k_{ij}\pa \varphi_i \bar \pa \varphi_j + 
\pa  \btchi_1 \bar \pa  \btpsi_1 e^{ \b( R_1-\varphi_1)} + 
\pa  \btchi_n \bar \pa \btpsi_n e^{ \b( R_n-\varphi_{n-2})} + \pa \nu \bar \pa \eta + \pa \eta  \bar \pa \nu \nonu \\ 
&+& 
{{1}\o {2(n+1)}} \( n\pa R_1 \bar \pa R_1 + n\pa R_n \bar \pa R_n +
 \pa R_1 \bar \pa R_n + \pa R_n \bar \pa R_1 \) - V_0^{CAT}
\label{lagr}
\er\
where 
\br
V_0^{CAT}&=& {{\mu^2}\o {\b^2}}\( \sum_{i=1}^{n-2} e^{-\b k_{ij} \varphi_j}+ e^{\b (\varphi_1 + \varphi_{n-2} -\eta )} (1 + 
\b^2  \btpsi_1  \btchi_1 e^{\b (R_1 -
\varphi_{1})})(1 + \b^2 \btpsi_n  \btchi_n e^{\b (R_n -
\varphi_{n-2})})  \right. \nonu \\
&-& \left. (n-1) e^{-{{\b\eta}\o {n-1}} }\). 
\nonu
\er
Within this context the ungauged IM (\ref{1.6}) (i.e.  (\ref{lagr}) with $\eta=\eta_0=$ const) describes the off-critical
(i.e. relevant deformation \cite{review}) 
and renormalization group flow properties  of strings on $AdS_3\otimes S_3 \otimes T_{n-2}$-target space, when $R_n$ and
$\varphi_l$ are taken imaginary and $\psi_n, \chi_n$ as complex conjugated to each other. The target space metric
$g_{MN}(X)$ for 
   gauge fixed IMs (\ref{1.2}) and 
(\ref{1.8}) represents both horizons and singularity and therefore could be used in the
 study of  the off-critical behaviour of 4-D black-holes (and black-strings) with
certain additional flat directions $X_l = \varphi_l$.

The  exact quantum spectrum of 2-d solitons is known to be crucial in the description of 
the strong coupling behaviour (and
S-duality properties) of the corresponding quantum integrable models \cite{fat}.  The problem to be solved before 
any attempt for
exact quantization of the considered dyonic IMs is to answer the question: Whether and for which 2-d IMs the semiclassical
spectrum survives the renormalization.
  The  answer is known  for sine-Gordon (sG) and the abelian affine Toda models, where the only renormalization is
of the coupling constant and indeed the semiclassical spectrum is exact.  The quantum consistency of the Lund-Regge model    
\cite{millet}, $U(1)$-dyonic IM (\ref{1.1}) (see Sect 7.2 of ref. \cite{eletric}) and T-selfdual Fateev's IMs \cite{fat}
requires that new counterterms to be added to the original classical Lagrangians.  Since the same counterterms are present
in the corresponding conformal quantum theories, i.e. the gauged $G_0/G_0^0$-WZW models (\ref{2.1}) \cite{tsey} one believes
that they cannot change the soliton masses.  But they  do change the form of the $U(1)$ currents and hence they might shift
the semiclassical values of the electric charges.  Recently the thermodynamic Bethe ansatz analysis \cite{fring}
of  the Homogeneous sine-Gordon models has confirmed  the exactness of their semiclassical mass spectra.  The answer of the
question whether such statement 
takes also place for multicharged dyonic IMs requires the construction of their exact  S-matrices, etc.
  It exists however a
strong  hint that the $n\rightarrow \infty$ limit of the semiclassical ($Q_1, Q_n$) 1-soliton masses 
\br
M_{j_{el}, j_{\varphi}} (n\rightarrow \infty ) = {{\mu}\o {\b_0^2}}|\b_0^2 Q_{mag} - Q_1^{el} + Q_n^{el} |
\nonu
\er
is exact.  The reason is that in the large $n$ limit the counterterms become proportional to the complicated $\psi_a,
\chi_a$ kinectic term and thus contribute to the $\b_0^2$ renormalization only.

 \sect{Appendix.  Two-Soliton Solution}
 
 The generic $4-$ vertex solution of equations (\ref{5.12})  can be extracted from eqs.
 (\ref{5.10}),(\ref{5.11}) for $n=3$ and the constant $G_0$ group element
\br
g^{(1)}&=& e^{b_2\bar F_{11}(\gamma_2)} e^{b_1\bar F_{11}^{\dagger}(\gamma_1)}e^{b_4\tilde F_{11}^{\dagger}(\gamma_4)} e^{b_3\tilde
F_{11}(\gamma_3)} \nonu \\
\er
 where $\bar F_{11}(\gamma),\tilde F_{11}(\gamma)$ are given by eqs. 
 (\ref{4.13}) (for $n=3$). By calculating  
 the matrix elements (\ref{5.10}) 
 we obtain the following explicit form of the $G_0-\tau-$functions\footnote{We have determined such solution
 using the Mathematica program of ref. \cite{math}}
 (and therefore of $r_a, q_a$ and $u$ according to eqs. (\ref{5.11})):
\br
\tau_0 & = &  1 + b_1 b_2 c_{12} \rho_1  \rho_2 +
b_3 b_4 c_{34} \rho_3  \rho_4 \nonu \\
& + &
c_{1234} \rho_1  \rho_2 \rho_3  \rho_4 b_1 b_2 b_3 b_4
\nonu \\
\tau_2 & = & 1 +b_1 b_2 d_{12} \rho_1  \rho_2   +
b_3 b_4 d_{34} \rho_3  \rho_4 \nonu \\
& + &
d_{1234}\rho_1  \rho_2 \rho_3  \rho_4 b_1 b_2 b_3 b_4\nonu \\
\tau_{n-1, n} & = & (g_{3}+b_1 b_2 g_{123}\rho_1  \rho_2)\rho_3 b_3   \nonu \\
\tau_{2n} & = & (i_{4}+b_1 b_2 i_{124}\rho_1  \rho_2)\rho_4 b_4\nonu \\
\tau_{12} & = & (e_{2}+b_3 b_4 e_{234}\rho_3  \rho_4)\rho_2 b_2 \nonu \\
\tau_{1, n-1} & = & (f_{1}+b_3 b_4 f_{134}\rho_3  \rho_4)\rho_1 b_1 \nonu \\
\label{A.1}
\er
where
\br
\rho_1(\gamma_1)=e^{-x\gamma_1+t\gamma_1^2}, \quad  \rho_2(\gamma_2)=e^{x\gamma_2-t\gamma_2^2}
\nonumber
\er
\br
\rho_3(\gamma_3)=e^{-x\gamma_3+t\gamma_3^2}, \quad  \rho_4(\gamma_4)=e^{x\gamma_4-t\gamma_4^2}
\label{A.2}
\er
and,  
\br 
c_{12}= {{2\g_1 \g_2^2}\o {{(\g_1-\g_2 )^2(\g_1 +
\g_2)}}}, \quad c_{34}= {{2\g_3^2 \g_4}\o {{(\g_3-\g_4 )^2(\g_3 +
\g_4)}}}\nonumber 
\er 
\br
d_{12}= {{2\g_1^2 \g_2}\o {{(\g_1-\g_2 )^2(\g_1 + \g_2)}}},
\quad d_{34}= {{2\g_3 \g_4^2}\o {{(\g_3-\g_4 )^2(\g_3 + \g_4)}}} \nonumber
\er
\br
c_{1234}= {{4\g_1 \g_2^2 \g_3^2 \g_4 (\g_1-\g_3)(\g_2 -\g_4)}\o {{(\g_1+\g_2 )(\g_1 - \g_2)^2(\g_2-\g_3 )(\g_1-\g_4 )
(\g_3+\g_4 )(\g_3-\g_4 )^2}}}
\er
\br
d_{1234}= {{4\g_1^2 \g_2 \g_3 \g_4^2 (\g_1-\g_3)(\g_2 -\g_4)}\o {{(\g_1+\g_2 )(\g_1 - \g_2)^2(\g_2-\g_3 )(\g_1-\g_4 )
(\g_3+\g_4 )(\g_3-\g_4 )^2}}}
\er
\br
e_2&=& \sqrt{2} \g_2,
\quad e_{234}= {{2\sqrt{2}\g_2\g_3 \g_4^2 (\g_2 - \g_4)}\o {{(\g_2-\g_3 )(\g_3+\g_4 )(\g_3 - \g_4)^2}}}\nonumber \\
f_1&=& \sqrt{2} \g_1,
\quad f_{134}= {{2\sqrt{2}\g_1\g_3^2 \g_4 (\g_1 - \g_3)}\o {{(\g_1-\g_4 )(\g_3 + \g_4)(\g_3 - \g_4)^2}}}\nonumber \\
g_3&=& \sqrt{2} \g_3,
\quad g_{123}= {{2\sqrt{2}\g_1^2 \g_2 \g_3  (\g_1 - \g_3)}\o {{(\g_1+\g_2 )(\g_2 - \g_3)(\g_1 - \g_2)^2}}}\nonumber \\
i_4&=& \sqrt{2} \g_4,
\quad i_{124}= {{2\sqrt{2}\g_1 \g_2^2 \g_4  (\g_2 - \g_4)}\o {{(\g_1+\g_2 )(\g_1 - \g_4)(\g_1 - \g_2)^2}}}
\label{A.3}
\er
It is worthwhile to note again that the $4$-vertex solution for arbitrary $q>2$
integer $A_3^{(1)}(p=2, q)$ model have the form identical to the $q=2$ one (\ref{A.1}), (\ref{A.3}) with $\rho_i(\gamma_i)$ replaced by, say
\be
\rho_1^{(q)}(\gamma)=e^{-x\gamma+t_q\gamma^q}, etc.
\ee
For the specific choice of $\g_i$, namely
\br
\g_1 = -e^{B_1-i \a_1} = - \g_2^{*}, \quad \g_3 =-e^{B_2-i \a_2} = - \g_4^{*}
\nonu
\er
the above four vertex solution has real energy and momenta and thus represents charged two-soliton solution, where $B_i
(v_i = th (B_i))$ are the two independent rapidities (velocities) of the 2-soliton and $\a_i$ are related to the $Q_1, Q_n$
charges.  As it was demonstrated in Sect. 5.2, the limit when $B_2 = B_1$, $\a_2 = \a_1$ leads to the charged 1-soliton
solution (\ref{5.13}).

{\bf Acknowledgements} \\

We are grateful to L.A. Ferreira and H. Aratyn for discussions and 
 Fapesp, Unesp and
CNPq for  financial support.


\begin{thebibliography}{99}


\bibitem{kov} A. Kovner, M. Shiffman and A. Smilga, Phys. Rev. {\bf D56}, (1997),7978-7989, also 
 G. Dvali and M. Shiffman, Phys. Lett. {\bf B396},(1997),  hep-th/9612128;
A. Hanany and K. Hori, \NPB{513}{1998}{119} hep-th/9706089.

\bibitem{wit1}E. Witten, \NPB{507}{1997}{658}

\bibitem{wit2}E. Witten,
\NPB{160}{1979}{461}

\bibitem{review} J. Polchinski and M. Strassler, ``The string dual of a confining four dimensional Gauge Theory'',
hepth/0003136; O. Ahrony, S. Gubser, J. Maldacena, H. Ooguri and Y. Oz, ``Large $N$ Field Theory, String Theory and
Gravity'', hepth/9905111, \PR{323}{2000}{183}   


\bibitem{gau}J. Gauntlett, D. Tong, P.  Townsend, 
 ``Supersymmetric intersecting domain walls in massive hyperkahler
 sigma models'',  hep-th/0007124, \PRD{63}{2001}{085001};
 J. Gauntlett, R. Portugues, D. Tong, P.  Townsend, ``D-brane solitons 
 in supersymmetric models''\PRD{63}{2001}{085002}, 
  hep-th/0008221; 
 



\bibitem{lez-sav}  A. N. Leznov, M. V. Saveliev, Group Theoretical Methods for 
Integration of
Nonlinear Dynamical Systems, Progress in Physics, Vol. 15 (1992), Birkhauser
Verlag, Berlin; 
L.A. Ferreira, J.L. Miramontes and J.S. Guillen, \NPB{449}{1995}{631};\\
  

\bibitem{eletric} J.F. Gomes, E.P. Gueuvoghlanian, G.M. Sotkov and A.H.
Zimerman, `` Electrically  charged  topological solitons'' \NPB{606}{2001}{441}, hepth 0007169 
 
\bibitem{dyonic}
J.F. Gomes,  E.P. Gueuvoghlanian, G.M. Sotkov and A.H. Zimerman, 
``Dyonic Integrable Models'',\NPB{598}{2001}{615} also in hepth/0011187. 
 
\bibitem{bon} O. Babelon and L. Bonora, \PLB{244}{1990}{220}

\bibitem{liao} H.C. Liao, D. Olive and N. Turok, 
Phys. Lett. {\bf 298B} (1993) 95; D. Olive, N. Turok and 
J. Underwood \NPB{409}{1993}{509}; \NPB{401}{1993}{663}; T.J. Hollowood, \NPB{384}{1992}{523}


\bibitem{dw} J.F. Gomes, E.P. Gueuvoghlanian, G.M. Sotkov and A.H.
Zimerman, `` Self-Dual Domain Walls'', in preparation



\bibitem{lund} F. Lund and T. Regge, \PRD{14}{1976}{1524};F. Lund, \PRL{38}{1977}{1175};
B.S. Getmanov, JETP Lett. {\bf 25}(1977)119;
N. Dorey and T.J. Hollowood, \NPB{440}{1995}{215};





\bibitem{chodos} A. Chodos, \PRD{21}{1980}{2818}; 
\bibitem{miramontes} J.L. Miramontes, \NPB{547}{1999}{623}

\bibitem{aratyn} H. Aratyn, L.A. Ferreira, J.F. Gomes and A.H. Zimerman,
Phys. Lett {\bf B 254} (1991) 372; L.A. Ferreira, J.F. Gomes, A. Schwimmer and A.H. Zimerman,
Phys. Lett {\bf B 274 } (1992) 65



\bibitem{aratyn1} H. Aratyn, L.A. Ferreira, J.F. Gomes and A.H. Zimerman, J.
Phys. {\bf A31}, (1998) 9483;
H. Aratyn, L.A. Ferreira, J.F. Gomes and A.H. Zimerman,
Supersymmetry and Integrable models (Lecture Notes in Physics Vol. 502), Ed. H.
Aratyn et. al. (Berlin, Springer), p. 197-210, (1998)



\bibitem{tdual}J.F. Gomes, E.P. Gueuvoghlanian, G.M. Sotkov and
 A.H. Zimerman, ``T-duality of axial and vector dyonic integrable models'',
 hep-th/0007116, \AoP{289}{2001}{232}
 
 

 
 \bibitem{aratyn2} H. Aratyn, L.A. Ferreira, J.F. Gomes and A.H. Zimerman, J.
Phys. {\bf A33}, (2000), L331
 
 
 
 \bibitem{marian}D. Fioravanti and M.
Stanishkov, \NPB{591}{2000}{685}. hepth/0005158
 
 
\bibitem{cabrera}I. Cabrera-Carnero,  J.F. Gomes, E.P. Gueuvoghlanian, G.M. Sotkov and
 A.H. Zimerman, ``Non Abelian Toda Models and  Constrained KP Hierarchies'',
 To appear in the VII International Wigner Symposium, Maryland (2001),
 hepth/0109117
 
\bibitem{emil}E. Nissimov and S. Pacheva, ``Symmetries  of Supersymmetric Integrable Hierarchies of KP type''
 nlin.si/0102010. 
 
\bibitem{ora} J. Balog, L. Feher , L. O'Raifeartaigh, P. Forgacs and  A. Wipf,  Ann. of Phys. 
{\bf 203}, (1990), 76; L. Feher , L. O'Raifeartaigh, P. Ruelle, I. Tsutsui and A. Wipf, \PR{222}{1992}{1}


 
\bibitem{annals}
J.F. Gomes,  G.M. Sotkov and A.H. Zimerman, ``Parafermionic reductions 
of WZNW model'',
hepth  9803234,  Ann. of Phys.{\bf 274}, (1999), 289-362


\bibitem{pousa}C.R. Fernandez-Pousa, M.V. Gallas, T.J. Hollowood and J.L. Miramontes,
\NPB{484}{1997}{609}; \NPB{499}{1997}{673};
C.R. Fernandez-Pousa and J.L. Miramontes,\NPB{518}{1998}{745}








 
\bibitem{dubna}J.F. Gomes, E.P. Gueuvoghlanian, G.M. Sotkov and A.H.
Zimerman, ``Classical Integrability of Non Abelian Affine Toda Models, 
to appear in the Proc. of the XXIII
International Colloquium on Group Theoretical Methods in
Physics, Ed. G. Pogosyan et. al., Dubna (2000), hepth/0010257



\bibitem{wit3}E. Witten, \PLB{86}{1979}{283}



\bibitem{polchinski}J. Polchinski, String Theory, Vol. I,   Cambridge Univ. Press (1998)



\bibitem{babelon}O. Babelon and D. Bernard, 
Int. J. Mod. Phys. {\bf A8}(1993)507


\bibitem{coleman}S. Coleman, \PRD{11}{1975}{2088}





\bibitem{aratnp} H. Aratyn, C.P. Constantinides, L.A. Ferreira, J.F. Gomes and
A.H. Zimerman, Nucl. Phys. {\bf B406}(1993),727 


\bibitem{math}A.G. Bueno, L.A. Ferreira and A.V. Razumov, ``Confinement and Soliton Solutions  in the 
 SL(3) Toda Model coupled to Matter Fields'',  hep-th/0105078 


\bibitem{fat}V.A. Fateev, \NPB{479}{1996}{594}

\bibitem{tsey} A. Tseytlin, \NPB{399}{1993}{601}; \NPB{411}{1994}{509}
\bibitem{fring}O.A. Castro-Alvaredo, A. Fring, C. Korff and J.L. Miramontes, \NPB{575}{2000}{535};
 O.A. Castro-Alvaredo and  A. Fring, \NPB{604}{2001}{367} 
\bibitem{millet} H. J. de Vega and J.M. Maillet, \PRD{28}{1983}{1441}

\end{thebibliography}
\end{document}